\documentclass[twocolumn,pre,twoside,groupedaddress,floatfix,showpacs]{revtex4}

\usepackage{amsmath}
\usepackage{amssymb}
\usepackage{graphicx}
\usepackage{dcolumn}
\usepackage{bm}
\SetSymbolFont{largesymbols}{bold}{OMX}{txex}{b}{n}
\usepackage{color}
\usepackage{enumitem}

\begin{document}

\title{Wetting behavior of a colloidal particle trapped at a composite
       liquid--vapor interface of a binary liquid mixture}

\author{Hyojeong Kim}
\email{hyojeong@is.mpg.de}
\author{Lothar Schimmele}
\author{S. Dietrich}
\affiliation
{
   Max-Planck-Institut f\"ur Intelligente Systeme, 
   Heisenbergstr.\ 3,
   70569 Stuttgart,
   Germany
}
\affiliation
{
   IV. Institut f\"ur Theoretische Physik,
   Universit\"at Stuttgart,
   Pfaffenwaldring 57,
   70569 Stuttgart,
   Germany
}

\date{\today}

\begin{abstract}
A partially miscible binary liquid mixture, composed of $A$ and $B$ particles, is considered theoretically under conditions for which a stable $A$-rich liquid phase
is in thermal equilibrium with the vapor phase. The $B$-rich liquid is metastable.
The liquids and the thermodynamic conditions are chosen such, that the interface
between the $A$-rich liquid and the vapor contains an intervening
wetting film of the $B$-rich phase. 
In order to obtain information about the large-scale fluid structure around a colloidal particle, 
which is trapped at such a composite liquid--vapor interface, three related and linked wetting phenomena
at planar liquid--vapor, wall--liquid, and wall--vapor interfaces are studied analytically,
using classical density functional theory in conjunction with the sharp-kink approximation
for the number density profiles of the $A$ and $B$ particles.
If in accordance with the so-called mixing rule the strength of the $A$--$B$ 
interaction is given by the geometric mean of the strengths of the $A$--$A$ and $B$--$B$ interactions,
and similarly the ratio between the wall--$A$ and the wall--$B$ interaction,
the scenario, in which the colloid is enclosed by a film of the $B$-rich liquid, can be excluded.
Up to six distinct wetting scenarios are possible,
if the above mixing rules for the fluid--wall and for the fluid--fluid interactions are relaxed.
The way the space of system parameters is divided into domains corresponding to the six scenarios, and which
of the domains actually appear, depends on the signs of the deviations from the mixing rule
prescriptions. Relevant domains, corresponding, e.g., to the scenario in which the colloid is 
enclosed by a film of the $B$-rich liquid, emerge,
if the ratio between the strengths of the wall--$A$ and the wall--$B$ interactions
is reduced as compared to the mixing rule prescription, or if the strength of the 
$A$--$B$ interaction is increased to values above the one from the mixing rule prescription.
The range, within which the contact angle may vary inside the various domains, is also studied.
\end{abstract}


\maketitle

\setcounter{secnumdepth}{1}
\section{\label{section:introduction}Introduction}
Interfaces involving fluids, i.e., fluid--fluid and fluid--wall (solid) interfaces, are very common and thus their study has received much interest for decades (see, e.g., Refs.~\cite{Evans1979,Stephan2019}). 
In this context wetting transitions are of particular interest~\cite{Cahn1977}. 
Numerous studies have been devoted to the classification of the wetting behavior at individual, planar fluid--fluid and fluid--solid interfaces, including binary liquid mixtures~\cite{Telodagama1983,Tarazona1983a,Tarazona1983b,Hadji1985,Dietrich1986,Dietrich1989,Getta1993,Mukherjee2020}. 
Wetting in more complicated surface geometries~\cite{Cheng1990,Napiorkowski1992,Dobbs1992,Osborn1995,Gil1997,Bieker1998,Rejmer1999}~and at chemically inhomogeneous surfaces~\cite{Koch1995,Rascon2001,Malijevsky2017,Pospisil2019}~have been extensively studied as well.

Wetting-induced or fluid-mediated, effective interactions between spherical particles~\cite{Bauer2000,Okamoto2013}~located inside a fluid are related topics, too. In the context of fluid interfaces, further studies are devoted to cylindrical particles approaching the interface between two coexisting liquid phases in binary liquid mixtures close to the critical point~\cite{Law2014}~or to colloidal particle located at the interface between two fluids~\cite{Bresme1998,Bresme1999}.
We also mention a practical example in which a meniscus acts as a capillary filter for colloidal particles. The thickness of the liquid film on the surface of a solid object determines the effectivity of the filter, as demonstrated in recent experimental studies~\cite{Sauret2019,Dincau2019}.

Despite numerous investigations concerning wetting of liquid--vapor and fluid--solid
interfaces in binary liquid mixtures, a number of seemingly simple questions remain unanswered. 
For instance, considering a partially miscible binary liquid mixture, 
composed of $A$ and $B$ particles, one may think of the following scenario:
A stable $A$-rich liquid phase in equilibrium and in contact with the vapor phase allows for the formation of 
a composite liquid--vapor interface containing an intervening film of the $B$-rich liquid phase, which is metastable in bulk. This scenario can be realized by properly selecting the liquids 
and by tuning the thermodynamic conditions.
For such a setup one can pose the question what kind of fluid structures emerge 
if such a composite liquid--vapor interface meets a solid wall, for instance the one provided
by a large colloidal particle trapped at such an interface. A film of the $B$-rich phase 
at the liquid--vapor interface for instance could surround the colloidal particle completely.
Alternatively, it could extend into the liquid $\alpha$ phase only, or only into the colloid--vapor interface, 
or it could disappear completely around the colloid.
Here we are not interested in the detailed, molecular structure in close vicinity of
the three-phase contact line; instead we focus on the large-scale structures.
Therefore, we can make use of studies for extended liquid--vapor and wall--fluid interfaces 
(see, e.g., Ref.~\cite{Dietrich1986})
and combine the results of those in order to develop a picture describing the entire scenario
around a colloid at a composite liquid--vapor interface.
In order to obtain analytical expressions we use a reduced version of classical 
density functional theory (DFT) in which only the long-ranged van der Waals type of interactions 
are taken into account explicitly and in which the number density profiles are assumed to vary only steplike between their corresponding bulk values, which is known as the so-called sharp-kink approximation (see, e.g., Ref.~\cite{Dietrich1986}).
In order to reduce the number of cases to be considered, we assume certain inequalities
for the number densities of the two components forming the three phases. They are
satisfied for typical partially miscible liquids at liquid--vapor coexistence. 
The analytical results allow us to find the domains in the space of system parameters
which correspond to the various conceivable wetting scenarios. 
We also analyze whether there are connections between these wetting domains and the 
contact angle, considered to be either smaller or larger than 90$^\mathrm{o}$.
The aim of the present study is also to provide a basis for a broader
understanding of capillarity induced interactions between colloidal particles
and eventually to design the self-assembly of colloidal particles at 
liquid--vapor interfaces.

\section{\label{section:model}Model}

We consider a region of the bulk phase diagram of a binary liquid mixture, composed of $A$ and $B$ particles,
in which the vapor phase ($\gamma$ phase) coexists with a stable $A$-rich liquid phase 
($\alpha$ phase), whereas the $B$-rich liquid phase ($\beta$ phase) is metastable. 
We focus on the special situation in which the interface between the coexisting phases 
(i.e., the $\alpha$--$\gamma$ interface)
is composite in the sense that between the $\alpha$ phase and the vapor a film of the $\beta$ phase intrudes. 
The thermodynamic state is taken to be only slightly off $\alpha$--$\beta$ coexistence, such that the
thickness of the $\beta$ film is larger than a few molecular diameters and the $\beta$ film can be treated
like a genuine $\beta$ phase.  

\begin{figure}[!t]
   \includegraphics[trim={0 1.5cm 0 0.5cm}, width=0.4\textwidth]{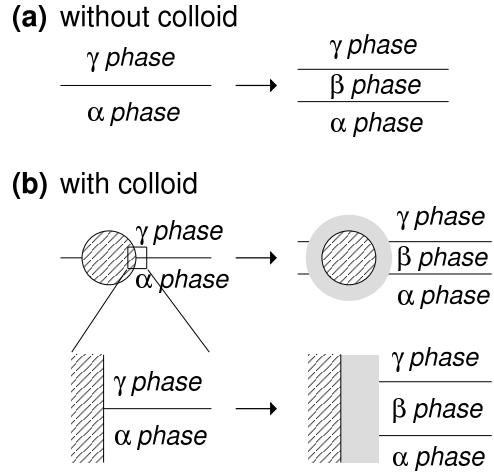}
   \caption{
      A binary liquid mixture exhibits three fluid bulk phases: vapor ($\gamma$), 
      an $A$-rich liquid ($\alpha$), and a $B$-rich liquid ($\beta$).
      (a) Liquid--vapor interface in a binary liquid mixture which is wetted by 
          the $\beta$ phase.
      (b) The composite liquid--vapor interface from (a) meets 
          the surface of a colloidal particle.
          In the gray area various interfacial structures are conceivable.
          The second row in (b) magnifies the structure within the square of
          the first row.
   }
   \label{fig:concept}
\end{figure}

Here, we analyze the kind of fluid structures which form if the composite
liquid--vapor interface meets a solid wall.
In the present study we address the simple situation in which the composite liquid--vapor interface
meets a \textit{planar} solid wall, and we explore which of the conceivable 
wall--liquid and wall--vapor structures
are compatible with the aforementioned composite liquid--vapor interface.
We consider the wall--$\alpha$-liquid interface, which is either wetted by a film 
of the $\beta$ phase or, alternatively, is a plain interface without any wetting film, the wall--vapor interface, which
is either wetted by a $\beta$ film or by an $\alpha$ film or is a plain interface without
a wetting film. The issue as to what happens to the $\beta$ film of the 
composite liquid--vapor interface once it meets the solid wall, can be resolved if one knows which
combination of interfacial structures is realized for a given case. 
The type of structure which emerges depends on the fluid--fluid and the fluid--wall interactions as well as on the thermodynamic state. These parameters can be varied
within certain limits imposed by the presupposed liquid--vapor interfacial structure. The results directly apply to the case in which the wall is
the curved surface of a colloid, provided the radius 
of the colloidal particle is sufficiently large so that curvature effects are negligible. 
The configurations are sketched in Fig.~\ref{fig:concept}.
For reasons of simplicity, there the liquid--vapor interface meets the wall at an angle of 90$^\mathrm{o}$; the actual
angle is given by Young's local contact angle.  

In order to rephrase the issue, it is our goal to find out which wall--vapor interfacial structure and which 
wall--liquid (wall--$\alpha$) interfacial structure are realized together for a certain fluid 
at given thermodynamic conditions and for a given wall, under the proposition that for the chosen fluid 
and the given thermodynamic conditions the $\alpha$--vapor interface is a composite one with an intervening film 
of phase $\beta$. Moreover, we want to identify various domains in the space of system parameters, 
each of which can be related to a particular combination of the structures of the three interfaces
involved.
In order to proceed we make use of previous results \cite{Dietrich1986} in which wetting of individual
planar interfaces has been studied based on a simplified version of classical density functional theory (DFT).
In these studies long-ranged van der Waals type of interactions are treated explicitly within a mean-field
approximation whereas other contributions to the free energy are treated within a local-density approximation. Furthermore,
the so-called sharp-kink approximation has been used according to which the number densities of the $A$ and $B$ particles
are considered to be piece-wise constant and to vary discontinuously, at the interface positions, between their respective
bulk values. The corresponding analytic expressions, which can be derived based on these
approximations, contain the interaction parameters as well as 
the equilibrium number densities of the $A$ and $B$ particles in the various phases. 
These equilibrium number densities follow from minimizing the bulk free energy with respect 
to the bulk number densities.
The equilibrium number densities depend on both the fluid--fluid interaction parameters as
well as on the thermodynamic state (i.e., the thermodynamic variables).  
We do not try to express the equilibrium number densities in terms of the thermodynamic variables 
and the fluid--fluid interaction parameters, as this would require to introduce a specific
expression for the free energy contribution, which is local in the densities.
Instead, we introduce variables which combine interaction parameters and densities, and we identify
various domains in the space of these parameters. In addition, we use knowledge and plausible
assumptions concerning certain inequalities between the two number densities characterizing each of the 
phases and inequalities between the number densities in different phases.

For the long-ranged part of the fluid--fluid interaction we choose the Lennard-Jones potential
$\tilde{w}_{ij}(\mathbf{r})=4\epsilon_{ij}
  \left[\left(a_{ij}/r\right)^{12}
             -\left(a_{ij}/r\right)^{6}\right]\, ,$
or rather, in order to avoid spurious singularities in certain expressions, a modified (shifted) 
version of it:
$\tilde{w}_{ij}(\mathbf{r})=4\epsilon_{ij}
                      \left[\left(\dfrac{a_{ij}}
                                  {r+a_{ij}}\right)^{12}
                           -\left(\dfrac{a_{ij}}
                                  {r+a_{ij}}\right)^{6}
                                                     \right]\, ,$
where $\epsilon_{ij}$ and $a_{ij}$ represent the interaction strengths and the length parameters
for the interaction between the $i$ and $j$ components, respectively.
For reasons of simplicity we assume that all length parameters are equal ($a_{AA}=a_{AB}=a_{BB}$).
Thus the parameters describing the fluid--fluid interaction are the three interaction strengths
$\epsilon_{AA}$, $\epsilon_{AB}$, and $\epsilon_{BB}$. The interaction strength $\epsilon_{AB}$ between unlike particles
is expressed in terms of those between the two sorts of like particles:
$\epsilon_{AB}=\xi_{f}\sqrt{\epsilon_{AA}\epsilon_{BB}}$ 
for the fluid--fluid interactions
\cite{Stephan2019,Dietrich1989,Delhommelle2001,Boda2008}. 
The case of $\xi_{f}=1$ is called the 
strict mixing rule \cite{Berthelot1889,Hansen1976,Allen1989}; 
simplified expressions for the dispersion forces give rise to the
strict mixing rule. Similar expressions can be put forward for the strengths of the interactions 
between a wall particle and a fluid particle of sort $A$ or $B$, respectively; i.e., 
$\epsilon_{\text{w}A}=\xi_{\text{w}A}\sqrt{\epsilon_{\text{ww}}\epsilon_{AA}}$ and
$\epsilon_{\text{w}B}=\xi_{\text{w}B}\sqrt{\epsilon_{\text{ww}}\epsilon_{BB}}$. 
Introducing $\xi_{\text{w}} = \xi_{\text{w}A}/\xi_{\text{w}B}$, the ratio between the wall--$A$ and the wall--$B$
interaction can be expressed as 
$\epsilon_{\text{w}A}/\epsilon_{\text{w}B} = \xi_{\text{w}}\sqrt{\epsilon_{AA}/\epsilon_{BB}}$.
Again, the case $\xi_{\text{w}} = 1$ is called the strict mixing rule for the
fluid--wall interaction.
We first assume that the strict mixing rules apply to both the fluid--fluid
and the fluid--wall interactions, which cuts in half the dimension of the parameter
space. We then identify domains in the space of reduced parameters such that
each domain represents a particular combination of the structures of the three interfaces involved.
Next, we relax the strict mixing rules and keep track of the consequences for the map of 
domains.
Before presenting in the next section the map of domains, 
in the next two subsections we provide the presently available results for the three different
interfaces and express them in a form which is suitable for our discussion.

\setcounter{secnumdepth}{2}

\subsection{\label{subsection:ff_wetting}
         Planar liquid--vapor interface}

In order to determine the conditions for which a planar 
$\alpha$--$\gamma$ interface is wetted by a film of the $\beta$ phase,
we introduce a wetting parameter $W_{\alpha\beta\gamma}$
defined as
\begin{equation}
   W_{\alpha\beta\gamma}:=\sigma_{\alpha\beta\gamma}-\sigma_{\alpha\gamma} \, ,
   \label{eq:wet_para1}
\end{equation}
where $\sigma_{\alpha\beta\gamma}$ is the surface free energy (surface tension) 
of a composite configuration in which an intervening wetting film of the $\beta$ phase occurs
at the $\alpha$--$\gamma$ interface; $\sigma_{\alpha\gamma}$ is the surface free 
energy (surface tension) of a plain configuration in which such a film is absent. 
If $W_{\alpha\beta\gamma}<0$, the stable configuration is the one in which the 
$\alpha$--$\gamma$ interface is wetted by a film of the $\beta$ phase. Otherwise, if
$W_{\alpha\beta\gamma}>0$, the liquid ($\alpha$ phase) and the vapor ($\gamma$ phase) 
are in direct spatial contact and the $\alpha$--$\gamma$ interface is a plain one without
an intruding $\beta$ film. 

We now follow the discussion in Ref.~\cite{Dietrich1986} and first separate
the grand canonical potential functional $\Omega$ into a bulk (b) and a surface (s) contribution:
\begin{equation}
   \Omega[\{\rho_{i}(\mathbf{r})\},T,\{\mu_i\}]
      =V\Omega_{\text{b}}+S\Omega_{\text{s}} \, ,
\end{equation} 
where $i=A\text{ and }B$ denote the two types of fluid particles and
$V$ and $S$ denote the volume and the surface area, respectively; $T$ is the temperature, and $\mu_i$ is the chemical potential of species $i$.
In the next step, the surface contribution to the grand canonical potential
for a composite configuration of the $\alpha$--$\gamma$ interface with an intruding film of
the $\beta$ phase can be written as 
\begin{equation}
   \Omega_{\text{s}}^{\alpha\beta\gamma}(l)
                     =l(\Omega^{\beta}-\Omega^{\gamma})
                      +\omega_{\alpha\beta\gamma}(l)
                      +\sigma_{\alpha\beta}
                      +\sigma_{\beta\gamma} \, .
   \label{eq:omega_surf}
\end{equation} 
The first term in Eq.~(\ref{eq:omega_surf}) is 
the bulk free energy needed to replace a slab of thickness $l$ 
of the $\gamma$ phase by the $\beta$ phase.
The second term in Eq.~(\ref{eq:omega_surf}) is 
the correction to the surface free energy due to the finite thickness $l$
of the slab. The terms $\sigma_{\alpha\beta}$ and $\sigma_{\beta\gamma}$
denote the surface tensions of plain interfaces
between the bulk phases $\alpha$, $\beta$ 
and $\beta$, $\gamma$, respectively.

In thermal equilibrium the thickness $l$ of the slab of the $\beta$ phase attains its equilibrium value
$l_{\alpha\beta\gamma}$. 
The surface tension of the composite configuration, i.e., the first term in Eq.~(\ref{eq:wet_para1}),
is determined in terms of this equilibrium configuration by
\begin{align}
\notag
   \sigma_{\alpha\beta\gamma}
   =&\min_{\{\rho_{i}(\mathbf{r})\}} 
          \Omega_{\text{s}}^{\alpha\beta\gamma}
                 [\{\rho_{i}(\mathbf{r})\},T,\{\mu_{i}\}]\\
\notag
   =&\min_{l}\Omega_{\text{s}}^{\alpha\beta\gamma}
                   (l,T,\{\mu_{i}\})\\
   =&\,\Omega_{\text{s}}^{\alpha\beta\gamma}
           (l_{\alpha\beta\gamma},T,\{\mu_{i}\})
   \label{eq:equil_cond1}
\end{align}
and 
\begin{equation}
   \dfrac{\partial{\Omega_{\text{s}}^{\alpha\beta\gamma}}(l)}
         {\partial{l}}\bigg{|}_{l=l_{\alpha\beta\gamma}}=0 \, .
   \label{eq:equil_cond2}
\end{equation} 
(Note that $\Omega_{\text{s}}^{\alpha\beta\gamma}[\{\rho_{i}(\mathbf{r})\}]$
and $\Omega_{\text{s}}^{\alpha\beta\gamma}(l)$
are distinct functionals and functions, respectively, as indicated by their arguments.)
It should be stressed that $l_{\alpha\beta\gamma}$ is a solution of
Eq.~(\ref{eq:equil_cond2}) corresponding to a minimum of the free energy. Furthermore,
the subdivision of the free energy used in Eq.~(\ref{eq:omega_surf})
is applicable only if $l_{\alpha\beta\gamma}$ is sufficiently large such
that the $\beta$ film can be treated like a piece of genuine $\beta$ phase. 
By inserting Eqs.~(\ref{eq:omega_surf}) and~(\ref{eq:equil_cond1}) 
into Eq.~(\ref{eq:wet_para1}),
the wetting parameter defined in Eq.~(\ref{eq:wet_para1}) for
the $\alpha$--$\gamma$ interface can be expressed as
\begin{align}
\notag
W_{\alpha\beta\gamma}=&\,\, l_{\alpha\beta\gamma}(\Omega^{\beta}-\Omega^{\gamma})
                        +\omega_{\alpha\beta\gamma}(l_{\alpha\beta\gamma})\\
                     &+\,\sigma_{\alpha\beta}
                        +\sigma_{\beta\gamma}
                        -\sigma_{\alpha\gamma} \, .
   \label{eq:wet_para2}
\end{align}
Further, using Eq.~(\ref{eq:omega_surf}) one can rewrite
Eq.~(\ref{eq:equil_cond2}) as  
\begin{equation}
   \Omega^{\beta}-\Omega^{\gamma}
   =-\dfrac{\partial{\omega_{\alpha\beta\gamma}(l)}}
           {\partial{l}}\bigg{|}_{l=l_{\alpha\beta\gamma}} \, .
   \label{eq:delta_omega}
\end{equation}
From here onwards the sharp-kink approximation is used. Within this
approximation the following expressions are obtained for
$\omega_{\alpha\beta\gamma}(l)$ and the interfacial
tensions of the various plain interfaces:
\begin{align}
   \omega_{\alpha\beta\gamma}(l)
     &=-\sum_{i,j}(\rho_{i,\alpha}-\rho_{i,\beta})
                  (\rho_{j,\beta}-\rho_{j,\gamma})
                  \int_{l}^{\infty}dy \, t_{ij}(y) \, ,
   \label{eq:omega_abg}\\
   \sigma_{\alpha\beta}
     &=-\dfrac{1}{2}\sum_{i,j}(\rho_{i,\alpha}-\rho_{i,\beta})
                             (\rho_{j,\alpha}-\rho_{j,\beta})
                             \int_{0}^{\infty}dy \, t_{ij}(y) \, ,
   \label{eq:sigma_ab}\\
   \sigma_{\beta\gamma}
     &=-\dfrac{1}{2}\sum_{i,j}(\rho_{i,\beta}-\rho_{i,\gamma})
                             (\rho_{j,\beta}-\rho_{j,\gamma})
                             \int_{0}^{\infty}dy \, t_{ij}(y) \, ,
   \label{eq:sigma_bg}
\end{align} 
and
\begin{equation}
   \sigma_{\alpha\gamma}
    =-\dfrac{1}{2}\sum_{i,j}(\rho_{i,\alpha}-\rho_{i,\gamma})
                            (\rho_{j,\alpha}-\rho_{j,\gamma})
                            \int_{0}^{\infty}dy \, t_{ij}(y) \, ,
   \label{eq:sigma_ag}
\end{equation} 
respectively, where
\begin{equation*}
   t_{ij}(y)=\int_{y}^{\infty} dx\int d^{2}\mathbf{r}_{||}\,
             \tilde{w}_{ij}
             \left[(\mathbf{r}_{||}^{2}+x^{2})
                   ^{\frac{1}{2}}\right] \, .
\end{equation*}
Here, $i$ and $j$ denote the species $A$ and $B$ forming the binary liquid mixture
and $\rho_{i,\zeta}$ is the number density of particles of species $i$ in the bulk phase
$\zeta=\alpha,\,\beta,\,\text{and }\gamma$.

\textit{Regular} analytic expressions are obtained by replacing in Eqs.~(\ref{eq:sigma_ab})-(\ref{eq:sigma_ag})
-- for the long-ranged part $\tilde{w}_{ij}(\mathbf{r})$
of the fluid--fluid interaction -- the simple Lennard-Jones potential
by its shifted version:
\begin{equation}
   \tilde{w}_{ij}(\mathbf{r})=4\epsilon_{ij}
                      \left[\left(\dfrac{a_{ij}}
                                  {r+a_{ij}}\right)^{12}
                           -\left(\dfrac{a_{ij}}
                                  {r+a_{ij}}\right)^{6}
                                                     \right] \, ,
   \label{eq:modi1_lj}
\end{equation}  
where $\epsilon_{ij}$ and $a_{ij}$ denote the strength of the interaction
energy and the length parameters for the interactions between the components $i$ and $j$.
In order to evaluate Eq.~(\ref{eq:omega_abg}) we use the additional assumption
$l\gg a_{ij}$, which is justified by the above requirement of a sufficiently 
thick $\beta$ film.
Eventually the following results are obtained:
\begin{align}
\notag
   \omega_{\alpha\beta\gamma}(l)
     =\pi\sum_{i,j}\bigg\{&\epsilon_{ij}a_{ij}^{4}
          (\rho_{i,\alpha}-\rho_{i,\beta})
          (\rho_{j,\beta}-\rho_{j,\gamma})\\
          &\times\left[\dfrac{1}{3}\left(\dfrac{a_{ij}}{l}\right)^{2}
               -\dfrac{4}{5}\left(\dfrac{a_{ij}}{l}\right)^{3}
                                                      \right]\bigg\}\, ,
   \label{eq:omega_abg2}
\end{align}
\begin{align}
   \sigma_{\alpha\beta}
     &=\dfrac{13}{132}\pi\sum_{i,j}\epsilon_{ij}a_{ij}^{4}
                             (\rho_{i,\alpha}-\rho_{i,\beta})
                             (\rho_{j,\alpha}-\rho_{j,\beta}) \, ,
   \label{eq:sigma_ab2}\\
   \sigma_{\beta\gamma}
     &=\dfrac{13}{132}\pi\sum_{i,j}\epsilon_{ij}a_{ij}^{4}
                             (\rho_{i,\beta}-\rho_{i,\gamma})
                             (\rho_{j,\beta}-\rho_{j,\gamma}) \, ,
   \label{eq:sigma_bg2}
\end{align} 
and
\begin{equation}
   \sigma_{\alpha\gamma}
    =\dfrac{13}{132}\pi\sum_{i,j}\epsilon_{ij}a_{ij}^{4}
                            (\rho_{i,\alpha}-\rho_{i,\gamma})
                            (\rho_{j,\alpha}-\rho_{j,\gamma}) \, .
   \label{eq:sigma_ag2}
\end{equation} 
Due to Eq.~(\ref{eq:omega_abg2}), 
we can rewrite Eq.~(\ref{eq:delta_omega}) as
\begin{align}
\notag
   \Omega^{\beta}-\Omega^{\gamma}
     =\pi\sum_{i,j}\bigg\{&\epsilon_{ij}a_{ij}^{3}
          (\rho_{i,\alpha}-\rho_{i,\beta})
          (\rho_{j,\beta}-\rho_{j,\gamma})\\
      &\times\left[\dfrac{2}{3}\left(\dfrac{a_{ij}}{l_{\alpha\beta\gamma}}\right)^{3}
          -\dfrac{12}{5}\left(\dfrac{a_{ij}}{l_{\alpha\beta\gamma}}\right)^{4}
                                                      \right]\bigg\} \, .
   \label{eq:delta_omega2}
\end{align} 
Using Eqs.~(\ref{eq:omega_abg2})-(\ref{eq:delta_omega2}), 
Eq.~(\ref{eq:wet_para2}) for $W_{\alpha\beta\gamma}$ 
can be rewritten as
\begin{equation} 
   W_{\alpha\beta\gamma}=\pi\sum_{i,j}\epsilon_{ij}a_{ij}^{4}
                           (\rho_{i,\beta}-\rho_{i,\alpha})
                           (\rho_{j,\beta}-\rho_{j,\gamma})
                           F_{ij}(l_{\alpha\beta\gamma}) \, ,
   \label{eq:wet_para3}
\end{equation}
where
\begin{equation}
   F_{ij}(l)=\dfrac{13}{66}
            -\left(\dfrac{a_{ij}}{l}\right)^{2}
            +\dfrac{16}{5}
             \left(\dfrac{a_{ij}}{l}\right)^{3} \, ,
   \label{eq:f1_ij}
\end{equation}   
and $l_{\alpha\beta\gamma}$ is the equilibrium thickness of 
the $\beta$ film intruding the $\alpha$--$\gamma$ interface. 

If the components $A$ and $B$ of the binary liquid mixture have 
equal molecular radii, the length parameters in Eq.~(\ref{eq:modi1_lj})
are all equal, i.e., $a_{AA}=a_{AB}=a_{BB}$.
In this case Eq.~(\ref{eq:wet_para3}) reduces to
\begin{equation}
   W_{\alpha\beta\gamma}=\pi {a}_{AA}^{4}
                           S_{\alpha\beta\gamma}
                           F_{AA}(l_{\alpha\beta\gamma}) \, ,
   \label{eq:wet_para4}
\end{equation}
where 
\begin{align}
\notag
      S_{\alpha\beta\gamma}
            =&\,\,(\rho_{A,\beta}-\rho_{A,\alpha})
              (\rho_{A,\beta}-\rho_{A,\gamma})\epsilon_{AA}\\
\notag
             &\,+\,(\rho_{A,\beta}-\rho_{A,\alpha})
              (\rho_{B,\beta}-\rho_{B,\gamma})\epsilon_{AB}\\
\notag
             &\,+\,(\rho_{B,\beta}-\rho_{B,\alpha})
              (\rho_{A,\beta}-\rho_{A,\gamma})\epsilon_{BA}\\
             &\,+\,(\rho_{B,\beta}-\rho_{B,\alpha})
              (\rho_{B,\beta}-\rho_{B,\gamma})\epsilon_{BB} \, .
   \label{eq:s_abg}
\end{align}
The sign of $F_{AA}(l_{\alpha\beta\gamma})$ in Eq.~(\ref{eq:wet_para4}) 
is always positive if $l_{\alpha\beta\gamma}>0$ (see Eq.~(\ref{eq:f1_ij}),
actually the relation $l_{\alpha\beta\gamma} \gg a$ should hold, otherwise there is no composite interface).
The sign of $W_{\alpha\beta\gamma}$ is therefore entirely determined by
the sign of $S_{\alpha\beta\gamma}$, 
which depends on the bulk number densities of the two species in the
various phases and on the three interaction strengths.
If $S_{\alpha\beta\gamma}<0$, the stable configuration is a composite
$\alpha$--$\gamma$ interface with an intruding $\beta$ film between
the $\alpha$ and the $\gamma$ phase.  
If $S_{\alpha\beta\gamma}>0$, 
the stable configuration is
a plain $\alpha$--$\gamma$ interface.

In order to obtain a statement concerning the sign of $S_{\alpha\beta\gamma}$, we
inspect the various contributions in Eq.~(\ref{eq:s_abg}). First,
all interaction strengths $\epsilon_{AA}$, $\epsilon_{AB}$, and $\epsilon_{BB}$
are positive. The prefactors may be positive or negative depending on
the relations between the various number densities. 
Since the $\beta$ phase is a liquid phase, the $\gamma$ phase is
vapor, and the number densities in the liquid phases are much higher than in
the vapor (at least away from the critical point), and $\rho_{B,\beta}-\rho_{B,\gamma}$ is always positive. 
Typically, $\rho_{A,\beta}-\rho_{A,\gamma}$ should be positive as well.
On the other hand, because by definition the $\beta$ phase is a $B$-rich
phase, it cannot be excluded that the component $A$ is very diluted
in the $\beta$ phase but present in a much higher concentration in
the vapor, such that $\rho_{A,\beta}-\rho_{A,\gamma}$ could be
negative. We exclude such exceptional cases from our discussion and
always assume in the following that $\rho_{A,\beta}-\rho_{A,\gamma}$
is positive.

Next, we consider the sign of $\rho_{A,\beta}-\rho_{A,\alpha}$ and
$\rho_{B,\beta}-\rho_{B,\alpha}$.
The possibility that $\rho_{A,\beta}-\rho_{A,\alpha} > 0$ and
$\rho_{B,\beta}-\rho_{B,\alpha} > 0$ would result in a positive sign
of $S_{\alpha\beta\gamma}$, irrespective of the values of the
interaction strengths. This case is not of interest to our present study,
because it excludes the occurrence of an intervening $\beta$ film
at the liquid--vapor interface. 
In the following we consider the case $\rho_{A,\beta}-\rho_{A,\alpha} < 0$
and $\rho_{B,\beta}-\rho_{B,\alpha} > 0$. In this case the sign of 
$S_{\alpha\beta\gamma}$ depends on the values of the interaction strengths
and the magnitude of the density differences. 
A case in which both $\rho_{A,\beta}-\rho_{A,\alpha} < 0$ and
$\rho_{B,\beta}-\rho_{B,\alpha} < 0$, which would always lead to a
negative sign of $S_{\alpha\beta\gamma}$, we consider as an untypical
case for a liquid--liquid mixture. In this case 
$\rho_{B,\beta} < \rho_{B,\alpha}$, i.e., the number density of the $B$
particles in the $B$-rich $\beta$ phase would be smaller than the number
density of the $B$ particles in the $A$-rich $\alpha$ phase. 
This would require that the total number density in the 
$\alpha$ phase is substantially higher than the one in the $\beta$
phase and that the $\beta$ phase is only marginally rich in $B$ particles
and the $\alpha$ phase only marginally rich in $A$ particles.
The remaining case, $\rho_{A,\beta}-\rho_{A,\alpha} >0$ and
$\rho_{B,\beta}-\rho_{B,\alpha} <0$, is not possible.
This can be seen as follows. The definition of the $A$-rich $\alpha$ phase
implies $\rho_{A,\alpha}>\rho_{B,\alpha}$. Next, we use the first of the
two conditions, i.e., $\rho_{A,\beta}-\rho_{A,\alpha}>0$, which leads to
the sequence $\rho_{A,\beta}>\rho_{A,\alpha}>\rho_{B,\alpha}$ of inequalities.
Finally, using the definition of the $B$-rich $\beta$ phase, i.e., 
$\rho_{B,\beta}>\rho_{A,\beta}$, one obtains 
$\rho_{B,\beta}>\rho_{A,\beta}>\rho_{A,\alpha}>\rho_{B,\alpha}$, i.e., 
$\rho_{B,\beta}-\rho_{B,\alpha}>0$, which is in contradiction to the second
of the two conditions, which means that both conditions cannot be satisfied
together. To conclude the above discussion, from here onwards the following inequalities
between the various number densities are assumed:
\begin{align}
\notag
   & \rho_{A,\alpha} > \rho_{A,\beta} \, , \quad 
     \rho_{B,\alpha} < \rho_{B,\beta} \, , \quad \\
   & \rho_{A,\beta} > \rho_{A,\gamma} \, , \quad 
     \mathrm{and} \quad \rho_{B,\beta} > \rho_{B,\gamma} \, .
   \label{eq:rel_densities}
\end{align}
  
Now, given the inequalities in Eq.~(\ref{eq:rel_densities}), the sign of $S_{\alpha\beta\gamma}$
is studied.   
First, the interaction parameter $\epsilon_{AB}$ between unlike particles is expressed 
in terms of the corresponding ones between like particles,
$\epsilon_{AA}$ and $\epsilon_{BB}$,
as 
\begin{equation}
   \epsilon_{AB}=\xi_{f}\sqrt{\epsilon_{AA}\epsilon_{BB}} \, ,
\label{eq:eps_AB}
\end{equation}
with $\xi_{f}>0$.
Next we introduce the dimensionless variable 
\begin{equation}
 X=\dfrac{\rho_{B,\beta}}{\rho_{A,\beta}}
   \sqrt{\dfrac{\epsilon_{BB}}{\epsilon_{AA}}} \, ,  
   \label{eq:X}
\end{equation}
which characterizes the relative strengths of the $A$--$A$ and the $B$--$B$
interactions, weighted according to the abundance of the two species in the
$\beta$ phase.
$X$ is always positive. 
Using Eq.~(\ref{eq:eps_AB}) and the dimensionless variable $X$, 
Eq.~(\ref{eq:s_abg}) can be expressed as
\begin{align}
   \notag
      S_{\alpha\beta\gamma}
       =\epsilon_{AA}\rho_{A,\beta}^{2}
         \bigg[&
          \left(1-\dfrac{\rho_{B,\alpha}}{\rho_{B,\beta}}\right)
          \left(1-\dfrac{\rho_{B,\gamma}}{\rho_{B,\beta}}\right)X^2\\
\notag
         &+\left(1-\dfrac{\rho_{B,\alpha}}{\rho_{B,\beta}}\right)
          \left(1-\dfrac{\rho_{A,\gamma}}{\rho_{A,\beta}}\right)
          \xi_f X\\
\notag
         &+\left(1-\dfrac{\rho_{A,\alpha}}{\rho_{A,\beta}}\right)
          \left(1-\dfrac{\rho_{B,\gamma}}{\rho_{B,\beta}}\right)
          \xi_f X\\
         &+\left(1-\dfrac{\rho_{A,\alpha}}{\rho_{A,\beta}}\right)
          \left(1-\dfrac{\rho_{A,\gamma}}{\rho_{A,\beta}}\right)
         \bigg] \, .
   \label{eq:s_abg2}
\end{align}
Based on Eq.~(\ref{eq:s_abg2}), the range of $X$ is determined
for which $S_{\alpha\beta\gamma}<0$ (see Appendix~\ref{sec:solve_s_ff}). 
Equivalently one can state that at a planar $\alpha$--$\gamma$ interface 
a $\beta$ film can occur if $X$ is in the range
\begin{align}
\notag
   0<X<&\,\sqrt{\left[\dfrac{\xi_{f}}{2}
                   \left(D_{\alpha\beta}
                   +D_{\beta\gamma}\right)\right]^{2}
           -D_{\alpha\beta}D_{\beta\gamma}}\\
       &+ \dfrac{\xi_{f}}{2}
        \left(D_{\alpha\beta}+D_{\beta\gamma}\right) \, ,
   \label{eq:s_abg_rng}
\end{align}
where 
\begin{equation*}
D_{\alpha\beta}={\left(\dfrac{\rho_{A,\alpha}}
                              {\rho_{A,\beta}}-1\right)}
                 \bigg/
                 {\left(1-\dfrac{\rho_{B,\alpha}}
                                {\rho_{B,\beta}}\right)}
\end{equation*} 
and 
\begin{equation*}
D_{\beta\gamma}={\left(1-\dfrac{\rho_{A,\gamma}}
                                {\rho_{A,\beta}}\right)}
                 \bigg/
                 {\left(\dfrac{\rho_{B,\gamma}}
                              {\rho_{B,\beta}}-1\right)} \, .
\end{equation*}
Otherwise, given the inequalities in Eq.~(\ref{eq:rel_densities}), 
no $\beta$ film can occur at the planar $\alpha$--$\gamma$ interface. 
In the case that the strict mixing rule applies ($\xi_{f} = 1$), 
the condition in Eq.~(\ref{eq:s_abg_rng}) for the occurrence of a $\beta$ film
reduces to (note that the inequalities in Eq.~(\ref{eq:rel_densities}) imply
$D_{\alpha\beta}>0$ and $D_{\beta\gamma}<0$)
$$ 0 <X < D_{\alpha\beta} \, . $$

\subsection{\label{subsection:wf_wetting} 
            Planar wall--fluid interfaces}

In analogy to Eq.~(\ref{eq:wet_para1}) we introduce three additional wetting parameters, 
the signs of which determine whether a configuration with an intruding wetting film at 
the wall--$\alpha$ or the wall--$\gamma$ interface, respectively, has
a lower free energy than the corresponding one without such a wetting film.
The configurations considered are a 
$\beta$ film wetting the wall--$\alpha$ interface, 
a $\beta$ film wetting the wall--$\gamma$ interface, 
and an $\alpha$ film wetting the wall--$\gamma$ interface.
The corresponding three wetting parameters 
$W_{\text{w}\beta\alpha}$, 
$W_{\text{w}\beta\gamma}$, 
and $W_{\text{w}\alpha\gamma}$ are given by
\begin{align}
   W_{\text{w}\beta\alpha}
                &=\sigma_{\text{w}\beta\alpha}
                -\sigma_{\text{w}\alpha} \, ,
   \label{eq:w_wba}\\
   W_{\text{w}\beta\gamma}
                &=\sigma_{\text{w}\beta\gamma}
                -\sigma_{\text{w}\gamma} \, , 
   \label{eq:w_wbg}
\end{align}
and
\begin{align}
   W_{\text{w}\alpha\gamma}
                &=\sigma_{\text{w}\alpha\gamma}
                -\sigma_{\text{w}\gamma} \, .
   \label{eq:w_wag}
\end{align}
Here, w represents the wall;
$\sigma_{\text{w}\beta\alpha}$ 
is the surface tension (surface free energy) of 
the wall--$\alpha$ interface which is wetted by a $\beta$ film;
$\sigma_{\text{w}\beta\gamma}$
and $\sigma_{\text{w}\alpha\gamma}$
are the surface tensions (surface free energies) of 
the wall--$\gamma$ interface which is wetted by 
a $\beta$ or an $\alpha$ film, respectively;
$\sigma_{\text{w}\alpha}$
and $\sigma_{\text{w}\gamma}$
are the surface tensions (surface free energies) of 
the wall--$\alpha$ and the wall--$\gamma$ interface, 
respectively, without any intruding wetting film.
If $W_{\text{w}\beta\alpha}<0$,  
the wall--$\alpha$ interface is wetted by the $\beta$ phase.
If $W_{\text{w}\beta\gamma}<0$, it is more favorable to
have a film of the $\beta$ phase at the wall--$\gamma$ interface 
than to have a direct wall--$\gamma$ contact without an
intruding wetting film. 
If $W_{\text{w}\alpha\gamma}<0$, a wetting film of the $\alpha$ phase
at the wall--$\gamma$ interface is more favorable than a direct wall--$\gamma$ 
contact. Additional considerations might be necessary in order to
decide whether wetting of a wall--$\gamma$ interface by a film of
the $\beta$ phase or by a film of the $\alpha$ phase renders the more favorable
configuration. 

By using the shifted Lennard-Jones potential 
(see Eq.~(\ref{eq:modi1_lj})) for the 
fluid--fluid interaction
$\tilde{w}_{ij}(\mathbf{r})$, 
and also for the fluid--wall interaction potential
$\tilde{v}_{i}(\mathbf{r})$, 
\begin{equation}
   \tilde{v}_{i}(\mathbf{r})
   =4\epsilon_{\text{w}i}
    \left[\left(\dfrac{a_{\text{w}i}}
                     {r+a_{\text{w}i}}\right)^{12}
         -\left(\dfrac{a_{\text{w}i}}
                     {r+a_{\text{w}i}}\right)^{6}\right] \, ,
   \label{eq:modi2_lj}
\end{equation}
with the length parameters $a_{\text{w}i}$ and the
strengths $\epsilon_{\text{w}i}$ of the wall--$i$ interactions,
the above wetting parameters 
can be expressed as
(see Appendix~\ref{sec:wf_w_para})
\begin{align}
\notag
   W_{\text{w}\beta\alpha}
                =\pi\sum_{i,j}(\rho_{i,\beta}
                               -\rho_{i,\alpha})
                  \big[&\,\epsilon_{ij}a_{ij}^{4}\rho_{j,\beta}
                        F_{ij}(l_{\text{w}\beta\alpha})\\
                 &\,-\,\delta_{ij}\epsilon_{\text{w}i}
                  a_{\text{w}i}^{4}\rho_{\text{w}}
                  F_{\text{w}i}(l_{\text{w}\beta\alpha})
                                                      \big] \, ,
   \label{eq:w_wba2}\\
\notag
   W_{\text{w}\beta\gamma}
                 =\pi\sum_{i,j}(\rho_{i,\beta}
                                -\rho_{i,\gamma})
                   \big[&\,\epsilon_{ij}a_{ij}^{4}\rho_{j,\beta}
                         F_{ij}(l_{\text{w}\beta\gamma})\\
                  &\,-\,\delta_{ij}\epsilon_{\text{w}i}
                   a_{\text{w}i}^{4}\rho_{\text{w}}
                   F_{\text{w}i}(l_{\text{w}\beta\gamma})
                                                      \big] \, ,
   \label{eq:w_wbg2}
\end{align}
and
\begin{align}
\notag
   W_{\text{w}\alpha\gamma}
                 =\pi\sum_{i,j}(\rho_{i,\alpha}
                                -\rho_{i,\gamma})
                   \big[&\,\epsilon_{ij}a_{ij}^{4}\rho_{j,\alpha}
                         \hat{F}_{ij}(l_{\text{w}\alpha\gamma})\\
                  &\,-\,\delta_{ij}\epsilon_{\text{w}i}
                   a_{\text{w}i}^{4}\rho_{\text{w}}
                   \hat{F}_{\text{w}i}(l_{\text{w}\alpha\gamma})
                                                      \big] \, ,
   \label{eq:w_wag2}
\end{align}
respectively,
where $i$, $j$ represent the fluid components $A$ and $B$, $\delta_{ij}$ is the 
Kronecker symbol, $F_{ij}(l)$ is defined in Eq.~(\ref{eq:f1_ij}),
and $F_{\text{w}i}(l)$ is defined by the same equation 
but with the length parameters $a_{\text{w}i}$.
The functions $\hat{F}_{ij}(l)$ and $\hat{F}_{\text{w}i}(l)$ in Eq.~(\ref{eq:w_wag2}) are
defined by 
\begin{equation}
   \hat{F}_{ij}(l)=\dfrac{13}{66}
                -\dfrac{1}{3}
                 \left(\dfrac{a_{ij}}{l}\right)^{2}
                +\dfrac{4}{5}
                 \left(\dfrac{a_{ij}}{l}\right)^{3} \, ;
   \label{eq:f2_ij}
\end{equation} 
the functional form of $\hat{F}_{\text{w}i}(l)$ is the same but with the length parameters $a_{\text{w}i}$.
Here 
$l_{\text{w}\beta\alpha}$ is the equilibrium thickness of 
the intruding $\beta$ film 
at the wall--$\alpha$ interface 
and $l_{\text{w}\beta\gamma}$
($l_{\text{w}\alpha\gamma}$) is the equilibrium thickness of 
the intruding $\beta$ ($\alpha$) film 
at the wall--$\gamma$ interface. (We only consider films with thicknesses much
larger than the length parameters of the interactions.) 

If the length parameters of all interactions, i.e., the ones between the components $A$ and $B$ of the binary liquid mixture and those between the two components and the wall, are all equal, i.e.,
$a_{AA}=a_{AB}=a_{BB}
=a_{\text{w}A}=a_{\text{w}B}$, 
Eqs.~(\ref{eq:w_wba2})-(\ref{eq:w_wag2}) reduce to
\begin{align}
   W_{\text{w}\beta\alpha}
   &=\pi{a}_{AA}^{4}S_{\text{w}\beta\alpha}
     F_{AA}(l_{\text{w}\beta\alpha}) \, ,
   \label{eq:w_wba3}\\
   W_{\text{w}\beta\gamma}
   &=\pi{a}_{AA}^{4}S_{\text{w}\beta\gamma}
     F_{AA}(l_{\text{w}\beta\gamma}) \, ,
   \label{eq:w_wbg3}
\end{align}
and
\begin{align}
   W_{\text{w}\alpha\gamma}
   &=\pi{a}_{AA}^{4}S_{\text{w}\alpha\gamma}
    \hat{F}_{AA}(l_{\text{w}\alpha\gamma}) \, ,
  \label{eq:w_wag3}
\end{align}
where 
\begin{align}
\notag
   S_{\text{w}\beta\alpha}
                =&\,\,(\rho_{A,\beta}-\rho_{A,\alpha})
                  (\epsilon_{AA}\rho_{A,\beta}
                  -\epsilon_{\text{w}A}\rho_{\text{w}})\\
\notag
                 &\,+\,(\rho_{A,\beta}-\rho_{A,\alpha})
                  (\epsilon_{AB}\rho_{B,\beta})\\
\notag
                 &\,+\,(\rho_{B,\beta}-\rho_{B,\alpha})
                  (\epsilon_{BA}\rho_{A,\beta})\\
                 &\,+\,(\rho_{B,\beta}-\rho_{B,\alpha})
                  (\epsilon_{BB}\rho_{B,\beta}
                  -\epsilon_{\text{w}B}\rho_{\text{w}}) \, ,
   \label{eq:s_wba}\\
\notag
   S_{\text{w}\beta\gamma}
                 =&\,\,(\rho_{A,\beta}-\rho_{A,\gamma})
                   (\epsilon_{AA}\rho_{A,\beta}
                   -\epsilon_{\text{w}A}\rho_{\text{w}})\\
\notag
                  &\,+\,(\rho_{A,\beta}-\rho_{A,\gamma})
                   (\epsilon_{AB}\rho_{B,\beta})\\
\notag
                  &\,+\,(\rho_{B,\beta}-\rho_{B,\gamma})
                   (\epsilon_{BA}\rho_{A,\beta})\\
                  &\,+\,(\rho_{B,\beta}-\rho_{B,\gamma})
                   (\epsilon_{BB}\rho_{B,\beta}
                   -\epsilon_{\text{w}B}\rho_{\text{w}}) \, ,
   \label{eq:s_wbg}
\end{align}
and 
\begin{align}
\notag
   S_{\text{w}\alpha\gamma}
                 =&\,\,(\rho_{A,\alpha}-\rho_{A,\gamma})
                   (\epsilon_{AA}\rho_{A,\alpha}
                   -\epsilon_{\text{w}A}\rho_{\text{w}})\\
\notag
                  &\,+\,(\rho_{A,\alpha}-\rho_{A,\gamma})
                   (\epsilon_{AB}\rho_{B,\alpha})\\
\notag
                  &\,+\,(\rho_{B,\alpha}-\rho_{B,\gamma})
                   (\epsilon_{BA}\rho_{A,\alpha})\\
                  &\,+\,(\rho_{B,\alpha}-\rho_{B,\gamma})
                   (\epsilon_{BB}\rho_{B,\alpha}
                   -\epsilon_{\text{w}B}\rho_{\text{w}}) \, .
   \label{eq:s_wag}
\end{align}

The signs of the various wetting parameters
in Eqs.~(\ref{eq:w_wba3})-(\ref{eq:w_wag3}) and thus the structures of the
wall--$\alpha$ and the wall--vapor interfaces are determined by the
signs of the various quantities $S$ given 
by Eqs.~(\ref{eq:s_wba})-(\ref{eq:s_wag}),
because $F_{AA}$ and $\hat{F}_{AA}$ are positive within the
range of film thicknesses of interest (see Eqs.~(\ref{eq:f1_ij}) and~(\ref{eq:f2_ij})). 
The signs of the various quantities $S$ depend on various differences between 
bulk densities
and on the strengths of the interactions.
As discussed above we only consider binary liquid mixtures and conditions such
that the inequalities in Eq.~(\ref{eq:rel_densities})
between the number densities of the
two species in different phases hold.
In order to identify the regions in the parameter space corresponding to a negative or 
a positive sign of the various wetting parameters we use the notation
already introduced in the description of our model.
We write $\epsilon_{AB}=\xi_{f}\sqrt{\epsilon_{AA}\epsilon_{BB}}$,
$\epsilon_{\text{w}A}=\xi_{\text{w}A}\sqrt{\epsilon_{\text{ww}}\epsilon_{AA}}$, and
$\epsilon_{\text{w}B}=\xi_{\text{w}B}\sqrt{\epsilon_{\text{ww}}\epsilon_{BB}}$.
Introducing $\xi_{\text{w}} = \xi_{\text{w}A}/\xi_{\text{w}B}$ the ratio 
between the wall--$A$ and the wall--$B$ interaction can be expressed as
$\epsilon_{\text{w}A}/\epsilon_{\text{w}B} = \xi_{\text{w}}\sqrt{\epsilon_{AA}/\epsilon_{BB}}$.
The case $\xi_{f}=1$ ($\xi_{\text{w}} = 1$) is called the strict fluid--fluid (fluid--wall) mixing rule.

We now consider the reduced space 
spanned by the two dimensionless variables
$X=(\rho_{B,\beta}/\rho_{A,\beta})
     \sqrt{\epsilon_{BB}/\epsilon_{AA}}$
(Eq.~(\ref{eq:X})) and
\begin{equation}
 Y=\,\dfrac{\rho_{\text{w}}}{\rho_{A,\beta}}
     \dfrac{\epsilon_{\text{w}A}}{\epsilon_{AA}}  \, ,
\label{eq:Y}
\end{equation} 
characterizing the parameter space of the system; one has $X>0$ and $Y>0$. 
$X$ characterizes the relative strengths of the $A$--$A$ and
the $B$--$B$ interactions in the $\beta$ phase, $Y$ gives the relative strengths of 
the wall--$A$ interaction and the $A$--$A$ interaction in the $\beta$ phase.
The two additional parameters
$\xi_{f}$ and $\xi_{\text{w}}$ eventually determine how the reduced parameter space
($X$,$Y$) is subdivided into various `wetting domains'. 
We divide the ($X$,$Y$) parameter space, for each wall--fluid interface
separately, into regions within which the wall is wetted by an intruding
phase (wet state) and regions within which there is no intruding phase (non-wet state) (see Appendix~\ref{sec:solve_s_wf}).

The planar wall--$\alpha$ interface is wetted by a $\beta$ film
($S_{\text{w}\beta\alpha}<0$) if
\begin{align}
   \notag
   0<X<&\,\xi_{\text{w}}D_{\alpha\beta} \quad \mathrm{and} \quad\\
\notag
   0<Y<&\,X+1+\left(\xi_{\text{w}}-1\right)
           \dfrac{X}{\left(X-\xi_{\text{w}}
           D_{\alpha\beta}\right)}\left(X+1\right)\\
   &+\,\xi_{\text{w}}\left(\xi_{f}-1\right)
     \left(1-D_{\alpha\beta}\right)
           \dfrac{X}{\left(X-\xi_{\text{w}}
           D_{\alpha\beta}\right)}
   \label{eq:s_wba_rng1}
\end{align}
or if
\begin{align}
   \notag
   \qquad X>&\,\xi_{\text{w}}D_{\alpha\beta} \quad \mathrm{and} \quad\\
\notag
   \qquad Y>&\,X+1+\left(\xi_{\text{w}}-1\right)
           \dfrac{X}{\left(X-\xi_{\text{w}}
           D_{\alpha\beta}\right)}\left(X+1\right)\\
   &+\,\xi_{\text{w}}\left(\xi_{f}-1\right)
          \left(1-D_{\alpha\beta}\right)
          \dfrac{X}{\left(X-\xi_{\text{w}}
          D_{\alpha\beta}\right)}
   \label{eq:s_wba_rng2}   
\end{align}
(with $D_{\alpha\beta}$ as given below Eq.~(\ref{eq:s_abg_rng})).
Otherwise, the wall is in direct contact with the $\alpha$ phase without
an intruding film of the $\beta$ phase.

A wall--vapor (wall--$\gamma$) interface which is wetted by an intervening
$\beta$ film is more favorable than a wall--$\gamma$ interface without
any wetting film ($S_{\text{w}\beta\gamma}<0$) if    
\begin{align}
   \notag
   X>&\,0  \quad \mathrm{and} \quad\\
\notag
   Y>&\,X+1
   +\left(\xi_{\text{w}}-1\right)
     \dfrac{X}{\left(X-\xi_{\text{w}}
               D_{\beta\gamma}\right)}\left(X+1\right)\\
   &+\,\xi_{\text{w}}\left(\xi_{f}-1\right)
     \left(1-D_{\beta\gamma}\right)
     \dfrac{X}{\left(X-\xi_{\text{w}}D_{\beta\gamma}\right)}
   \label{eq:s_wbg_rng}   
\end{align}
(with $D_{\beta\gamma}$ as given below Eq.~(\ref{eq:s_abg_rng})).
Otherwise, the wall which is in direct contact with the vapor ($\gamma$ phase), i.e., without a $\beta$ film gives rise to a lower free energy.

A wall--vapor (wall--$\gamma$) interface which is wetted by an intruding
$\alpha$ film is more favorable than a wall--$\gamma$ interface without
any wetting film ($S_{\text{w}\alpha\gamma}<0$) if 
\begin{align}
   \notag
   X>&\,0  \quad \mathrm{and}  \quad\\
\notag
   Y>&\,\dfrac{\rho_{B,\alpha}}{\rho_{B,\beta}}X
     +\dfrac{\rho_{A,\alpha}}{\rho_{A,\beta}}\\
\notag
   &+\,\left(\xi_{\text{w}}-1\right)
     \dfrac{X}{\left(X-\xi_{\text{w}}D_{\alpha\gamma}\right)}
     \left(\dfrac{\rho_{B,\alpha}}{\rho_{B,\beta}}X
    +\dfrac{\rho_{A,\alpha}}{\rho_{A,\beta}}\right)\\
   &+\,\xi_{\text{w}}\left(\xi_{f}-1\right)
     \left(\dfrac{\rho_{A,\alpha}}{\rho_{A,\beta}}
    -\dfrac{\rho_{B,\alpha}}{\rho_{B,\beta}}
     D_{\alpha\gamma}\right)
     \dfrac{X}{\left(X-\xi_{\text{w}}D_{\alpha\gamma}\right)} \, ,
   \label{eq:s_wag_rng}   
\end{align}
where  
$$D_{\alpha\gamma}={\left(\dfrac{\rho_{A,\alpha}}
                               {\rho_{A,\beta}}
                        -\dfrac{\rho_{A,\gamma}}
                               {\rho_{A,\beta}}\right)}
                  \bigg/
                  {\left(\dfrac{\rho_{B,\gamma}}
                               {\rho_{B,\beta}}
                        -\dfrac{\rho_{B,\alpha}}
                               {\rho_{B,\beta}}\right)} \, .$$
Otherwise, the wall which is in direct contact with the vapor ($\gamma$ phase), i.e., without an $\alpha$ film, leads to a lower free energy. 
In regions, in which wetting of the wall--$\gamma$ interface with
both a film of the $\beta$ phase and a film of the $\alpha$ phase is
more favorable than a configuration without any wetting film, 
a direct comparison of these two wetting scenarios is required.
This is discussed below. 

\section{\label{section:discussion}Discussion}

\subsection{\label{subsection:dis_poss_wet_dom_near_col}
                   Colloid particle at a composite $\alpha$--$\beta$--$\gamma$ interface:
                   possible wetting scenarios }

\begin{figure*}[!t]
   \includegraphics[trim={0.5cm 1.5cm 1cm 0.5cm},width=0.27\textwidth]{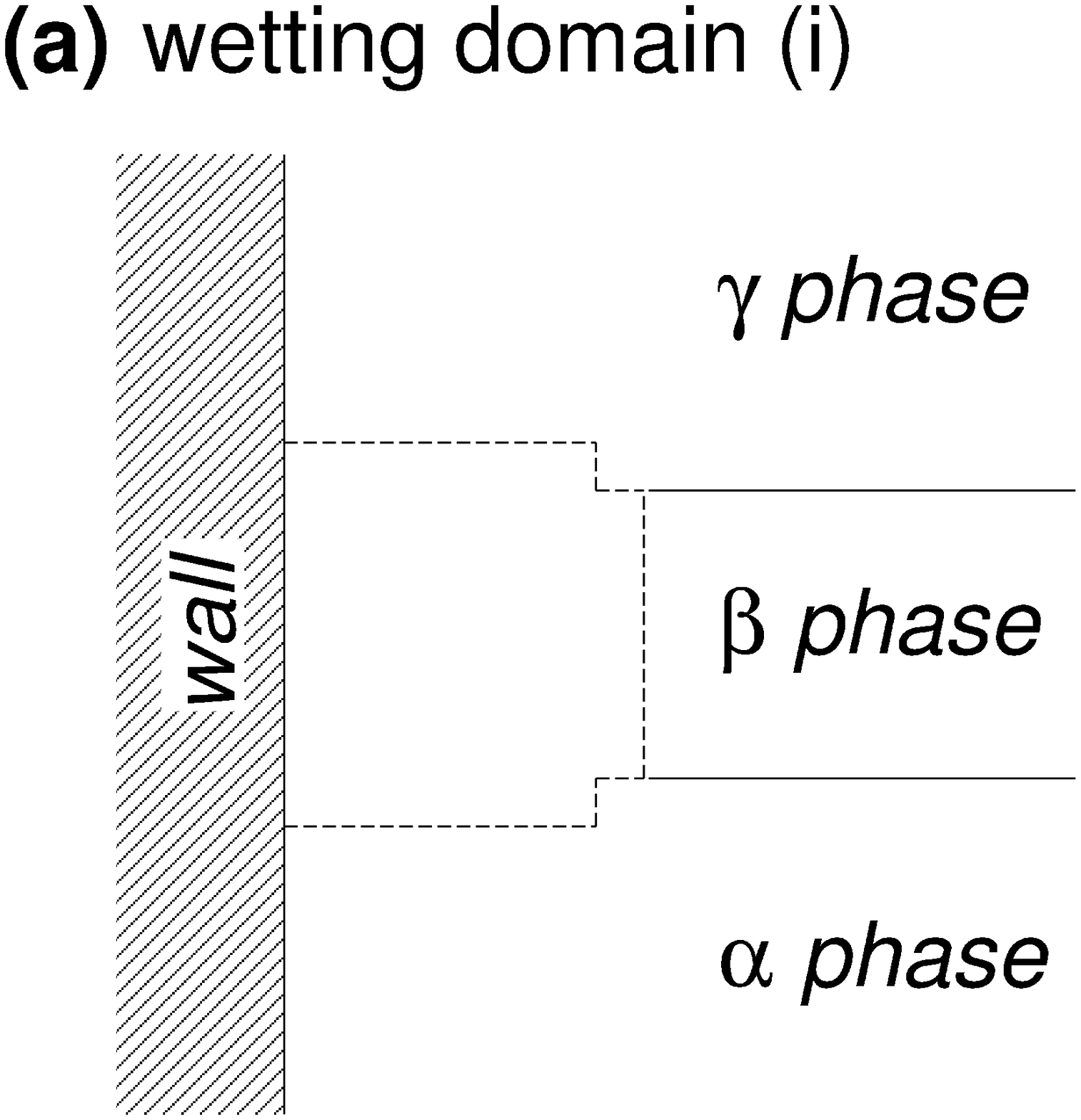}
   \includegraphics[trim={0.5cm 1.5cm 1cm 0.5cm},width=0.27\textwidth]{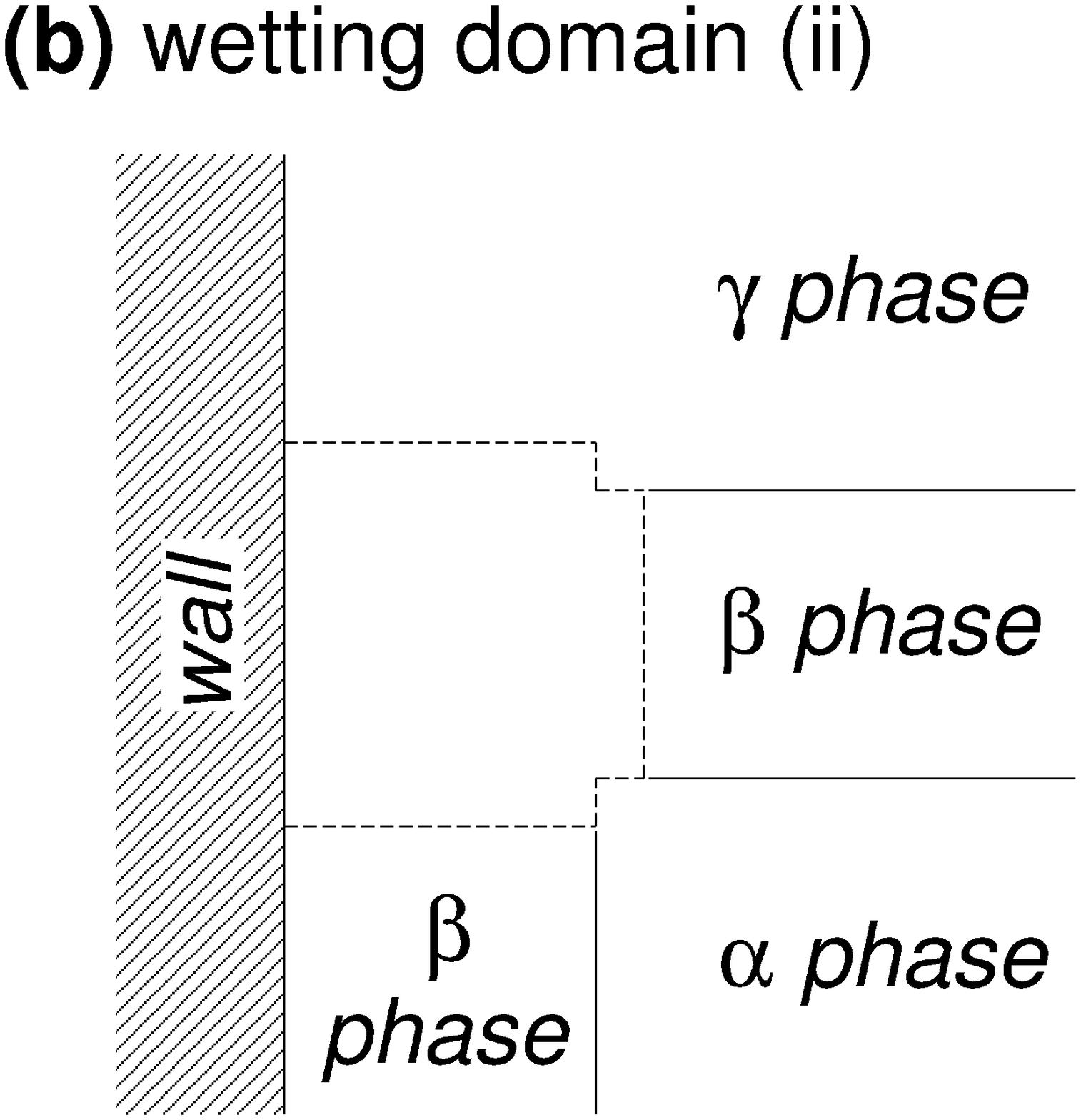}
   \includegraphics[trim={0.5cm 1.5cm 1cm 0.5cm},width=0.27\textwidth]{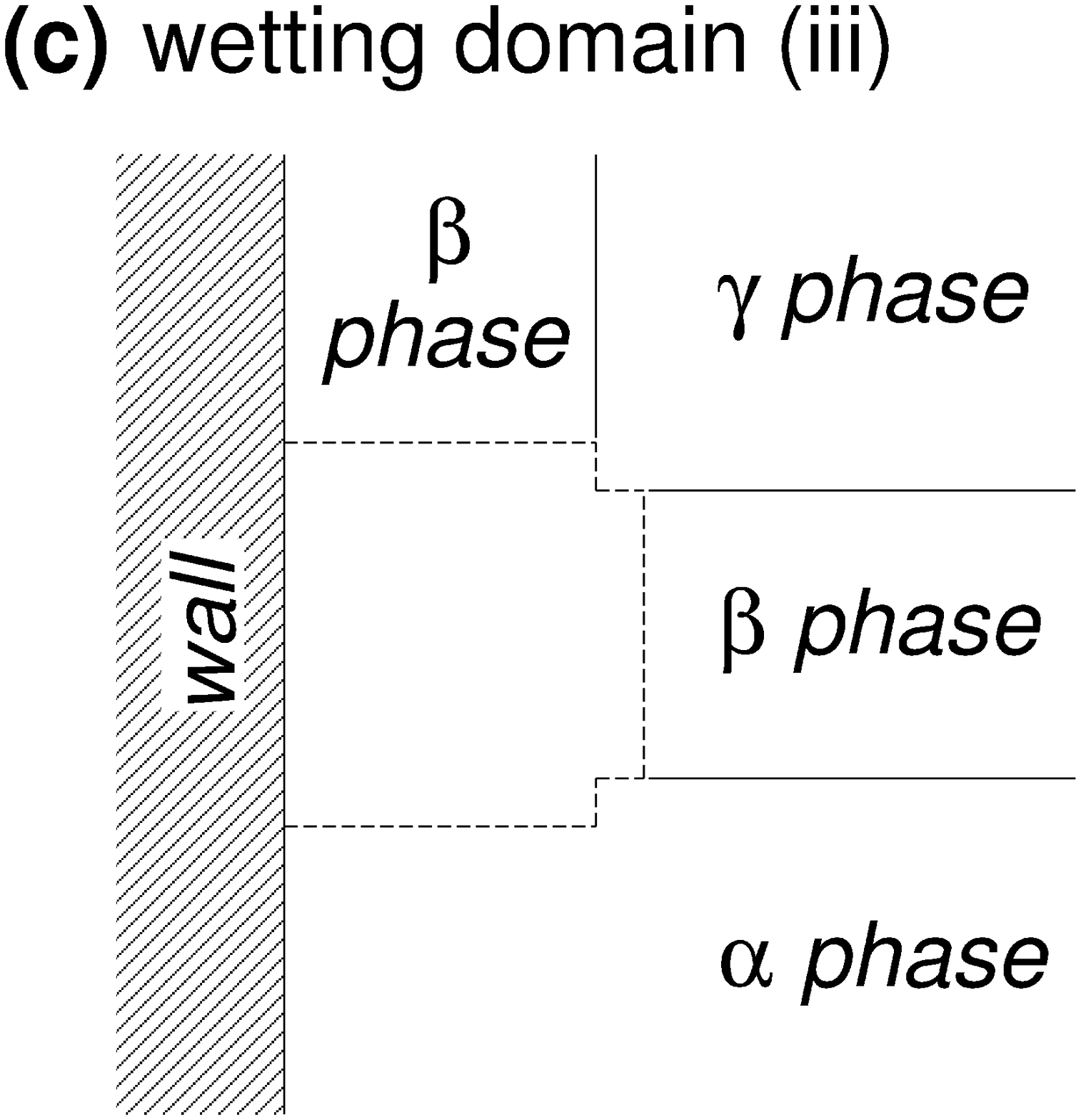}
   \includegraphics[trim={0.5cm 7.5cm 1cm 0.5cm},width=0.27\textwidth]{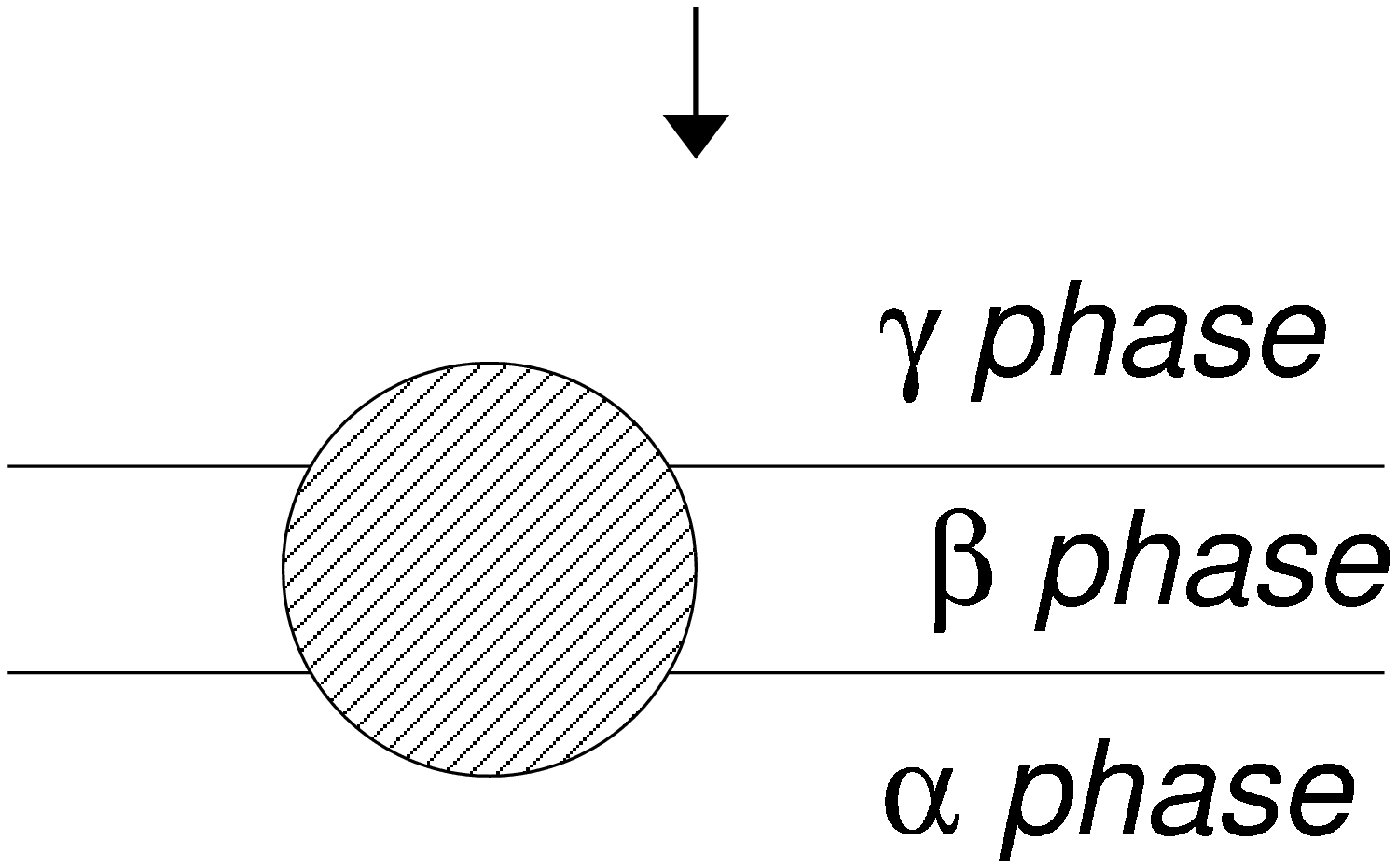}
   \includegraphics[trim={0.5cm 7.5cm 1cm 0.5cm},width=0.27\textwidth]{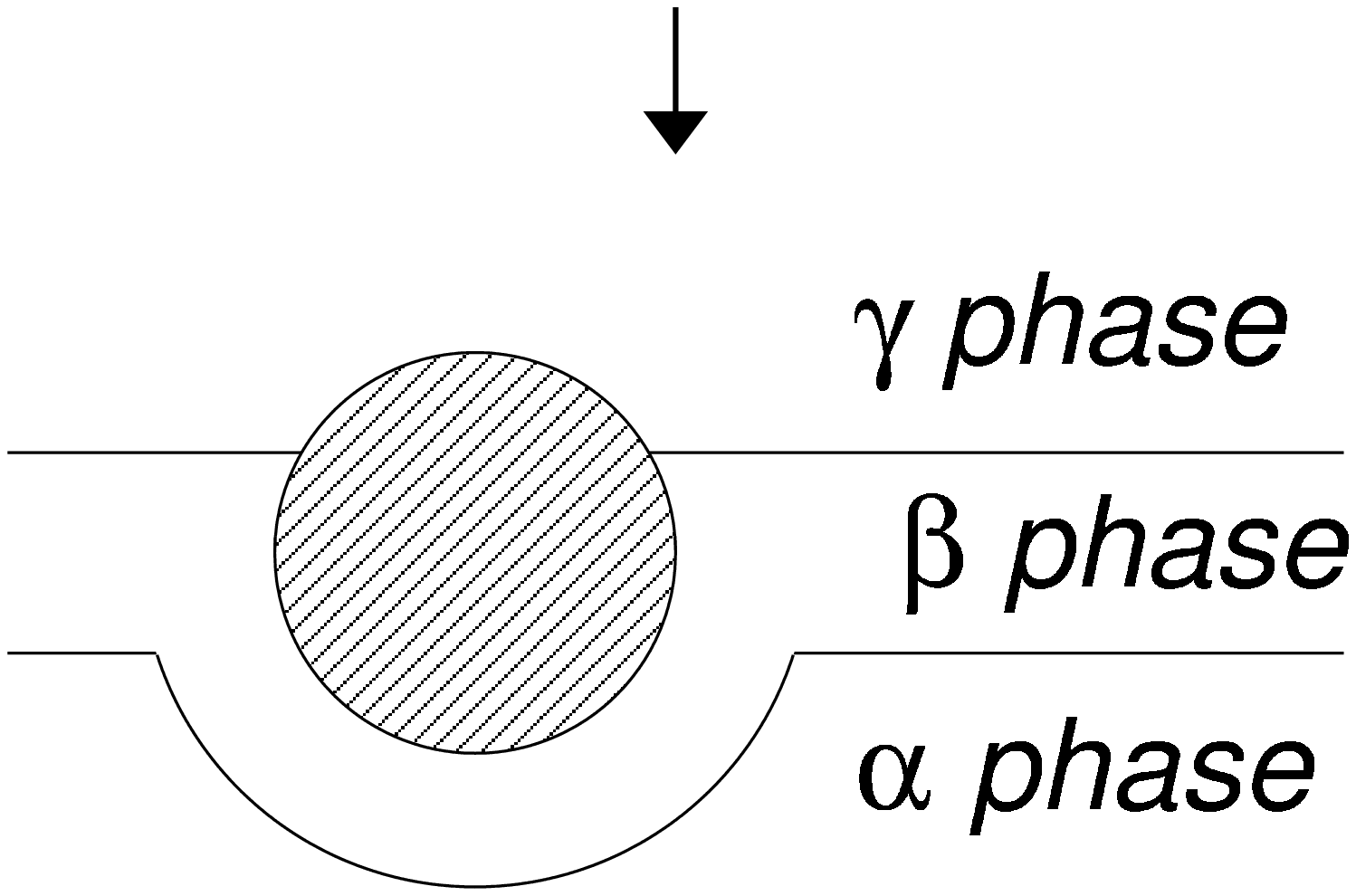}
   \includegraphics[trim={0.5cm 7.5cm 1cm 0.5cm},width=0.27\textwidth]{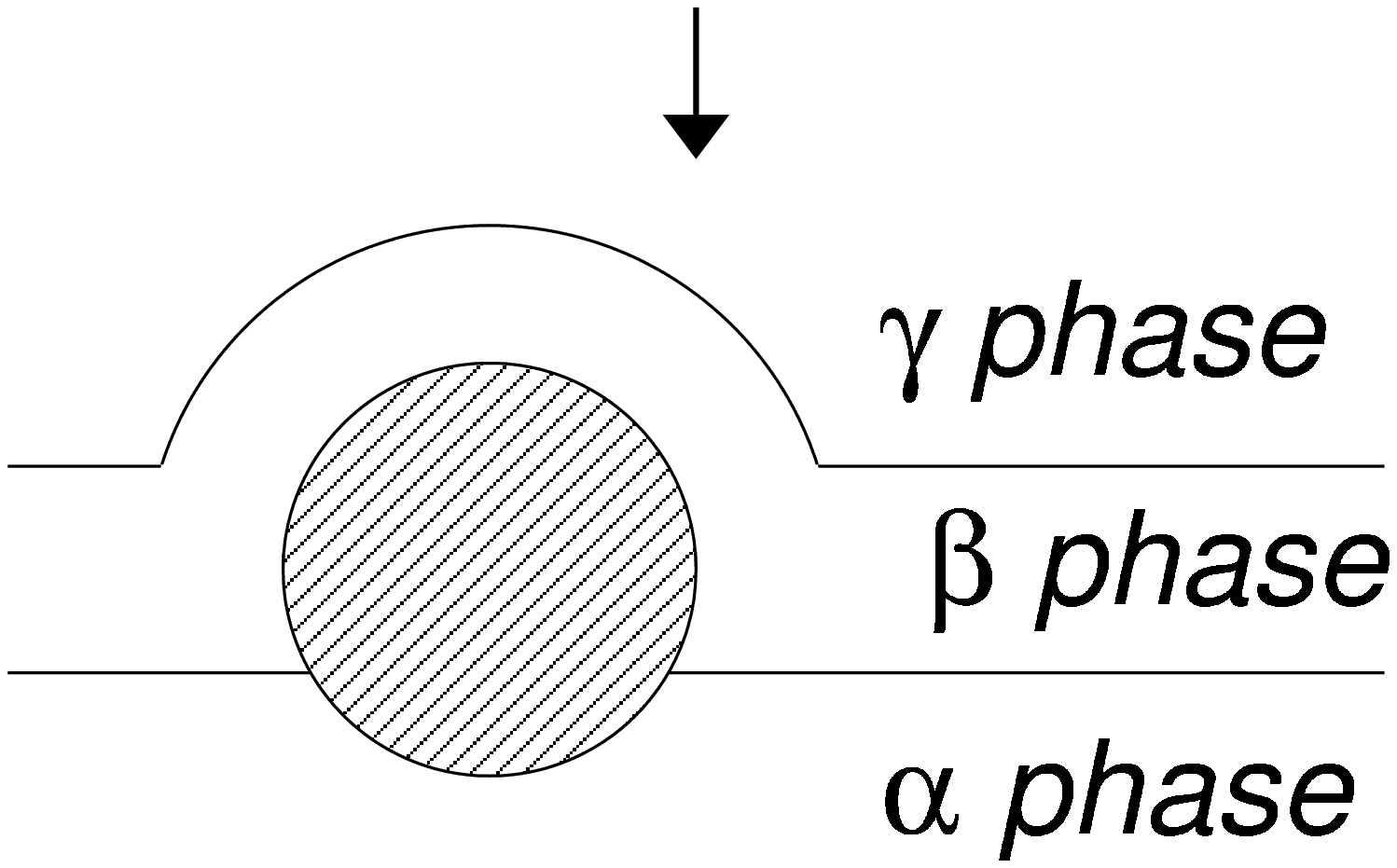}
   \includegraphics[trim={0.5cm 1.5cm 1cm 0.5cm},width=0.27\textwidth]{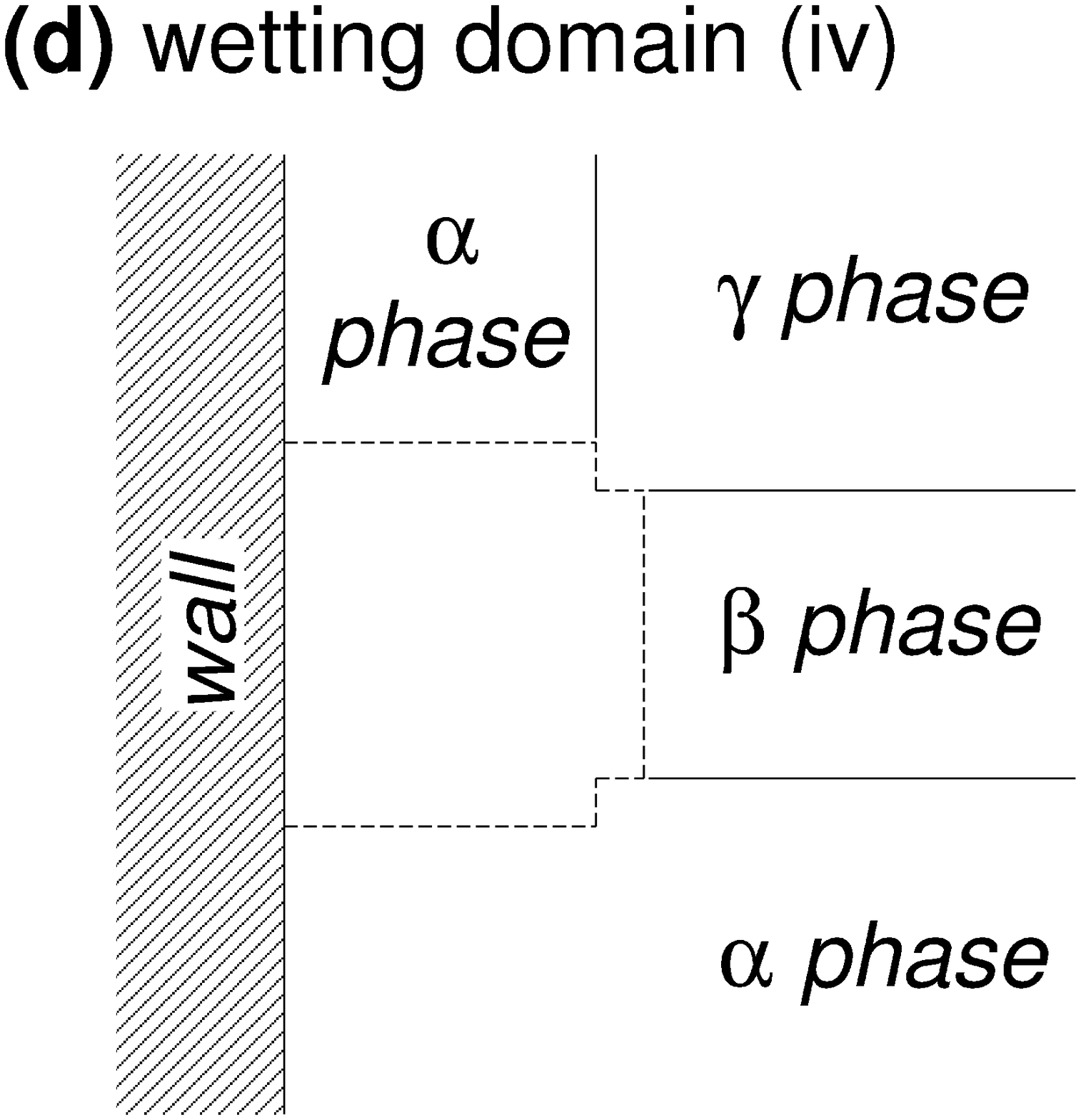}
   \includegraphics[trim={0.5cm 1.5cm 1cm 0.5cm},width=0.27\textwidth]{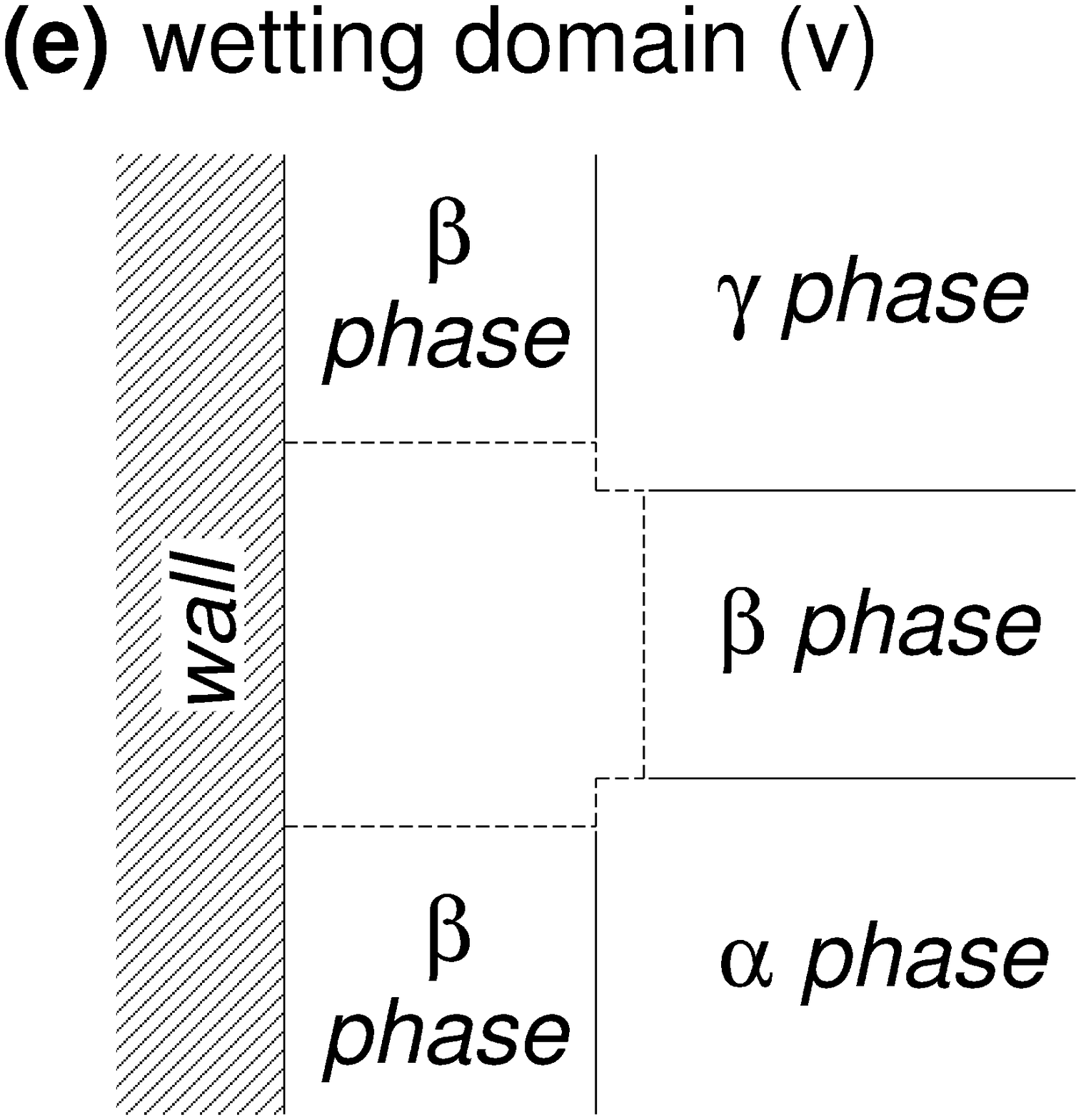}
   \includegraphics[trim={0.5cm 1.5cm 1cm 0.5cm},width=0.27\textwidth]{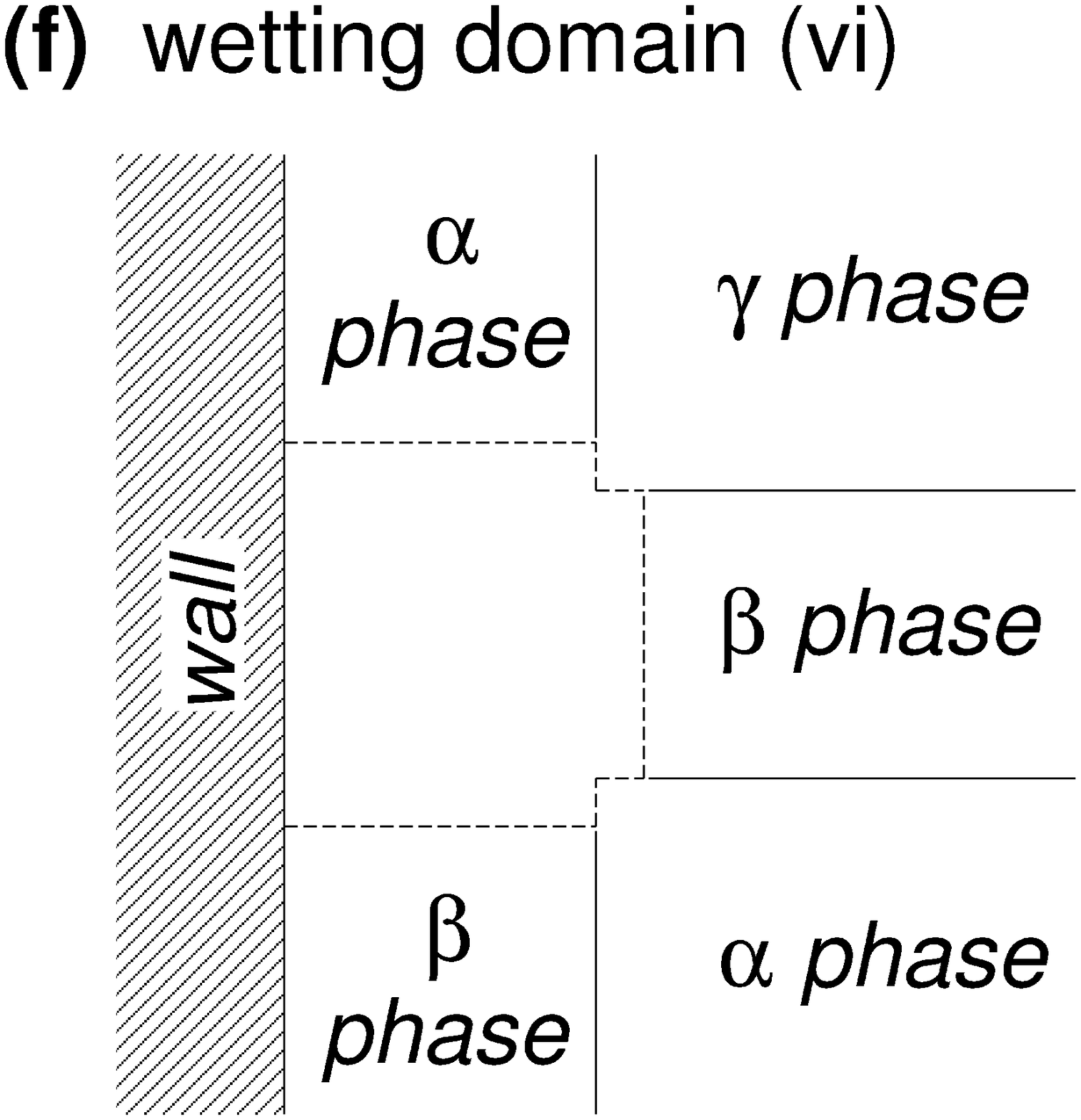}
   \includegraphics[trim={0.5cm 7.5cm 1cm 0.5cm},width=0.27\textwidth]{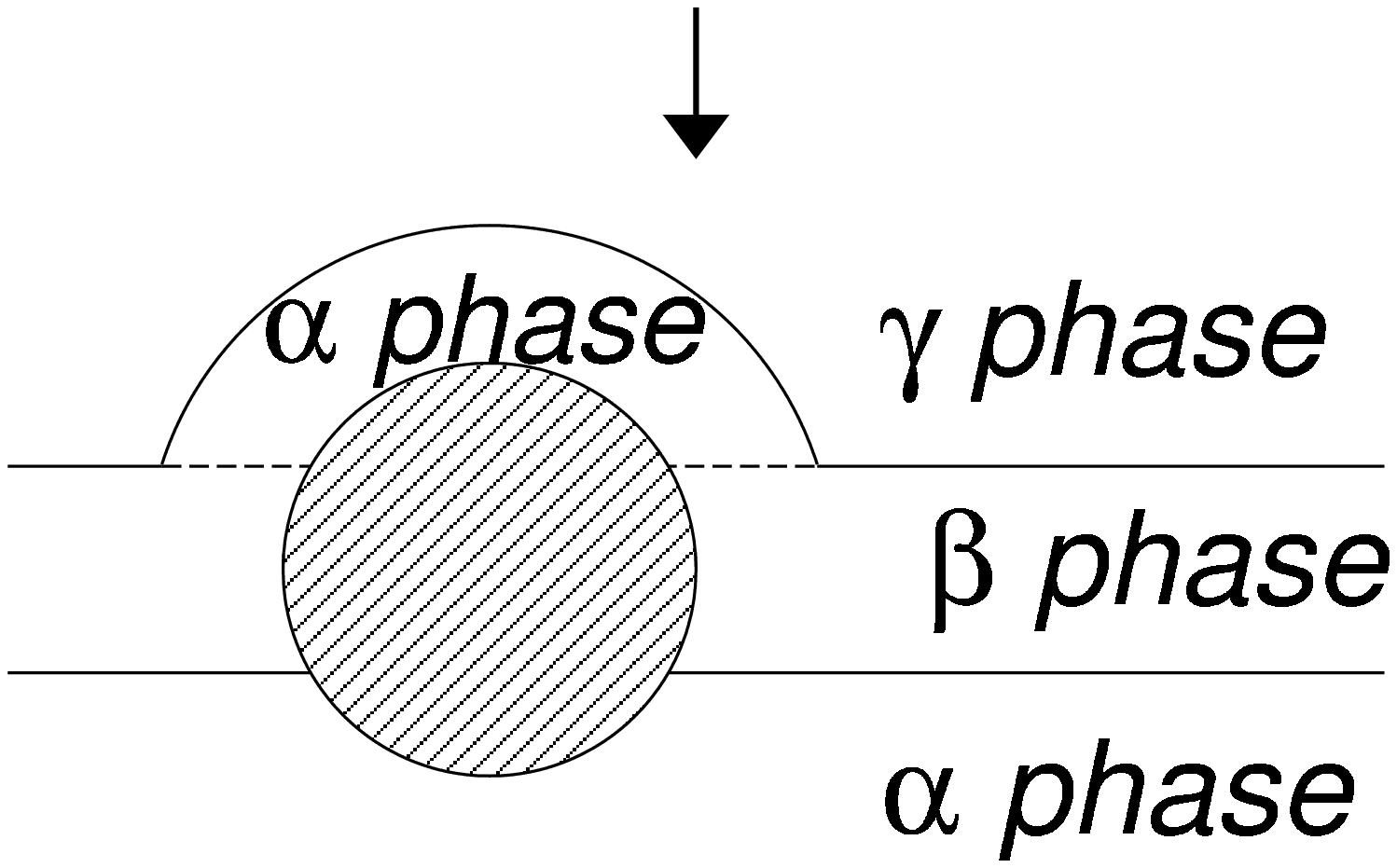}
   \includegraphics[trim={0.5cm 7.5cm 1cm 0.5cm},width=0.27\textwidth]{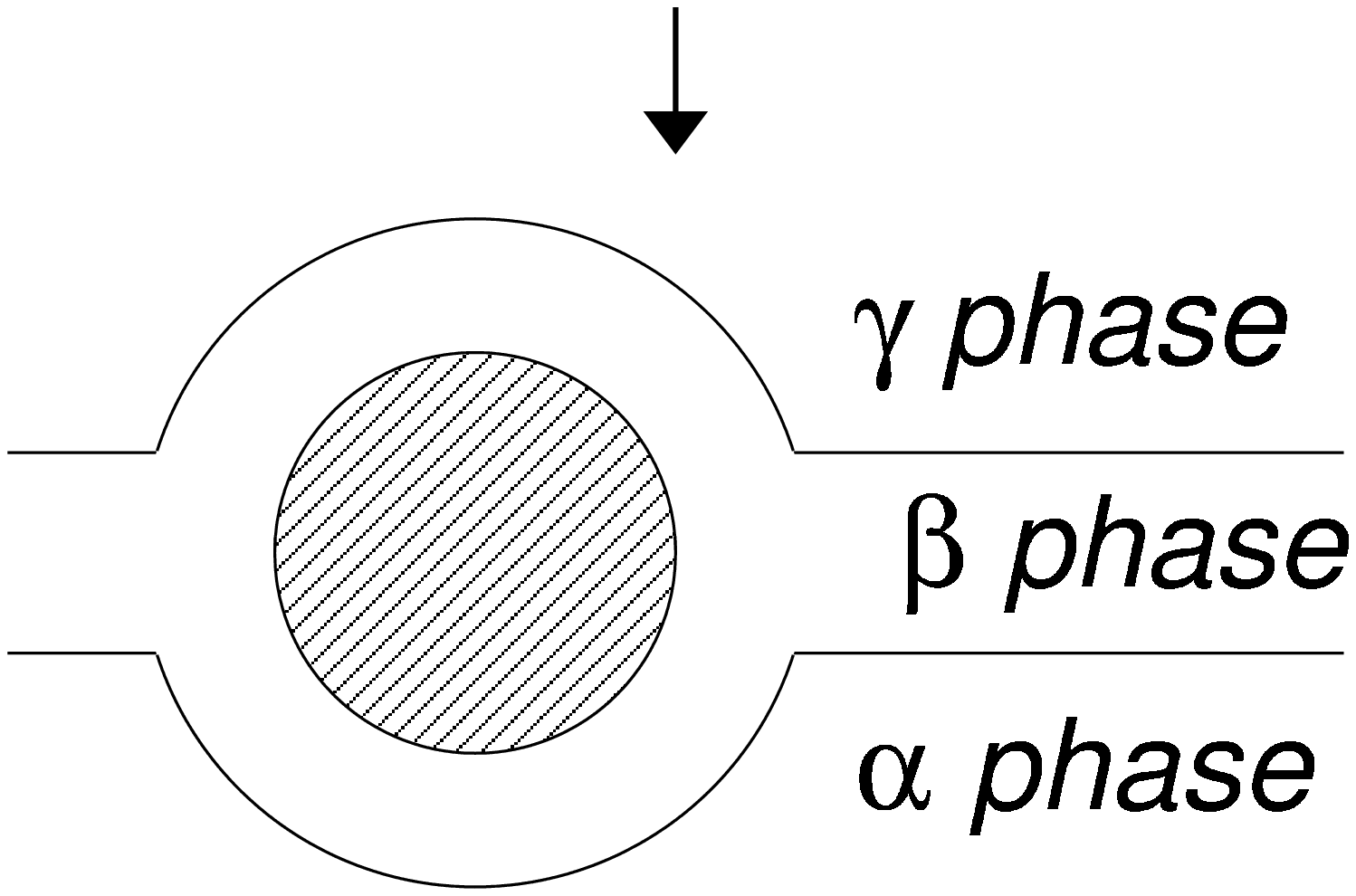}
   \includegraphics[trim={0.5cm 7.5cm 1cm 0.5cm},width=0.27\textwidth]{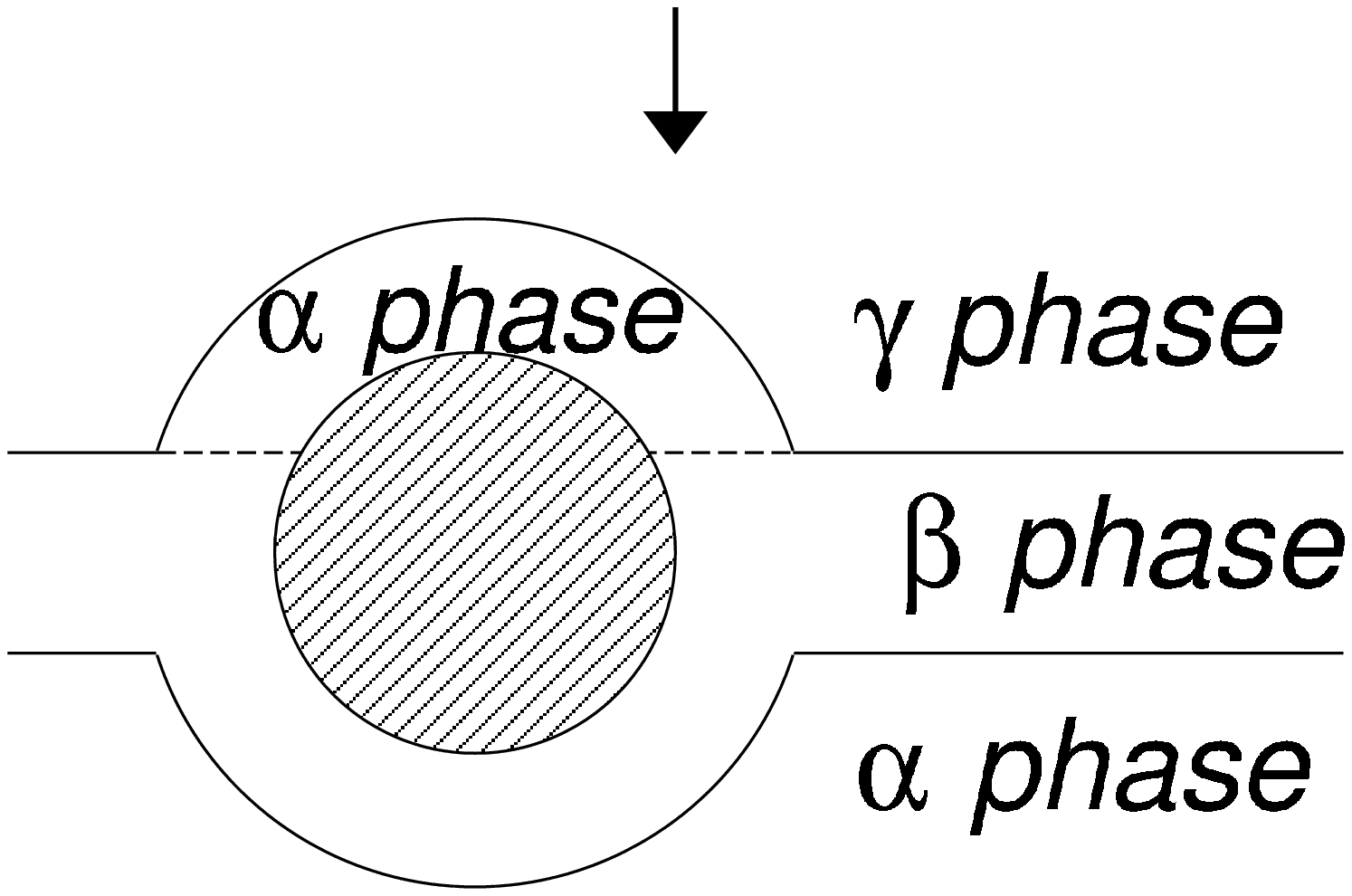}
   \caption{
      Possible wetting scenarios around a colloid floating at a composite 
      $\alpha$--$\gamma$ interface with an intruding $\beta$ film.
      (a) Wetting domain (i):
          the $\beta$ film terminates near the wall.
      (b) Wetting domain (ii):
          the $\beta$ film extends down into the wall--$\alpha$-liquid interface. 
      (c) Wetting domain (iii):
          the $\beta$ film extends up into the wall--$\gamma$ interface.  
      (d) Wetting domain (iv):
          the $\beta$ film terminates near the wall, and a 
          film of the $\alpha$ phase forms at the wall--vapor interface.
      (e) Wetting domain (v):
          the $\beta$ film extends into the wall--$\alpha$-liquid interface and
	  into the wall--vapor interface, surrounding the colloid entirely.
      (f) Wetting domain (vi):
          the $\beta$ film extends into the wall--$\alpha$-liquid interface, and  
	  a wetting film of the $\alpha$ phase forms at the wall--vapor interface. 
      Within our approach 
      the fluid structure around the three-phase contact line 
      (indicated by dashed lines) remains unknown.
   }
   \label{fig:wetting_domain}
\end{figure*}

In Fig.~\ref{fig:wetting_domain} we sketch the possible wetting scenarios around 
a colloid floating at a composite $\alpha$--$\gamma$ interface with a
$\beta$ film intruding between the adjacent phases. In addition we depict the simplified system actually 
studied, which is a composite $\alpha$--$\gamma$ interface meeting 
a planar wall, instead of a curved wall as provided by the surface of a colloid.
The information obtained in the previous chapter on the individual interfacial wetting 
problems in binary liquid mixtures, can now be combined in order to assign to
each of the six scenarios sketched in Fig.~\ref{fig:wetting_domain} a domain
in the parameter space ($X$,$Y$).

The images shown in Fig.~\ref{fig:wetting_domain} are simplified in several respects.
The $\alpha$--$\gamma$ interface meets the wall at Young's contact angle;
for simplicity in the drawings we use a contact angle of $90^o$. 
In general the film thickness at the wall--vapor (wall--$\gamma$) interface
is different from the one at the wall--liquid (wall--$\alpha$) interface and
not equal as drawn in Fig.~\ref{fig:wetting_domain}. 
From the information on the individual interfaces, 
only the gross features can be deduced. The fluid structure
in the spatial region close to the core of the three-phase contact
line cannot be determined within the present approach but requires a fully fledged density functional calculation.

The first possible wetting scenario is shown in Fig.~\ref{fig:wetting_domain}(a).
In this case the $\beta$ film at the $\alpha$--$\gamma$ interface neither
extends up into the wall--vapor (wall--$\gamma$) interface nor down
into the wall--$\alpha$-liquid interface. The $\beta$ film just terminates
near the wall. This scenario is called wetting domain (i).

The wetting scenarios shown in Figs.~\ref{fig:wetting_domain}(b)-(d)
(wetting domains (ii)-(iv))
are characterized by a thick wetting film at just one of the two wall--fluid interfaces,
either at the wall--$\alpha$ or at the wall--$\gamma$ interface. 
In the wetting domain (ii) the $\beta$ film at the $\alpha$--$\gamma$ interface
extends down into the wall--$\alpha$ interface
whereas no wetting film is present at the wall--$\gamma$ interface.
In the wetting domain (iii) the $\beta$ film extends up into the wall--$\gamma$ interface
whereas no wetting film is present at the wall--$\alpha$-liquid interface.
In the wetting domain (iv) the $\beta$ film at the $\alpha$--$\gamma$ interface
terminates near the wall, but a film of the $\alpha$ phase intrudes
at the wall--$\gamma$ interface. 

If wetting films are present at both wall--fluid interfaces,
two scenarios are conceivable. 
In the wetting domain (v) the $\beta$ film at the $\alpha$--$\gamma$ interface
extends into the wall--$\alpha$-liquid interface as well as into the wall--$\gamma$ 
interface; a colloid would be completely surrounded by a film of the $\beta$ phase
(see Fig.~\ref{fig:wetting_domain}(e)).
In the wetting domain (vi) the $\beta$ film at the $\alpha$--$\gamma$ interface
extends into the wall--$\alpha$-liquid interface, but at the wall--$\gamma$
interface a wetting film of the $\alpha$ phase forms 
(see Fig.~\ref{fig:wetting_domain}(f)). 

\subsection{\label{subsection:dis_strict_mix_rule}
                   Fluid--fluid and fluid--wall interactions exhibiting the strict mixing rules }

The relations between the strengths of the fluid--fluid
and the fluid--wall interactions imposed by the strict mixing
rules lead to a number of simplifications. The strict mixing
rule $\xi_{f}=1$ for the fluid--fluid interactions together
with the inequalities in Eq.~(\ref{eq:rel_densities}) between the number
densities constrain the range of $X$, which we have to consider, to
$0<X<D_{\alpha\beta}$.
Otherwise there is no $\beta$ film at
the $\alpha$--$\gamma$ interface (see Eq.~(\ref{eq:s_abg_rng})).
We do not pursue this latter case.
  
Second, the condition for the formation of a $\beta$ film 
at a wall--$\alpha$ interface simplifies if the strict mixing
rules for the fluid--fluid as well as for the fluid--wall interactions
are valid ($\xi_{f}=1$ and $\xi_{\text{w}}=1$). Within the range $0<X<D_{\alpha\beta}$
one obtains from Eq.~(\ref{eq:s_wba_rng1})
$$0<Y<X+1 \, .$$

Third, for the wall--$\gamma$ interface, the two distinct comparisons 
which have been made (see Eqs.~(\ref{eq:s_wbg_rng}) and~(\ref{eq:s_wag_rng}))
also simplify if the strict mixing rules
$\xi_{f}=1$ and $\xi_{\text{w}}=1$ apply.

From Eq.~(\ref{eq:s_wbg_rng}) one finds that the formation of a 
$\beta$ film is more favorable than having a plain wall--$\gamma$ 
interface without an interleaving film if
 $$Y>X+1 \, .$$ 
The formation of an $\alpha$ film at the wall--$\gamma$ interface is
favored with respect to a plain wall--$\gamma$ interface
without an interleaving film
if
 $$Y>\dfrac{\rho_{B,\alpha}}{\rho_{B,\beta}}X
   +\dfrac{\rho_{A,\alpha}}{\rho_{A,\beta}} \, ,$$
which follows from Eq.~(\ref{eq:s_wag_rng}) and the strict mixing rules.

Within the range $0<X<D_{\alpha\beta}$, on which we can focus here,
the inequality
 $$\dfrac{\rho_{B,\alpha}}{\rho_{B,\beta}}X
   +\dfrac{\rho_{A,\alpha}}{\rho_{A,\beta}}>X+1$$ 
is satisfied.
This inequality leads to the following sequence of wetting scenarios
at the wall--$\gamma$ interface. 
\begin{itemize}
   \item $0<Y<X+1$: both a $\beta$ wetting film and an $\alpha$ wetting
                    film can be excluded and a plain wall--$\gamma$ interface
                    is the preferred structure. 
   \item $X+1<Y<\dfrac{\rho_{B,\alpha}}{\rho_{B,\beta}}X
               +\dfrac{\rho_{A,\alpha}}{\rho_{A,\beta}}$:
                    a $\beta$ wetting film is the preferred structure;
                    a plain wall--$\gamma$ interface and wetting
                    by the $\alpha$ phase can be excluded. 
   \item $Y>\dfrac{\rho_{B,\alpha}}{\rho_{B,\beta}}X
           +\dfrac{\rho_{A,\alpha}}{\rho_{A,\beta}}$: 
                    wetting by the $\alpha$ phase becomes possible in
                    addition to wetting by the $\beta$ phase; 
                    a plain wall--$\gamma$ interface can be excluded.  
\end{itemize}

In the range  
$Y>\dfrac{\rho_{B,\alpha}}{\rho_{B,\beta}}X
  +\dfrac{\rho_{A,\alpha}}{\rho_{A,\beta}}$,
one still has to find another boundary which determines whether
wetting of the wall--$\gamma$ interface by a film of the $\beta$
phase or by a film of the $\alpha$ phase is the preferred configuration.
In order to find this boundary we introduce a further wetting parameter, 
$W_{\text{w}\beta(\alpha)\gamma} 
=\sigma_{\text{w}\beta\gamma}
-\sigma_{\text{w}\alpha\gamma}$,
which is the difference between the surface tensions of a wall--$\gamma$ interface 
with an intruding $\beta$ film and of one with an intruding $\alpha$ film.  
If $W_{\text{w}\beta(\alpha)\gamma}>0$, 
the wall--$\gamma$ interface is wetted by the $\alpha$ phase. 
Otherwise, an intruding $\beta$ phase wets the wall--$\gamma$ interface.  
By using Eqs.~(\ref{eq:w_wbg}) and~(\ref{eq:w_wag}), 
$W_{\text{w}\beta(\alpha)\gamma}$ can be written as
$W_{\text{w}\beta\gamma}-W_{\text{w}\alpha\gamma}$.
Again we make the simplifying assumption that all length parameters are equal:
$a_{AA}=a_{AB}=a_{BB}
=a_{\text{w}A}=a_{\text{w}B}$. 
Under this condition, 
by using Eqs.~(\ref{eq:w_wbg3}) and~(\ref{eq:w_wag3}) one finds
\begin{align}
\notag
   W_{\text{w}\beta(\alpha)\gamma}
     &=\pi{a}_{AA}^{4}\left[S_{\text{w}\beta\gamma}
                            F_{AA}(l_{\text{w}\beta\gamma})
                           -S_{\text{w}\alpha\gamma}
                            \hat{F}_{AA}(l_{\text{w}\alpha\gamma})
                                                      \right]\\
   &=: \pi{a}_{AA}^{4} K(X,Y) \, .
   \label{eq:w_wb(a)g}
\end{align}
Expressing $S_{\text{w}\beta\gamma}$ and $S_{\text{w}\alpha\gamma}$, as given in
Eqs.~(\ref{eq:s_wbg2_app}) and~(\ref{eq:s_wag2_app}), one finds 
\begin{widetext}
\begin{align}
\notag
   K(X,Y)=\epsilon_{AA}\rho_{A,\beta}^{2}
          \bigg\{&\left[\left(1-\dfrac{\rho_{B,\gamma}}
                                      {\rho_{B,\beta}}\right)X
                       +1-\dfrac{\rho_{A,\gamma}}
                                      {\rho_{A,\beta}}\right]
                 \left(X+1-Y\right)
                  F_{AA}(l_{\text{w}\beta\gamma})\\
                 &\,-\left[\left(\dfrac{\rho_{B,\alpha}}
                                     {\rho_{B,\beta}}
                              -\dfrac{\rho_{B,\gamma}}
                                     {\rho_{B,\beta}}\right)X
                        +\dfrac{\rho_{A,\alpha}}
                                     {\rho_{A,\beta}}
                              -\dfrac{\rho_{A,\gamma}}
                                     {\rho_{A,\beta}}
                   \right]
                 \left(\dfrac{\rho_{B,\alpha}}{\rho_{B,\beta}}X
                        +\dfrac{\rho_{A,\alpha}}{\rho_{A,\beta}}-Y
                   \right)\hat{F}_{AA}(l_{\text{w}\alpha\gamma})
                                                       \bigg\} \, .
   \label{eq:k(x,y)}
\end{align}
\end{widetext}
The sign of $W_{\text{w}\beta(\alpha)\gamma}$ is determined by the sign of $K(X,Y)$.
Accordingly, for $K(X,Y)>0$ a configuration with an intruding $\alpha$ film at the 
wall--$\gamma$ interface is more stable than one with an intruding $\beta$ film.
For $K(X,Y)<0$ the configuration with an intruding $\beta$ film becomes more stable.
(However, for $Y < X + 1$ a plain wall--$\gamma$ interface without any wetting
film is the configuration preferred most.)
The condition $K(X,Y)<0$ can be rewritten as
\begin{widetext}
\begin{align}
\notag
   P_{K(X,Y)}\times Y>
   &\left[\left(1-\dfrac{\rho_{B,\gamma}}
                               {\rho_{B,\beta}}\right)X
                +1-\dfrac{\rho_{A,\gamma}}
                               {\rho_{A,\beta}}\right]
           \left(X+1\right)F_{AA}(l_{\text{w}\beta\gamma})\\
  &\,-\left[\left(\dfrac{\rho_{B,\alpha}}{\rho_{B,\beta}}
                      -\dfrac{\rho_{B,\gamma}}{\rho_{B,\beta}}
                                                     \right)X
                +\dfrac{\rho_{A,\alpha}}{\rho_{A,\beta}}
                -\dfrac{\rho_{A,\gamma}}{\rho_{A,\beta}}
                                                       \right]
           \left(\dfrac{\rho_{B,\alpha}}{\rho_{B,\beta}}X
                +\dfrac{\rho_{A,\alpha}}{\rho_{A,\beta}}
                \right)\hat{F}_{AA}(l_{\text{w}\alpha\gamma}) \, ,
   \label{eq:k(x,y)_nega}
\end{align}
where
\begin{align}
   P_{K(X,Y)}=\left[\left(1-\dfrac{\rho_{B,\gamma}}
                                   {\rho_{B,\beta}}\right)X
                    +1-\dfrac{\rho_{A,\gamma}}
                                   {\rho_{A,\beta}}\right]
                    F_{AA}(l_{\text{w}\beta\gamma})
              -\left[\left(\dfrac{\rho_{B,\alpha}}{\rho_{B,\beta}}
                          -\dfrac{\rho_{B,\gamma}}{\rho_{B,\beta}}
                     \right)X
                    +\dfrac{\rho_{A,\alpha}}{\rho_{A,\beta}}
                          -\dfrac{\rho_{A,\gamma}}{\rho_{A,\beta}}
              \right]\hat{F}_{AA}(l_{\text{w}\alpha\gamma}) \, .
   \label{eq:p_k(x,y)}
\end{align}
\end{widetext}

In order to determine the sign of $P_{K(X,Y)}$ within the relevant
range of $X$ values it is convenient to rewrite Eq.~(\ref{eq:p_k(x,y)}) as 
\begin{align}
\notag
   P_{K(X,Y)}=&\left(1-\dfrac{\rho_{B,\alpha}}
                             {\rho_{B,\beta}}\right)
               \left(X-D_{\alpha\beta}\right)
               F_{AA}(l_{\text{w}\beta\gamma})\\
\notag
              &-\left(\dfrac{\rho_{B,\alpha}}{\rho_{B,\beta}}
                    -\dfrac{\rho_{B,\gamma}}{\rho_{B,\beta}}
               \right)
               \left(X-D_{\alpha\gamma}\right)\\
              &\,\quad\times\left[\hat{F}_{AA}(l_{\text{w}\alpha\gamma})
                    -F_{AA}(l_{\text{w}\beta\gamma})\right] \, .
   \label{eq:p_k(x,y)2}
\end{align}
Given the inequalities in Eq.~(\ref{eq:rel_densities}) between the number densities
we have $\rho_{B,\alpha}/\rho_{B,\beta}<1$, $D_{\alpha\beta}>0$,
$\rho_{B,\alpha}/\rho_{B,\beta}-\rho_{B,\gamma}/\rho_{B,\beta}>0$,
and $D_{\alpha\gamma}<0$.
Moreover, we have 
$0<F_{AA}(l_{\text{w}\beta\gamma})
  <\hat{F}_{AA}(l_{\text{w}\alpha\gamma})$.
As a result, $P_{K(X,Y)}$ is negative in the range 
$0<X<D_{\alpha\beta}$. 

The inequality 
in Eq.~(\ref{eq:k(x,y)_nega}) can be rewritten 
as $Y<Y_{K(X,Y)}$, where the separatrix $Y_{K(X,Y)}$ between
the $\beta$ phase and the $\alpha$ phase wetting of the wall--vapor (wall--$\gamma$) interface
is given by (see Eq.~(\ref{eq:p_k(x,y)2}))
\begin{widetext}
\begin{align}
    Y_{K(X,Y)}
     =\dfrac{\rho_{B,\alpha}}{\rho_{B,\beta}}X
     +\dfrac{\rho_{A,\alpha}}{\rho_{A,\beta}}
     +\left[\left(1-\dfrac{\rho_{B,\gamma}}
                               {\rho_{B,\beta}}\right)X
                +1-\dfrac{\rho_{A,\gamma}}
                         {\rho_{A,\beta}}\right]
     \left[\left(1-\dfrac{\rho_{B,\alpha}}
                          {\rho_{B,\beta}}\right)X
           +1-\dfrac{\rho_{A,\alpha}}
                          {\rho_{A,\beta}}\right]
            \dfrac{F_{AA}(l_{\text{w}\beta\gamma})}
                  {P_{K(X,Y)}} \, .
   \label{eq:y_k(x,y)}
\end{align}
\end{widetext}
In the $X$ range of interest, i.e., $0<X<D_{\alpha\beta}$,
$Y_{K(X,Y)}$ is always located above the straight line
$Y=\dfrac{\rho_{B,\alpha}}{\rho_{B,\beta}}X
  +\dfrac{\rho_{A,\alpha}}{\rho_{A,\beta}}$, which
in turn is located above $Y = X + 1$.  
Interestingly, the three curves $Y_{K(X,Y)}$, $Y=X+1$, and   
$Y=\dfrac{\rho_{B,\alpha}}{\rho_{B,\beta}}X
  +\dfrac{\rho_{A,\alpha}}{\rho_{A,\beta}}$
meet at the same point
($X=D_{\alpha\beta}$,$\,Y=D_{\alpha\beta}+1$), at that $X$ value above which
the $\beta$ wetting film at the $\alpha$--$\gamma$ interface ceases
to exist. 

We now consider the special case of three-phase ($\alpha$--$\beta$--$\gamma$)
coexistence. In this case one has both
$l_{\text{w}\beta\gamma}\rightarrow\infty$ and 
$l_{\text{w}\alpha\gamma}\rightarrow\infty$
(see Appendix~\ref{sec:equil_wet_film_thick}),
and
$F_{AA}(l_{\text{w}\beta\gamma})
=\hat{F}_{AA}(l_{\text{w}\alpha\gamma})
=\dfrac{13}{66}$.
Thus the sign of $K(X,Y)$ depends only on 
the sign of 
$S_{\text{w}\beta\gamma}-S_{\text{w}\alpha\gamma}$. 
By inspecting this expression, one finds that wetting of the 
wall--$\gamma$ interface by the $\beta$ phase is more favorable than wetting by the $\alpha$ phase
($0<X<D_{\alpha\beta}$) if
\begin{equation}
    Y<\dfrac{\rho_{B,\alpha}}{\rho_{B,\beta}}X
    +\dfrac{\rho_{A,\alpha}}{\rho_{A,\beta}}
    +\left[
       \left(1-\dfrac{\rho_{B,\gamma}}{\rho_{B,\beta}}\right)X
       +1-\dfrac{\rho_{A,\gamma}}{\rho_{A,\beta}}
     \right] \, .
   \label{eq:k(x,y)_nega_tpc}
\end{equation}
The curve represented by the right hand side of the inequality in Eq.~(\ref{eq:k(x,y)_nega_tpc}) 
is, for $0<X<D_{\alpha\beta}$, always located above the curve $Y_{K(X,Y)}$.
Thus, the separatrix $Y_{K(X,Y)}$ must be located in the interval given by
\begin{align*}
   &\,\dfrac{\rho_{B,\alpha}}{\rho_{B,\beta}}X
  +\dfrac{\rho_{A,\alpha}}{\rho_{A,\beta}} \, < \, Y_{K(X,Y)} \,\\
  &\,\,< \, \dfrac{\rho_{B,\alpha}}{\rho_{B,\beta}}X
  +\dfrac{\rho_{A,\alpha}}{\rho_{A,\beta}}
  +\left[\left(1-\dfrac{\rho_{B,\gamma}}
                       {\rho_{B,\beta}}\right)X
        +1-\dfrac{\rho_{A,\gamma}}
                       {\rho_{A,\beta}}\right] \, .
\end{align*}

Accordingly, in the $X$ range of interest, i.e., $0<X<D_{\alpha\beta}$,
the wetting behavior at the wall--$\gamma$ interface 
can now be classified as follows:
\begin{itemize}
   \item $0<Y<X+1$: the wall--$\gamma$ interface is a plain one
                    without any intervening wetting film. 
   \item $X+1<Y<Y_{K(X,Y)}$: the wall--$\gamma$ interface is wetted by
                             a film of the $\beta$ phase.
   \item $Y>Y_{K(X,Y)}$: the wall--$\gamma$ interface is wetted by
                         a film of the $\alpha$ phase.
\end{itemize}

\begin{figure}[!t]
   \includegraphics[trim={-0.5cm 0.5cm -0.5cm 0.5cm}, width=0.4\textwidth]{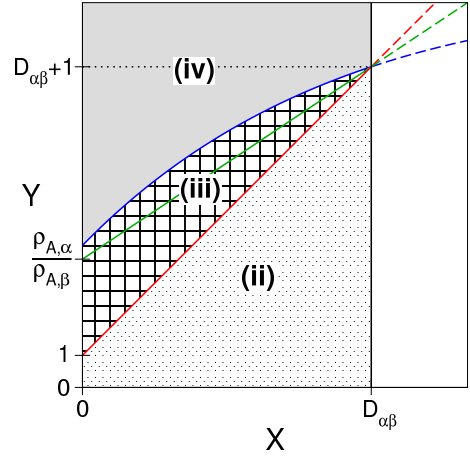}
   \caption{
      Wetting domains in the space ($X$,$Y$) of system parameters, in the case that
      the fluid--fluid and the fluid--wall interactions obey the strict mixing rules.
      Further, the inequalities in Eq.~(\ref{eq:rel_densities}) between the number densities
      are assumed to hold. The domains (ii), (iii), and (iv) correspond to the scenarios 
      depicted in Figs.~\ref{fig:wetting_domain}(b)-(d).
      The domains (i), (v), and (vi) are not realized in this case.
      Only to the left of the black line ($X=D_{\alpha\beta}$), the desired composite
      $\alpha$--$\gamma$ interface with an intruding $\beta$ film can occur.
      Below the red line ($Y=X+1$), the wall--$\alpha$ interface is wetted by a
      film of the $\beta$ phase whereas the wall--$\gamma$ interface is a plain
      one without an intervening wetting film.  
      Between the straight red line and the curved blue line ($Y=Y_{K(X,Y)}$, see Eq.~(\ref{eq:y_k(x,y)})), 
      a $\beta$ film occurs at the wall--$\gamma$ interface, whereas the wall--$\alpha$ interface is
      a plain one. Above the blue line the wall--$\gamma$ interface is wetted by 
      an $\alpha$ film. The curved blue line ($Y=Y_{K(X,Y)}$) is always located above the straight green line
      $Y=\dfrac{\rho_{B,\alpha}}{\rho_{B,\beta}}X + \dfrac{\rho_{A,\alpha}}{\rho_{A,\beta}}$.
      All curves meet at the special point ($X=D_{\alpha\beta}$,$\,Y=D_{\alpha\beta}+1$) 
      (see the main text).  
      The dashed lines extend the domain boundaries into the region where
      the $\alpha$--$\gamma$ interface is a plain one without an intruding 
      $\beta$ film; this region is beyond the interest of the present study.
      The curves correspond to the choices $\rho_{A,\alpha}/\rho_{A,\beta}=4$,
      $\rho_{B,\alpha}/\rho_{B,\beta}=2/3$, which implies $D_{\alpha\beta}=9$,
      $F_{AA}(l_{\text{w}\beta\gamma})=13/132$, and
      $\hat{F}_{AA}(l_{\text{w}\alpha\gamma})=13/66$.
   }
   \label{fig:strict}
\end{figure}

Finally, the information about wetting at the three individual interfaces can be combined
in order to deliminate a domain in the space ($X$,$Y$) of system parameters, 
which correspond to a certain wetting scenario as sketched in Fig.~\ref{fig:wetting_domain}.  
We only have to consider the interval $0<X<D_{\alpha\beta}$, because only in this $X$ range
a composite $\alpha$--$\gamma$ interface with an intervening wetting film of the $\beta$ phase can occur.
This subspace is divided into three different domains as indicated in Fig.~\ref{fig:strict}.
In domain (ii) (see the corresponding scenario depicted in Fig.~\ref{fig:wetting_domain}(b)), 
below the straight line $Y = X + 1$ (red line in Fig.~\ref{fig:strict}),
the wall--$\alpha$ interface is wetted by the $\beta$ phase. 
On the other hand below $Y = X + 1$ no film of the $\beta$ phase 
can occur at the wall--$\gamma$ interface.
Thus this interface is a plain one without any wetting film. 
The domain (iii), which corresponds to the scenario depicted 
in Fig.~\ref{fig:wetting_domain}(c), is bounded from below by $Y = X + 1$ 
(straight red line in Fig.~\ref{fig:strict}) and from
above by $Y=Y_{K(X,Y)}$ (curved blue line in Fig.~\ref{fig:strict}). In this domain
the wall--$\alpha$ interface is a plain one without an intervening wetting film,
whereas the wall--$\gamma$ interface is wetted by a film of the $\beta$ phase.
The domain (iv) above the curve $Y=Y_{K(X,Y)}$ corresponds to the scenario shown in 
Fig.~\ref{fig:wetting_domain}(d). In this domain the wall--$\alpha$ interface is a plain one,
whereas the wall--$\gamma$ interface is wetted by a film of the $\alpha$ phase.
The remaining three scenarios depicted in Fig.~\ref{fig:wetting_domain} are
not realized if the strict mixing rules are imposed on both the strengths of the fluid--fluid 
and the fluid--wall interactions (the corresponding domains are absent in Fig.~\ref{fig:strict}). 
In particular, the scenario depicted in
Fig.~\ref{fig:wetting_domain}(e) cannot occur, featuring wetting films of the $\beta$ phase at all
three interfaces, i.e., with a $\beta$ film completely surrounding a colloid.   

Some of the lines shown in Fig.~\ref{fig:strict}, which separate the domains,
depend on the ratios of various number densities. These ratios are not independent of
$X$, but are also not uniquely determined by $X$; these ratios also depend on the
thermodynamic state. The fluid model used here is also not complete and does not allow one
to predict the number densities in the different phases at given thermodynamic conditions.
Thus the boundaries between the domains still depend parametrically on ratios between the number densities; we only made use of the inequalities in Eq.~(\ref{eq:rel_densities}).
Moreover, the equation defining the separatrix $Y_{K(X,Y)}$ implicitly depends
even on $Y$ via the equilibrium thicknesses $l_{\text{w}\beta\gamma}$ and
$l_{\text{w}\alpha\gamma}$ of the wetting films.
Nevertheless, strict statements about possible and impossible wetting scenarios can be made.  

\subsection{\label{subsection:dis_relax_wf_mix_rule}
            Relaxed mixing rule for the fluid--wall interactions 
            and strict mixing rule for the fluid--fluid interactions}
Here we consider the case in which the relations between the strengths of the fluid--fluid
interactions still follow the strict mixing rule $\xi_{f}=1$; this constraint, however,
is no longer imposed on the fluid--wall interactions (i.e., $\xi_{\text{w}}\neq 1$).  
Since the condition for the formation of a composite $\alpha$--$\gamma$ interface with an
intervening $\beta$ film is still the same as in the previous subsection, the parameter
space can again be constrained to the interval $0<X<D_{\alpha\beta}$.
The conditions for the formation of an intruding $\beta$ film at the wall--$\alpha$ interface
follow from Eqs.~(\ref{eq:s_wba_rng1}) and~(\ref{eq:s_wba_rng2}) and are given by
\begin{equation*}                 
   0<Y<Y_{\text{w}\beta\alpha}(X) \;\;\; \mathrm{for} \;\;\;
   0<X<\xi_{\text{w}}D_{\alpha\beta} \, 
\end{equation*}
with
\begin{equation}
      Y_{\text{w}\beta\alpha}(X) = X+1+(\xi_{\text{w}}-1)\dfrac{X}
           {(X-\xi_{\text{w}}D_{\alpha\beta})}(X+1) 
\label{eq:relax_wf_wba}
\end{equation} 
and
\begin{equation*}                 
   Y>Y_{\text{w}\beta\alpha}(X)  \;\;\; \mathrm{for} \;\;\;
   X>\xi_{\text{w}}D_{\alpha\beta} \, . 
\end{equation*} 
In the above conditions the following distinctions have to
be made in accordance with the magnitude of $\xi_{\text{w}}$.
\begin{enumerate}
 \item   For $0<\xi_{\text{w}}<1$,
         $Y_{\text{w}\beta\alpha}(X)$ is positive within
               the interval $0<X<\xi_{\text{w}}D_{\alpha\beta}$
               and negative within $\xi_{\text{w}}D_{\alpha\beta}<X<D_{\alpha\beta}$.
         Therefore a $\beta$ film wets the wall--$\alpha$ interface if the
         following conditions are fulfilled:
         \begin{itemize}
          \item[] $0<Y<Y_{\text{w}\beta\alpha}(X) \;\;\; \mathrm{for} \;\;\;
            0<X<\xi_{\text{w}}D_{\alpha\beta} \, ,$ 
          \item[] or
          \item[] $Y_{\text{w}\beta\alpha}(X)<0<Y \;\;\; \mathrm{for} \;\;\; \xi_{\text{w}}D_{\alpha\beta}<X<D_{\alpha\beta} \, .$
         \end{itemize}
   \item For $\xi_{\text{w}}>1$, 
         $Y_{\text{w}\beta\alpha}(X)$ is positive
                 in the whole range $0<X<D_{\alpha\beta}$ of interest.
         Therefore a $\beta$ film wets the wall--$\alpha$ interface if the
         following condition is satisfied:
         \begin{itemize}
            \item[] $ 0<Y<Y_{\text{w}\beta\alpha}(X)$
            \item[] $\mathrm{\quad\,\,\,\; for\,\,the\,\,whole\,\,range} \,\,\,
                    0<X<D_{\alpha\beta} \, .$ 
         \end{itemize}
\end{enumerate}

A wetting film of the $\beta$ phase at the wall--$\gamma$ interface is
more favorable than a plain interface without a wetting film if
(see Eq.~(\ref{eq:s_wbg_rng})) 
\begin{equation*}
   Y>Y_{\text{w}\beta\gamma}(X)
\end{equation*}
with 
\begin{equation}
   Y_{\text{w}\beta\gamma}(X) = X+1+(\xi_{\text{w}}-1)\dfrac{X}
         {(X-\xi_{\text{w}}D_{\beta\gamma})}(X+1) \, .
\label{eq:relax_wf_wbg}
\end{equation}
A wetting film of the $\alpha$ phase at the wall--$\gamma$ interface is
more favorable than a plain interface if (see Eq.~(\ref{eq:s_wag_rng}))
\begin{equation*}                 
   Y>Y_{\text{w}\alpha\gamma}(X) 
\end{equation*}
with
\begin{align}
\notag
   Y_{\text{w}\alpha\gamma}(X) =&\,\dfrac{\rho_{B,\alpha}}{\rho_{B,\beta}}X
    +\dfrac{\rho_{A,\alpha}}{\rho_{A,\beta}}\\
    &+\,(\xi_{\text{w}}-1)\dfrac{X}    
     {(X-\xi_{\text{w}}D_{\alpha\gamma})}
     \left(\dfrac{\rho_{B,\alpha}}{\rho_{B,\beta}}X
    +\dfrac{\rho_{A,\alpha}}{\rho_{A,\beta}}\right) \, . 
\label{eq:relax_wf_wag}
\end{align}
In order to figure out whether the configuration with an $\alpha$ film or the one
with a $\beta$ film at the wall--$\gamma$ interface is more favorable, 
one has to inspect the sign of $W_{\text{w}\beta(\alpha)\gamma}$ 
(see Eq.~(\ref{eq:w_wb(a)g})).
By applying Eqs.~(\ref{eq:s_wbg2_app}) 
and (\ref{eq:s_wag2_app}),
with $W_{\text{w}\beta(\alpha)\gamma} =: \pi{a}_{AA}^{4}L(X,Y)$, one has
\begin{widetext}
\begin{align*}
   L(X,Y)=\epsilon_{AA}\rho_{A,\beta}^{2}
          \bm{\,\bigg(}&\bigg\{
           \left[\left(1-\dfrac{\rho_{B,\gamma}}
                               {\rho_{B,\beta}}\right)X
                +1-\dfrac{\rho_{A,\gamma}}
                               {\rho_{A,\beta}}\right]
           \left(X+1\right)
          -\left[\left(1-\dfrac{\rho_{B,\gamma}}
                                  {\rho_{B,\beta}}\right)
                    \dfrac{X}{\xi_{\text{w}}}
                   +1-\dfrac{\rho_{A,\gamma}}
                                 {\rho_{A,\beta}}
              \right]Y\bigg\}F_{AA}(l_{\text{w}\beta\gamma})\\
\notag
          &-\bigg\{
            \left[\left(\dfrac{\rho_{B,\alpha}}{\rho_{B,\beta}}
                       -\dfrac{\rho_{B,\gamma}}{\rho_{B,\beta}}
                                                     \right)X
                 +\dfrac{\rho_{A,\alpha}}{\rho_{A,\beta}}
                      -\dfrac{\rho_{A,\gamma}}{\rho_{A,\beta}}
                                                       \right]
            \left(\dfrac{\rho_{B,\alpha}}{\rho_{B,\beta}}X
                 +\dfrac{\rho_{A,\alpha}}{\rho_{A,\beta}}
            \right)\\
        &\,\qquad-\left[\left(\dfrac{\rho_{B,\alpha}}{\rho_{B,\beta}}
                       -\dfrac{\rho_{B,\gamma}}{\rho_{B,\beta}}
                  \right)\dfrac{X}{\xi_{\text{w}}}
                 +\dfrac{\rho_{A,\alpha}}{\rho_{A,\beta}}
                       -\dfrac{\rho_{A,\gamma}}{\rho_{A,\beta}}
            \right]Y\bigg\}\hat{F}_{AA}(l_{\text{w}\alpha\gamma})
        \bm{\bigg)} \, .
\end{align*}
\end{widetext}
Wetting of the wall--$\gamma$ interface by the $\beta$ phase is preferred
as compared to wetting by the $\alpha$ phase, if $L(X,Y)<0$.
The condition $L(X,Y)<0$ can be expressed as
\begin{widetext}
\begin{align*}
   P_{L(X,Y)}\times Y>
      &\left[\left(1-\dfrac{\rho_{B,\gamma}}
                           {\rho_{B,\beta}}\right)X
            +1-\dfrac{\rho_{A,\gamma}}
                           {\rho_{A,\beta}}\right]
       \left(X+1\right)F_{AA}(l_{\text{w}\beta\gamma})\\
     &\,-\left[\left(\dfrac{\rho_{B,\alpha}}{\rho_{B,\beta}}
                  -\dfrac{\rho_{B,\gamma}}{\rho_{B,\beta}}
             \right)X
            +\dfrac{\rho_{A,\alpha}}{\rho_{A,\beta}}
                  -\dfrac{\rho_{A,\gamma}}{\rho_{A,\beta}}
             \right]
       \left(\dfrac{\rho_{B,\alpha}}{\rho_{B,\beta}}X
            +\dfrac{\rho_{A,\alpha}}{\rho_{A,\beta}}
       \right)\hat{F}_{AA}(l_{\text{w}\alpha\gamma}) \, ,
\end{align*}
where
\begin{align*}
   P_{L(X,Y)}
     =&\left[\left(1-\dfrac{\rho_{B,\gamma}}
                           {\rho_{B,\beta}}\right)
             \dfrac{X}{\xi_{\text{w}}}
            +1-\dfrac{\rho_{A,\gamma}}
                           {\rho_{A,\beta}}
       \right]F_{AA}(l_{\text{w}\beta\gamma})
      -\left[\left(\dfrac{\rho_{B,\alpha}}{\rho_{B,\beta}}
                  -\dfrac{\rho_{B,\gamma}}{\rho_{B,\beta}}
             \right)\dfrac{X}{\xi_{\text{w}}}
            +\dfrac{\rho_{A,\alpha}}{\rho_{A,\beta}}
                  -\dfrac{\rho_{A,\gamma}}{\rho_{A,\beta}}
       \right]\hat{F}_{AA}(l_{\text{w}\alpha\gamma})  \\
     =&\,\dfrac{1}{\xi_{\text{w}}}
       \bigg\{\left(1-\dfrac{\rho_{B,\alpha}}
                           {\rho_{B,\beta}}\right)
              \left(X-\xi_{\text{w}}D_{\alpha\beta}\right)
              F_{AA}(l_{\text{w}\beta\gamma})
      -\left(\dfrac{\rho_{B,\alpha}}{\rho_{B,\beta}}
                  -\dfrac{\rho_{B,\gamma}}{\rho_{B,\beta}}
             \right)
             \left(X-\xi_{\text{w}}D_{\alpha\gamma}\right)
             \left[\hat{F}_{AA}(l_{\text{w}\alpha\gamma})
                  -F_{AA}(l_{\text{w}\beta\gamma})\right]\bigg\} \, .
\end{align*}
\end{widetext}
The inequalities between the number densities (Eq.~(\ref{eq:rel_densities})) imply
$\rho_{B,\alpha}/\rho_{B,\beta}<1$, $D_{\alpha\beta}>0$,
$\rho_{B,\alpha}/\rho_{B,\beta}-\rho_{B,\gamma}/\rho_{B,\beta}>0$,
and $D_{\alpha\gamma}<0$.
Furthermore, we have 
$0<F_{AA}(l_{\text{w}\beta\gamma})<\hat{F}_{AA}(l_{\text{w}\alpha\gamma})$.

If $0<X<\xi_{\text{w}}D_{\alpha\beta}$, one has $P_{L(X,Y)}<0$.  
If $X>\xi_{\text{w}}D_{\alpha\beta}$, $P_{L(X,Y)}$ can be positive or negative.  
Which possibility prevails depends on the magnitude of $\xi_{\text{w}}$. 
\begin{enumerate} 
   \item If $0<\xi_{\text{w}}<1$, the inequalities
         $0<\xi_{\text{w}}D_{\alpha\beta}<D_{\alpha\beta}$ hold.
         For $0<X<\xi_{\text{w}}D_{\alpha\beta}$,  $P_{L(X,Y)}$ is negative. 
         Within the interval $\xi_{\text{w}}D_{\alpha\beta}<X<D_{\alpha\beta}$, 
         $P_{L(X,Y)}$ can be positive or negative. 
   \item If $\xi_{\text{w}}>1$, 
         one has $\xi_{\text{w}}D_{\alpha\beta}>D_{\alpha\beta}$.
         Thus $P_{L(X,Y)}<0$ in the whole range of $X$ values
         of interest, i.e., for $0<X<D_{\alpha\beta}$.
\end{enumerate} 

In the case $P_{L(X,Y)}<0$ and if $0<Y<Y_{L(X,Y)}$, 
an intruding $\beta$ film at the wall--$\gamma$ interface is
more favorable than a wetting film of the $\alpha$ phase; here
\begin{widetext}
\begin{align}
\notag
   Y_{L(X,Y)}=&\left[\left(1-\dfrac{\rho_{B,\gamma}}
                                   {\rho_{B,\beta}}\right)X
                    +1-\dfrac{\rho_{A,\gamma}}
                                   {\rho_{A,\beta}}
               \right]\left(X+1\right)
               \dfrac{F_{AA}(l_{\text{w}\beta\gamma})}
                     {P_{L(X,Y)}}\\
             &\,-\left[\left(\dfrac{\rho_{B,\alpha}}
                                 {\rho_{B,\beta}}
                          -\dfrac{\rho_{B,\gamma}}
                                 {\rho_{B,\beta}}\right)X
                    +\dfrac{\rho_{A,\alpha}}
                                 {\rho_{A,\beta}}
                          -\dfrac{\rho_{A,\gamma}}
                                 {\rho_{A,\beta}}
               \right]
               \left(\dfrac{\rho_{B,\alpha}}{\rho_{B,\beta}}X
                    +\dfrac{\rho_{A,\alpha}}{\rho_{A,\beta}}
               \right)
               \dfrac{\hat{F}_{AA}(l_{\text{w}\alpha\gamma})}
                     {P_{L(X,Y)}} \, .
   \label{eq:y_l(x,y)}
\end{align}
\end{widetext}
If $P_{L(X,Y)}>0$ and if $Y>Y_{L(X,Y)}$, 
the $\beta$ wetting film at the wall--$\gamma$ interface is preferred.

\begin{figure*}[!t]
   \includegraphics[trim={0 0.5cm -1cm 0.5cm},width=0.4\textwidth]{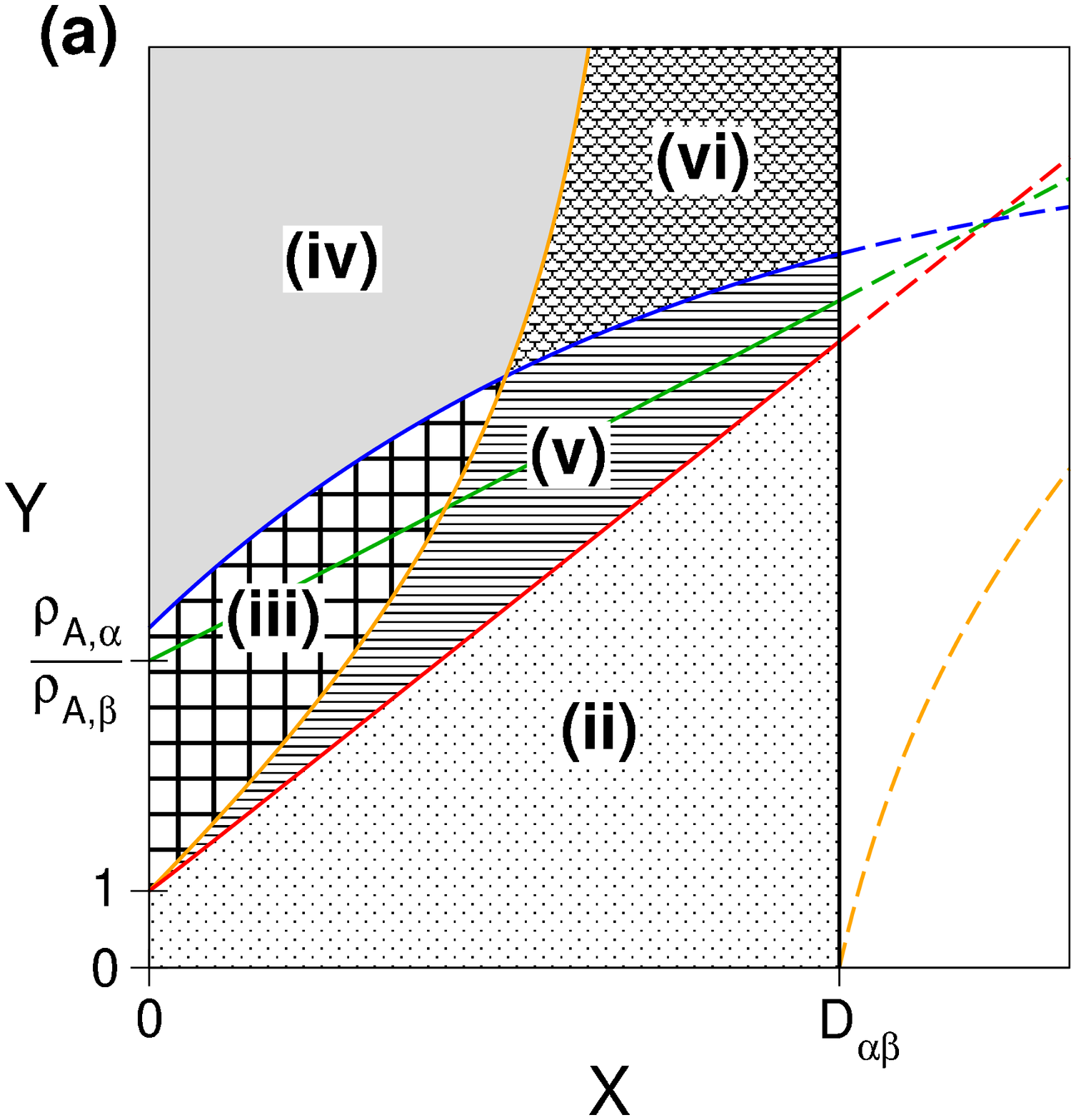}
   \includegraphics[trim={-1cm 0.5cm 0 0.5cm},width=0.4\textwidth]{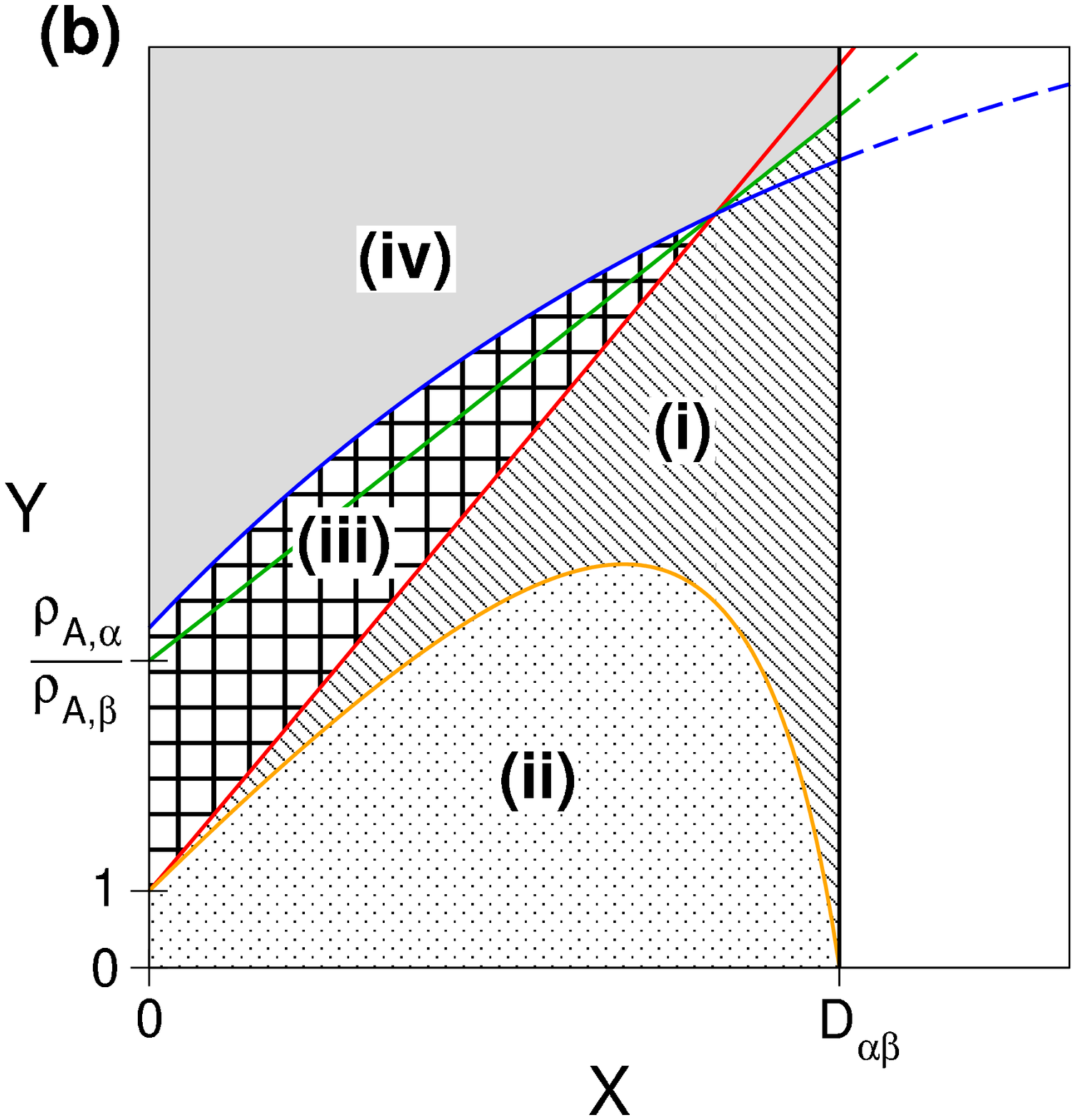}
   \caption{
      Wetting domains (i)-(vi) in the parameter space ($X$,$Y$) 
      (Fig.~\ref{fig:wetting_domain}), in the case that
      the fluid--fluid interactions obey the strict mixing rule, but the
      fluid--wall interactions are not constrained ($\xi_{\text{w}} \neq 1$).
      Further, the inequalities between the number densities
      (Eq.~(\ref{eq:rel_densities})) are assumed. 
      The domains correspond to the scenarios depicted 
      in Fig.~\ref{fig:wetting_domain}.
      Two distinct scenarios emerge, depending on whether (a) $0<\xi_{\text{w}}<1$
      or (b) $\xi_{\text{w}}>1$.
      Only to the left of the black line ($X=D_{\alpha\beta}$)
      one encounters the composite $\alpha$--$\gamma$ interface 
      of interest with an intruding $\beta$ film.
      Above the red line $Y_{\text{w}\beta\gamma}(X)$ (Eq.~(\ref{eq:relax_wf_wbg}))
      the wall--$\gamma$ interface is potentially wetted by a $\beta$ film.
      Above the green line $Y_{\text{w}\alpha\gamma}(X)$
      (Eq.~(\ref{eq:relax_wf_wag})) the wall--$\gamma$ interface
      is potentially wetted by an $\alpha$ film. The separatrix between
      wetting of this interface by the $\beta$ phase
      and wetting by the $\alpha$ phase is given by the blue line 
      $Y_{L(X,Y)}$ (Eq.~(\ref{eq:y_l(x,y)})). 
      Depending on whether the green line lies above the red line or below,
      wetting of the wall--$\gamma$ interface by an $\alpha$ film is found above 
      or below the blue line.
      Below the orange line $Y_{\text{w}\beta\alpha}(X)$ (Eq.~(\ref{eq:relax_wf_wba}))
      the wall--$\alpha$ interface is wetted by a film of the $\beta$ phase.
      The dashed lines extend the domain boundaries into the region within which
      the $\alpha$--$\gamma$ interface is a plain one without an intruding 
      $\beta$ film; this region is beyond the interest of 
      the present study.
      In particular, the dashed orange line in (a) represents 
      $Y_{\text{w}\beta\alpha}(X)$ for $X>D_{\alpha\beta}$. 
      $Y_{\text{w}\beta\alpha}(X)$ has a vertical asymptote and 
      changes sign at $X=\xi_{\text{w}}D_{\alpha\beta}$;
      it returns to positive values for $X>D_{\alpha\beta}$
      (see Eq.~(\ref{eq:relax_wf_wba})).
      The lines correspond to the choices 
      $\rho_{A,\alpha}/\rho_{A,\beta}=4$,
      $\rho_{B,\alpha}/\rho_{B,\beta}=2/3$,
      which implies $D_{\alpha\beta}=9$,
      $F_{AA}(l_{\text{w}\beta\gamma})=13/132$, and
      $\hat{F}_{AA}(l_{\text{w}\alpha\gamma})=13/66$.
      In (a) we have $\xi_{\text{w}}=0.8$ and in (b) $\xi_{\text{w}}=1.2$. 
   }
   \label{fig:relax_w}
\end{figure*}

The separatrix $Y=Y_{L(X,Y)}$ (Eq.~(\ref{eq:y_l(x,y)}))
is always located above 
the curve $Y_{\text{w}\alpha\gamma}(X)$ (Eq.~(\ref{eq:relax_wf_wag})) as long as 
$Y_{\text{w}\beta\gamma}(X) < Y_{\text{w}\alpha\gamma}(X)$ 
($Y_{\text{w}\beta\gamma}(X)$ is defined in Eq.~(\ref{eq:relax_wf_wbg})).
Otherwise, $Y_{L(X,Y)}$ is located below the line $Y_{\text{w}\alpha\gamma}(X)$. 
The intersection between $Y_{L(X,Y)}$ and $Y_{\text{w}\alpha\gamma}(X)$ lies on the curve
$Y_{\text{w}\beta\gamma}(X)$, i.e., the three curves have a common intersection.

By admitting deviations from the strict mixing rule for the fluid--wall interactions,
additional wetting scenarios may be realized as compared to those which are possible
in the case that the strict mixing rules apply to both the fluid--fluid and the 
fluid--wall interactions (see Fig.~\ref{fig:strict}).
The main reason for this is that in the latter case 
the domain boundary for wetting of the wall--$\alpha$ interface 
by the $\beta$ phase and the domain boundary for
wetting of the wall--$\gamma$ interface by the $\beta$ phase (red line in Fig.~\ref{fig:strict})
coincide, whereas these two boundaries are different (orange and red lines in
Fig.~\ref{fig:relax_w}) once the mixing rules for the fluid--wall interactions
are relaxed. 

If $0<\xi_{\text{w}}<1$ (i.e., the wall--$A$ interaction is weaker than the one prescribed by the strict mixing rule)
the two additional scenarios (v) and (vi) (Figs.~\ref{fig:wetting_domain}(e) and~(f)) become possible in their corresponding
domains in the parameter space ($X$,$Y$) (see Fig.~\ref{fig:relax_w}(a)).
In scenario (v) the $\beta$ wetting film extends into 
both the wall--$\alpha$-liquid interface and 
into the wall--$\gamma$ (wall--vapor) interface. 
In this case the surface of a colloidal particle floating 
at the $\alpha$--$\gamma$ interface would be fully covered 
by a film of the $\beta$ phase (see Fig.~\ref{fig:wetting_domain}(e)).
In scenario (vi) the wall--$\alpha$-liquid interface is wetted by a film of the $\beta$ phase,
whereas the wall--$\gamma$ interface is wetted by the $\alpha$ phase
(see Fig.~\ref{fig:wetting_domain}(f)).
 
If $\xi_{\text{w}}>1$ (i.e., the wall--$A$ interaction is stronger than the one prescribed by the strict mixing rule)
only the additional domain (i) appears (see Fig.~\ref{fig:relax_w}(b)).
In this scenario, the wall--$\alpha$-liquid and the wall--$\gamma$ interfaces are
both plain ones without an intruding wetting film (see Fig.~\ref{fig:wetting_domain}(a)).

\subsection{\label{subsection:dis_relax_ff_mix_rule}
            Relaxed mixing rule for the fluid--fluid interactions 
            and strict mixing rule for the fluid--wall interactions}

Here, we consider deviations from the strict mixing rule for the
fluid--fluid interactions, $\xi_{f}\neq1$, whereas the ratio
of the strengths of the interactions of the $A$ and $B$ particles 
with the wall is strictly fixed by the fluid--wall
mixing rule ($\xi_{\text{w}}=1$).
The condition for the formation of a composite $\alpha$--$\gamma$ interface with an
intervening $\beta$ film is now given by the general expression in Eq.~(\ref{eq:s_abg_rng}),
and thus the parameter space of interest is constrained to $0<X<X_{\xi_{f}}$,
with 
\begin{align}
\notag
   X_{\xi_{f}}=&\,\sqrt{\left[\dfrac{\xi_{f}}{2}
                     \left(D_{\alpha\beta}
                     +D_{\beta\gamma}\right)\right]^{2}
                     -D_{\alpha\beta}D_{\beta\gamma}}\\
              &+\dfrac{\xi_{f}}{2}
               \left(D_{\alpha\beta}+D_{\beta\gamma}\right) \, .
   \label{eq:relax_ff_abg}
\end{align}
Depending on whether $\xi_{f}<1$ or $\xi_{f}>1$, one has $X_{\xi_{f}}<D_{\alpha\beta}$
or $X_{\xi_{f}}>D_{\alpha\beta}$. 

The conditions for the formation of an intruding $\beta$ film at the wall--$\alpha$ interface
follow from Eqs.~(\ref{eq:s_wba_rng1}) and (\ref{eq:s_wba_rng2}) and can be expressed as
\begin{equation*}
   0<Y< \widetilde{Y}_{\text{w}\beta\alpha}(X) \,\,\, \mathrm{for} \,\,\, 0<X<D_{\alpha\beta} 
\end{equation*}
with
\begin{equation}
 \widetilde{Y}_{\text{w}\beta\alpha}(X) = X+1+\left(\xi_{f}-1\right)
            \left(1-D_{\alpha\beta}\right)
            \dfrac{X}{\left(X-D_{\alpha\beta}\right)}
\label{eq:relax_ff_wba}
\end{equation}
and
\begin{equation*}
   Y>\widetilde{Y}_{\text{w}\beta\alpha}(X) \,\,\, \mathrm{for} \,\,\, X>D_{\alpha\beta} \, . 
\end{equation*}
Within the interesting range $0<X<X_{\xi_{f}}$ of $X$, 
depending on the magnitude of $\xi_{f}$,
the following distinctions can be made in the above conditions:
\begin{enumerate}
   \item For $0<\xi_{f}<1$ and thus $X_{\xi_{f}}<D_{\alpha\beta}$, the wall--$\alpha$ interface
         is wetted by an intruding film of the $\beta$ phase if 
      \begin{itemize}
         \item[]  $0<Y<\widetilde{Y}_{\text{w}\beta\alpha}(X)$  
         \item[]  $\mathrm{\quad\,\,\,\;for\,\,the\,\,whole\,\,range} \,\,\,
                  0<X<X_{\xi_{f}}\, .$
      \end{itemize}
   \item For $\xi_{f}>1$ and thus for $X_{\xi_{f}}>D_{\alpha\beta}$, an intruding $\beta$ film at 
         the wall--$\alpha$ interface occurs if
      \begin{itemize}
         \item[] $0<Y< \widetilde{Y}_{\text{w}\beta\alpha}(X)   \,\,\, \mathrm{for} \,\,\,
                 0<X<D_{\alpha\beta}$
         \item[] or
         \item[] $Y>\widetilde{Y}_{\text{w}\beta\alpha}(X)
                 \,\,\, \mathrm{for} \,\,\,
                 D_{\alpha\beta}<X<X_{\xi_{f}} \, .$
      \end{itemize}
\end{enumerate}

From Eq.~(\ref{eq:s_wbg_rng}) one obtains that a wetting film of 
the $\beta$ phase at the wall--$\gamma$ interface is more favorable than 
a plain interface without a wetting film if
\begin{equation*}
   Y>\widetilde{Y}_{\text{w}\beta\gamma}(X)
\end{equation*}
with
\begin{equation}
 \widetilde{Y}_{\text{w}\beta\gamma}(X) = X+1+\left(\xi_{f}-1\right)
                  \left(1-D_{\beta\gamma}\right)
                  \dfrac{X}{\left(X-D_{\beta\gamma}\right)} \, .
\label{eq:relax_ff_wbg}
\end{equation}
From Eq.~(\ref{eq:s_wag_rng}) one finds that an intruding $\alpha$ film 
at the wall--$\gamma$ interface is more favorable than a plain interface if 
\begin{equation*}
   Y>\widetilde{Y}_{\text{w}\alpha\gamma}(X)
\end{equation*}
with
\begin{widetext}
\begin{align}
\widetilde{Y}_{\text{w}\alpha\gamma}(X) =\dfrac{\rho_{B,\alpha}}{\rho_{B,\beta}}X
              +\dfrac{\rho_{A,\alpha}}{\rho_{A,\beta}}
              +\left(\xi_{f}-1\right)
              \left(\dfrac{\rho_{A,\alpha}}{\rho_{A,\beta}}
              -\dfrac{\rho_{B,\alpha}}{\rho_{B,\beta}}
              D_{\alpha\gamma}\right)
              \dfrac{X}{\left(X-D_{\alpha\gamma}\right)} \, .
\label{eq:relax_ff_wag}
\end{align}
\end{widetext}
We recall that in addition we are interested only in $X$ values 
within the interval $0<X<X_{\xi_{f}}$.

In the parameter region, in which wetting of the wall--$\gamma$ interface 
both by the $\alpha$ and by the $\beta$ phase is more
favorable than a plain wall--$\gamma$ interface without any
wetting film, we still have to determine whether wetting
by a film of the $\alpha$ phase or of the $\beta$ phase is preferred.
This distinction hinges on the sign of 
$W_{\text{w}\beta(\alpha)\gamma} =: \pi{a}_{AA}^{4}M(X,Y)$
(see Eq.~(\ref{eq:w_wb(a)g})) with 
\begin{widetext}
\begin{align*}
   M(X,Y)=\epsilon_{AA}\rho_{A,\beta}^{2}
          \bm{\,\bigg(}&\bigg\{
          \left[\left(1-\dfrac{\rho_{B,\gamma}}
                               {\rho_{B,\beta}}\right)X
                +1-\dfrac{\rho_{A,\gamma}}
                               {\rho_{A,\beta}}\right]
           \left(X+1-Y\right)\\
         &\;\;\,\,+\left[\left(1-\dfrac{\rho_{B,\gamma}}
                               {\rho_{B,\beta}}\right)
                +\left(1-\dfrac{\rho_{A,\gamma}}
                               {\rho_{A,\beta}}\right)
           \right]\left(\xi_{f}-1\right)X\bigg\}
           F_{AA}(l_{\text{w}\beta\gamma})\\
         &-\bigg\{
          \left[\left(\dfrac{\rho_{B,\alpha}}{\rho_{B,\beta}}
                      -\dfrac{\rho_{B,\gamma}}{\rho_{B,\beta}}
                 \right)X
                +\dfrac{\rho_{A,\alpha}}{\rho_{A,\beta}}
                      -\dfrac{\rho_{A,\gamma}}{\rho_{A,\beta}}
                 \right]
           \left(\dfrac{\rho_{B,\alpha}}{\rho_{B,\beta}}X
                +\dfrac{\rho_{A,\alpha}}{\rho_{A,\beta}}
                -Y\right)\\
       &\,\qquad+\left[\left(\dfrac{\rho_{B,\alpha}}{\rho_{B,\beta}}
                    -\dfrac{\rho_{B,\gamma}}{\rho_{B,\beta}}
               \right)\dfrac{\rho_{A,\alpha}}{\rho_{A,\beta}}
              +\left(\dfrac{\rho_{A,\alpha}}{\rho_{A,\beta}}
                    -\dfrac{\rho_{A,\gamma}}{\rho_{A,\beta}}
               \right)
               \dfrac{\rho_{B,\alpha}}{\rho_{B,\beta}}
         \right]\left(\xi_{f}-1\right)X\bigg\}
        \hat{F}_{AA}(l_{\text{w}\alpha\gamma})\bm{\bigg)} \, .
\end{align*}
\end{widetext}
For $M(X,Y)>0$ wetting of the wall--$\gamma$ interface by an $\alpha$ film
is more favorable than wetting by a $\beta$ film; for $M(X,Y)<0$
wetting by the $\beta$ phase is preferred. The condition $M(X,Y)<0$ can be
rewritten as
\begin{widetext}
\begin{align*}
   P_{M(X,Y)}\times Y>
      &\,\bigg\{
     \left[\left(1-\dfrac{\rho_{B,\gamma}}
                          {\rho_{B,\beta}}\right)X
           +1-\dfrac{\rho_{A,\gamma}}
                          {\rho_{A,\beta}}\right]
      \left(X+1\right)
    +\left[\left(1-\dfrac{\rho_{B,\gamma}}
                          {\rho_{B,\beta}}\right)
           +\left(1-\dfrac{\rho_{A,\gamma}}
                          {\rho_{A,\beta}}\right)
      \right]\left(\xi_{f}-1\right)X\bigg\}
      F_{AA}(l_{\text{w}\beta\gamma})\\
     &\,-\bigg\{
     \left[\left(\dfrac{\rho_{B,\alpha}}{\rho_{B,\beta}}
                 -\dfrac{\rho_{B,\gamma}}{\rho_{B,\beta}}
            \right)X
           +\dfrac{\rho_{A,\alpha}}{\rho_{A,\beta}}
                 -\dfrac{\rho_{A,\gamma}}{\rho_{A,\beta}}
            \right]
      \left(\dfrac{\rho_{B,\alpha}}{\rho_{B,\beta}}X
           +\dfrac{\rho_{A,\alpha}}{\rho_{A,\beta}}\right)\\
    &\,\,\qquad+\left[\left(\dfrac{\rho_{B,\alpha}}{\rho_{B,\beta}}
                 -\dfrac{\rho_{B,\gamma}}{\rho_{B,\beta}}
            \right)\dfrac{\rho_{A,\alpha}}{\rho_{A,\beta}}
           +\left(\dfrac{\rho_{A,\alpha}}{\rho_{A,\beta}}
                 -\dfrac{\rho_{A,\gamma}}{\rho_{A,\beta}}
            \right)\dfrac{\rho_{B,\alpha}}{\rho_{B,\beta}}
      \right]\left(\xi_{f}-1\right)X\bigg\}
      \hat{F}_{AA}(l_{\text{w}\alpha\gamma}) \, ,
\end{align*}
\end{widetext}
with $P_{M(X,Y)}=P_{K(X,Y)}$, where $P_{K(X,Y)}$ is given by
Eqs.~(\ref{eq:p_k(x,y)}) and~(\ref{eq:p_k(x,y)2}).

\begin{figure*}[!t]
   \includegraphics[trim={0 0.5cm -1cm 0.5cm},width=0.4\textwidth]{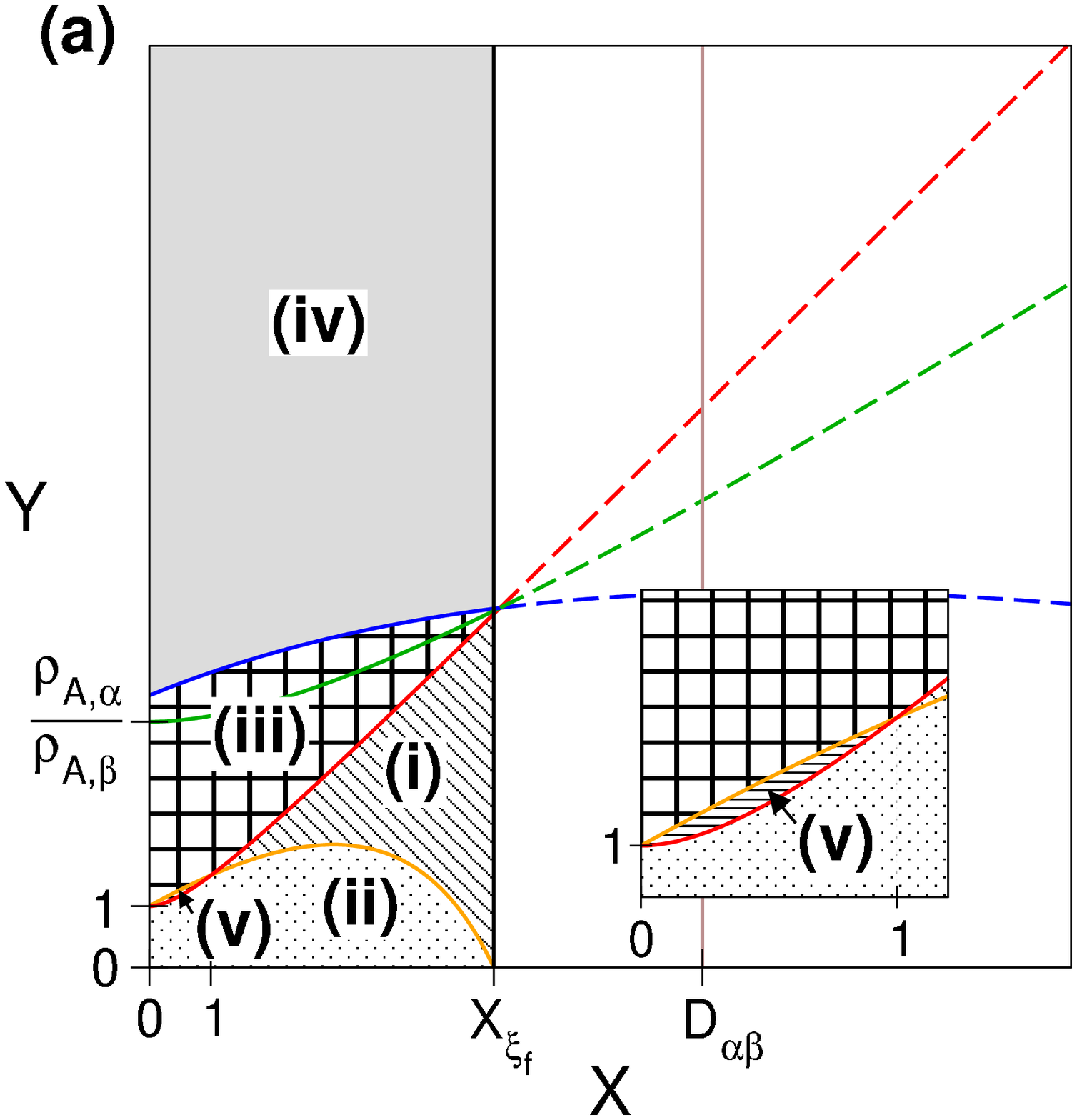}
   \includegraphics[trim={-1cm 0.5cm 0 0.5cm},width=0.4\textwidth]{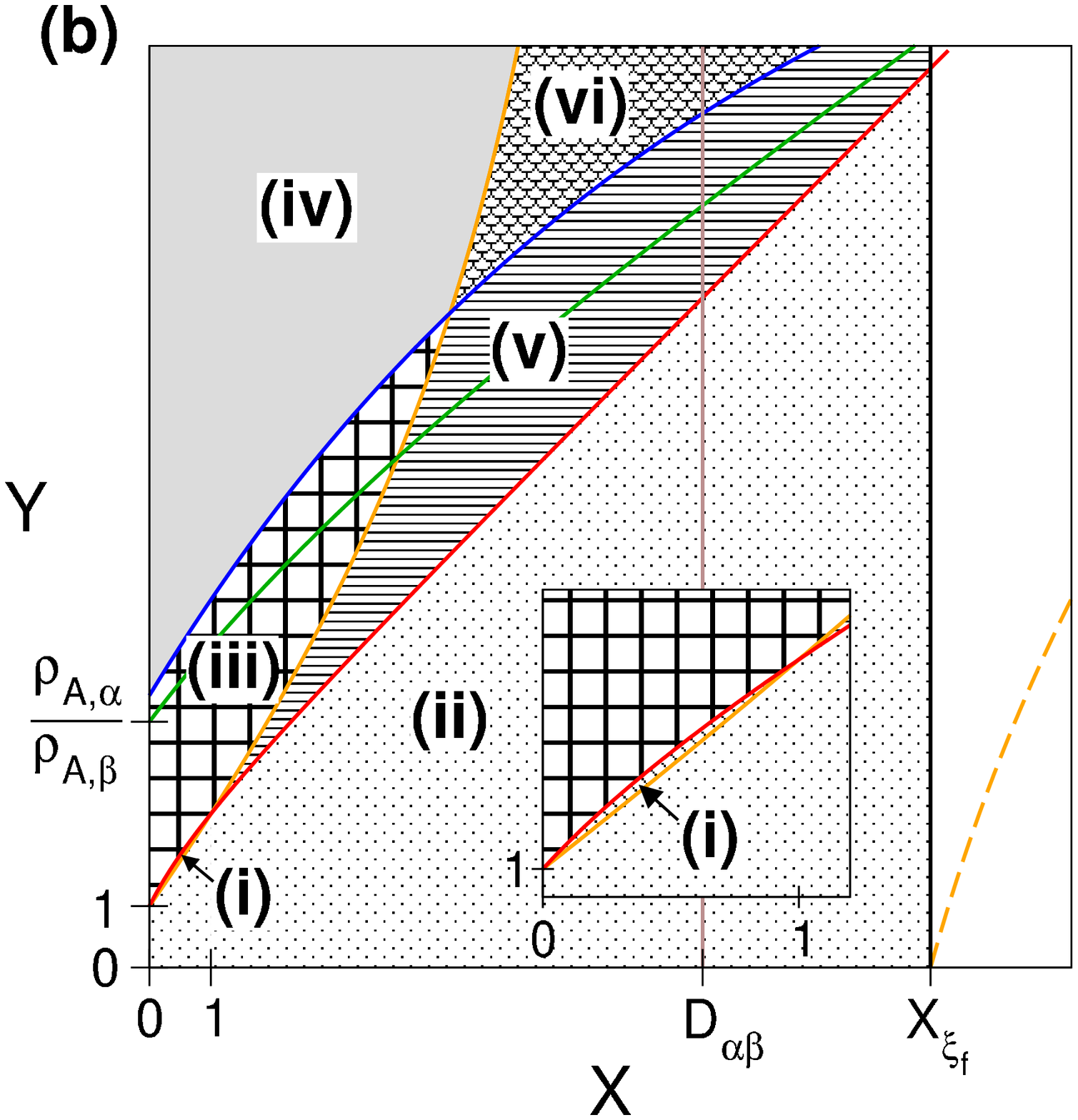}
   \caption{
      Wetting domains in the parameter space ($X$,$Y$), in the case that
      the fluid--wall interactions obey the strict mixing rule, but that the
      fluid--fluid interactions are not constrained ($\xi_{f} \neq 1$).
      Furthermore the inequalities in Eq.~(\ref{eq:rel_densities}) between the number densities
      are respected. The domains correspond to the wetting scenarios depicted in Fig.~\ref{fig:wetting_domain}.
      Two different scenarios emerge in addition, depending on whether (a) $0<\xi_{f}<1$
      or (b) $\xi_{f}>1$.
      Only to the left of the black line $X=X_{\xi_{f}}$ (Eq.~(\ref{eq:relax_ff_abg})), the desired composite
      $\alpha$--$\gamma$ interface with an intruding $\beta$ film can occur. 
      Above the red line $\widetilde{Y}_{\text{w}\beta\gamma}(X)$ (Eq.~(\ref{eq:relax_ff_wbg}))
      the wall--$\gamma$ interface is potentially wetted by a $\beta$ film. 
      Above the green line $\widetilde{Y}_{\text{w}\alpha\gamma}(X)$ (Eq.~(\ref{eq:relax_ff_wag})) 
      the wall--$\gamma$ interface is potentially wetted by an $\alpha$ film. 
      The separatrix between wetting by the $\beta$ phase and the $\alpha$ phase, respectively, 
      is given by the blue line
      $Y_{M(X,Y)}$ (Eq.~(\ref{eq:y_m(x,y)})).
      Depending on whether the green line lies above the red line or below,
      wetting of the wall--$\gamma$ interface by an $\alpha$ film is found above
      or below the blue line.
      Below the orange line $\widetilde{Y}_{\text{w}\beta\alpha}(X)$ (Eq.~(\ref{eq:relax_ff_wba}))
      the wall--$\alpha$-liquid interface is wetted by a film of the $\beta$ phase.
      The dashed lines extend the domain boundaries into the region within which
      the $\alpha$--$\gamma$ interface is a plain one 
      without an intruding $\beta$ film. 
      In particular, the dashed orange line in (b) represents 
      $\widetilde{Y}_{\text{w}\beta\alpha}(X)$ for $X>X_{\xi_{f}}$.
      The vertical asymptote of $\widetilde{Y}_{\text{w}\beta\alpha}(X)$, 
      at which in addition this function changes sign, is located at $X=D_{\alpha\beta}$.
      $\widetilde{Y}_{\text{w}\beta\alpha}(X)$
      returns to positive values for $X>X_{\xi_{f}}$
      (see Eq.~(\ref{eq:relax_ff_wba})).
      The brown vertical line shows $X=D_{\alpha\beta}$. 
      The lines are drawn using $\rho_{A,\alpha}/\rho_{A,\beta}=4$,
      $\rho_{B,\alpha}/\rho_{B,\beta}=2/3$, which implies $D_{\alpha\beta}=9$,
      $F_{AA}(l_{\text{w}\beta\gamma})=13/132$, and
      $\hat{F}_{AA}(l_{\text{w}\alpha\gamma})=13/66$.
      In (a) we have chosen $\xi_{f}=0.5$ and in (b) $\xi_{f}=1.5$ 
      so that $X_{\xi_{f}}=5.605$ in (a) and $X_{\xi_{f}}=12.709$ in (b).
      The insets are magnifications of the regions indicated by
      arrows in the main figures.
   }
   \label{fig:relax_f}
\end{figure*}

Depending on the magnitude of $\xi_{f}$, the following distinctions
can be made:
\begin{enumerate} 
   \item For $0<\xi_{f}<1$ and thus $X_{\xi_{f}}<D_{\alpha\beta}$,
         $P_{M(X,Y)}$ is negative in the whole range 
         $0<X<X_{\xi_{f}}$ of interest, because $P_{M(X,Y)}<0$ for $0<X<D_{\alpha\beta}$.
   \item For $\xi_{f}>1$ and thus $X_{\xi_{f}}>D_{\alpha\beta}$, two
         regions have to be distinguished: $P_{M(X,Y)}<0$ for $0<X<D_{\alpha\beta}$,
         whereas within $D_{\alpha\beta}<X<X_{\xi_{f}}$ $P_{M(X,Y)}$ can be positive or 
         negative.
\end{enumerate} 

For $P_{M(X,Y)}>0$ the wall--$\gamma$ interface
is wetted by a $\beta$ film only if $Y>Y_{M(X,Y)}$, with  
\begin{widetext}
\begin{align}
\notag
   Y_{M(X,Y)}=&\,\bigg\{
             \left[\left(1-\dfrac{\rho_{B,\gamma}}
                                  {\rho_{B,\beta}}\right)X
                   +1-\dfrac{\rho_{A,\gamma}}
                                  {\rho_{A,\beta}}
              \right]\left(X+1\right)
            +\left[\left(1-\dfrac{\rho_{B,\gamma}}
                                  {\rho_{B,\beta}}\right)
                   +\left(1-\dfrac{\rho_{A,\gamma}}
                                  {\rho_{A,\beta}}\right)
              \right]\left(\xi_{f}-1\right)X\bigg\}
              \dfrac{F_{AA}(l_{\text{w}\beta\gamma})}
                    {P_M(X,Y)}\\
\notag
             &\,-\bigg\{
             \left[\left(\dfrac{\rho_{B,\alpha}}{\rho_{B,\beta}}
                         -\dfrac{\rho_{B,\gamma}}{\rho_{B,\beta}}
                    \right)X
                   +\dfrac{\rho_{A,\alpha}}{\rho_{A,\beta}}
                         -\dfrac{\rho_{A,\gamma}}{\rho_{A,\beta}}
                    \right]
              \left(\dfrac{\rho_{B,\alpha}}{\rho_{B,\beta}}X
                   +\dfrac{\rho_{A,\alpha}}{\rho_{A,\beta}}
              \right)\\
            &\,\,\qquad+\left[\left(\dfrac{\rho_{B,\alpha}}{\rho_{B,\beta}}
                         -\dfrac{\rho_{B,\gamma}}{\rho_{B,\beta}}
                    \right)
                    \dfrac{\rho_{A,\alpha}}{\rho_{A,\beta}}
                   +\left(\dfrac{\rho_{A,\alpha}}{\rho_{A,\beta}}
                         -\dfrac{\rho_{A,\gamma}}{\rho_{A,\beta}}
                    \right)
                    \dfrac{\rho_{B,\alpha}}{\rho_{B,\beta}}
              \right]\left(\xi_{f}-1\right)X\bigg\}
              \dfrac{\hat{F}_{AA}(l_{\text{w}\alpha\gamma})}
                    {P_{M(X,Y)}} \, .
   \label{eq:y_m(x,y)}
\end{align}
\end{widetext}
If $P_{M(X,Y)}<0$, wetting of the wall--$\gamma$ interface by a $\beta$ film 
is preferred only if $0<Y<Y_{M(X,Y)}$.

For
$$\widetilde{Y}_{\text{w}\beta\gamma}(X) < \widetilde{Y}_{\text{w}\alpha\gamma}(X) \, ,$$
$Y_{M(X,Y)}$ (Eq.~(\ref{eq:y_m(x,y)})) lies always above 
$\widetilde{Y}_{\text{w}\alpha\gamma}(X)$.
Otherwise, $Y_{M(X,Y)}$ lies below $\widetilde{Y}_{\text{w}\alpha\gamma}(X)$
(Eq.~(\ref{eq:relax_ff_wag})). 
The intersection between $Y_{M(X,Y)}$ and $\widetilde{Y}_{\text{w}\alpha\gamma}(X)$ 
is located on the curve $\widetilde{Y}_{\text{w}\beta\gamma}(X)$ (Eq.~(\ref{eq:relax_ff_wbg})),
which means that $\widetilde{Y}_{\text{w}\alpha\gamma}(X)$, $\widetilde{Y}_{\text{w}\beta\gamma}(X)$,
and $Y_{M(X,Y)}$ have a common intersection point. 

In Fig.~\ref{fig:relax_f} we illustrate how the parameter space ($X$,$Y$) is subdivided into
domains, which are associated with the various wetting scenarios, in the case that deviations from the
strict mixing rule for the fluid--fluid interactions are admitted. 
In Fig.~\ref{fig:relax_f}(a) the division into domains for $0<\xi_{f}<1$ 
(i.e., the strength of the $A$--$B$ interaction is smaller than prescribed by the strict mixing rule)
is shown. Figure~\ref{fig:relax_f}(a) resembles closely Fig.~\ref{fig:relax_w}(b)
(i.e., deviations from the mixing rule concerning the fluid--wall interactions and $\xi_{\text{w}}>1$).
However, in the present case an additional domain associated with the wetting scenario (v)
appears, although only in a very small region of the parameter space.
In the case $\xi_{f}>1$ (see Fig.~\ref{fig:relax_f}(b)) one obtains a picture
which is very similar to the one shown in Fig.~\ref{fig:relax_w}(a), but with
an additional domain, occupying also only a very small region of the parameter space,
corresponding to scenario (i). 

\subsection{\label{subsection:contact_angles}Contact angles}

The equilibrium contact angle $\theta$, with which the 
liquid--vapor ($\alpha$--$\gamma$) interface meets the wall,
is a measurable observable.
It can be expressed via Young's equation, 
\begin{equation}
   \cos\theta = \dfrac{\sigma^{\text{eq}}_{\text{w}\gamma}
                               -\sigma^{\text{eq}}_{\text{w}\alpha}}
                               {\sigma^{\text{eq}}_{\alpha\gamma}} \, ,
   \label{eq:contact_angle_def}
\end{equation}
in terms of the interfacial tensions of the wall--$\gamma$, wall--$\alpha$, and $\alpha$--$\gamma$ interfaces.
The tensions correspond to the respective equilibrium structures.
The various wetting domains introduced above are characterized by 
combinations of interfacial structures at the wall--$\gamma$ and wall--$\alpha$ interfaces.
One may pose the question whether this is reflected by the possible values of the contact angle $\theta$.
For instance, it might be the case that 
in one domain the wall must be lyophilic (i.e., $\theta < \pi/2$)
whereas in another domain the wall must be lyophobic (i.e., $\theta > \pi/2$). 
However, it is also conceivable that in one domain 
both lyophilic and lyophobic behaviors are possible 
and that there is a dividing line, inside the domain, 
separating the two behaviors. 

Here we focus on the case in which the mixing rules apply to both the fluid--fluid
and the fluid--wall interactions. In this particular case the parameter space
is divided into three distinct wetting domains.
Now we relate Eq.~(\ref{eq:contact_angle_def}) to this case and to the three
wetting domains (ii), (iii), and (iv) by inserting the interfacial tensions for
the respective interfacial structures and by using the notation introduced
above. This leads to the three expressions
\begin{align*}
   \cos\theta_{\text{(ii)}}
      &=\dfrac{\sigma_{\text{w}\gamma}
              -\sigma_{\text{w}\beta\alpha}}
              {\sigma_{\alpha\beta\gamma}} \, ,\\
   \cos\theta_{\text{(iii)}}
      &=\dfrac{\sigma_{\text{w}\beta\gamma}
              -\sigma_{\text{w}\alpha}}
              {\sigma_{\alpha\beta\gamma}} \, ,
\end{align*}
and 
\begin{equation*}
   \cos\theta_{\text{(iv)}}
      =\dfrac{\sigma_{\text{w}\alpha\gamma}
             -\sigma_{\text{w}\alpha}}
             {\sigma_{\alpha\beta\gamma}} \, .
\end{equation*}
Here, 
$\theta_{\kappa}$ is the contact angle according
to Young's equation specialized to domain $\kappa$,
with $\kappa=\text{(ii), (iii), and (iv)}$.
If $\cos\theta_{\kappa}>0$, we have $\theta_{\kappa}<\dfrac{\pi}{2}$.
Otherwise, $\theta_{\kappa}>\dfrac{\pi}{2}$.
The sign of $\cos\theta_{\kappa}$ is determined by the numerators 
in the above expressions because $\sigma_{\alpha\beta\gamma}$ is positive. 
By using Eqs.~(\ref{eq:w_wba})-(\ref{eq:w_wag}), 
these numerators, called $\psi_{\kappa}$, 
can be expressed as follows:
\begin{align}
\notag
   \psi_{\text{(ii)}}=&\,\sigma_{\text{w}\gamma}
                        -\sigma_{\text{w}\beta\alpha}\\  
                     =&\,\sigma_{\text{w}\gamma}
                        -\sigma_{\text{w}\alpha}
                        -W_{\text{w}\beta\alpha} \, ,
   \label{eq:psi_w2}  
\end{align}	
\begin{align}   
\notag
   \psi_{\text{(iii)}}=&\,\sigma_{\text{w}\beta\gamma}
                         -\sigma_{\text{w}\alpha}\\  
                      =&\,W_{\text{w}\beta\gamma}
                         +\sigma_{\text{w}\gamma}
                         -\sigma_{\text{w}\alpha} \, ,
   \label{eq:psi_w3}
\end{align}
and
\begin{align}
\notag
   \psi_{\text{(iv)}}=&\,\sigma_{\text{w}\alpha\gamma}
                        -\sigma_{\text{w}\alpha}\\  
                     =&\,W_{\text{w}\alpha\gamma}
                        +\sigma_{\text{w}\gamma}
                        -\sigma_{\text{w}\alpha} \, .
   \label{eq:psi_w4}
\end{align}

Preliminary conclusions regarding the sign of $\cos\theta$ can be drawn 
based already on the sign of $\sigma_{\text{w}\gamma} - \sigma_{\text{w}\alpha}$
and our knowledge that $W_{\text{w}\beta\alpha}$, $W_{\text{w}\beta\gamma}$,
and $W_{\text{w}\alpha\gamma}$ are negative within the respective domain
for which Eqs.~(\ref{eq:psi_w2}),~(\ref{eq:psi_w3}), and~(\ref{eq:psi_w4})
are applicable. We find $\sigma_{\text{w}\gamma} - \sigma_{\text{w}\alpha} = 0$
if $Y = Y_{\text{ref}}(X)$ with
\begin{equation}
   Y_{\text{ref}}(X)=\dfrac{1}{2}
      \left[\left(\dfrac{\rho_{B,\alpha}}
                        {\rho_{B,\beta}}
                 +\dfrac{\rho_{B,\gamma}}
                        {\rho_{B,\beta}}\right)X
           +\dfrac{\rho_{A,\alpha}}
                        {\rho_{A,\beta}}
                 +\dfrac{\rho_{A,\gamma}}
                        {\rho_{A,\beta}}
      \right]
   \label{eq:y_ref}
\end{equation}
(see Appendix~\ref{sec:cond_for_sigma_wa_sigma_wg}). 
The difference $\sigma_{\text{w}\gamma} - \sigma_{\text{w}\alpha}$
is positive if $Y > Y_{\text{ref}}$,
and it is negative if $0<Y<Y_{\text{ref}}$.
Based on the inequalities in Eq.~(\ref{eq:rel_densities}) between the number densities,
we also know that $Y_{\text{ref}}(X)$ lies above $Y = X + 1$, 
which is the boundary between the domains (ii) and (iii) for small $X$. 
$Y_{\text{ref}}(X)$ intersects
$Y = X + 1$ at $X_{\text{int}}$ (see, c.f., Eq.~(\ref{eq:x_y_int})) 
and is located below $Y = X + 1$ for $X > X_{\text{int}}$.
It also follows that $Y_{\text{ref}}(X)$ is located below the green line 
in Fig.~\ref{fig:strict} and thus it is located below the domain (iv). 
From Eq.~(\ref{eq:psi_w4}) and the sign of $W_{\text{w}\alpha\gamma}$ we infer
that the line, above which $\theta_{\kappa}<\dfrac{\pi}{2}$, 
must be located above $Y_{\text{ref}}(X)$; only if this shift is unexpectedly large this boundary
would move up into domain (iv). Thus, it is very likely that domain (iv)
does not contain a boundary between lyophilic and lyophobic behavior 
so that the whole domain (iv) is linked to lyophilic walls. 
The knowledge acquired up to this point can be summarized as follows:
\begin{itemize}
   \item In the wetting domain (ii) one has $\theta_{\text{(ii)}}<\dfrac{\pi}{2}$  
         for $Y>Y_{\text{ref}}(X)$; the actual boundary $\theta_{\text{(ii)}}=\dfrac{\pi}{2}$
         is located below $Y_{\text{ref}}(X)$. 
   \item In the wetting domain (iii) one has $\theta_{\text{(iii)}}>\dfrac{\pi}{2}$
         within the interval $0<Y<Y_{\text{ref}}(X)$; the actual boundary $\theta_{\text{(iii)}}=\dfrac{\pi}{2}$
         is located above $Y_{\text{ref}}(X)$. 
   \item In the wetting domain (iv), it is likely that  
         $\theta_{\text{(iv)}}<\dfrac{\pi}{2}$ inside the entire domain.
\end{itemize}

\begin{figure*}[!t]
   \includegraphics[trim={0 0.5cm -1cm 0.5cm},width=0.4\textwidth]{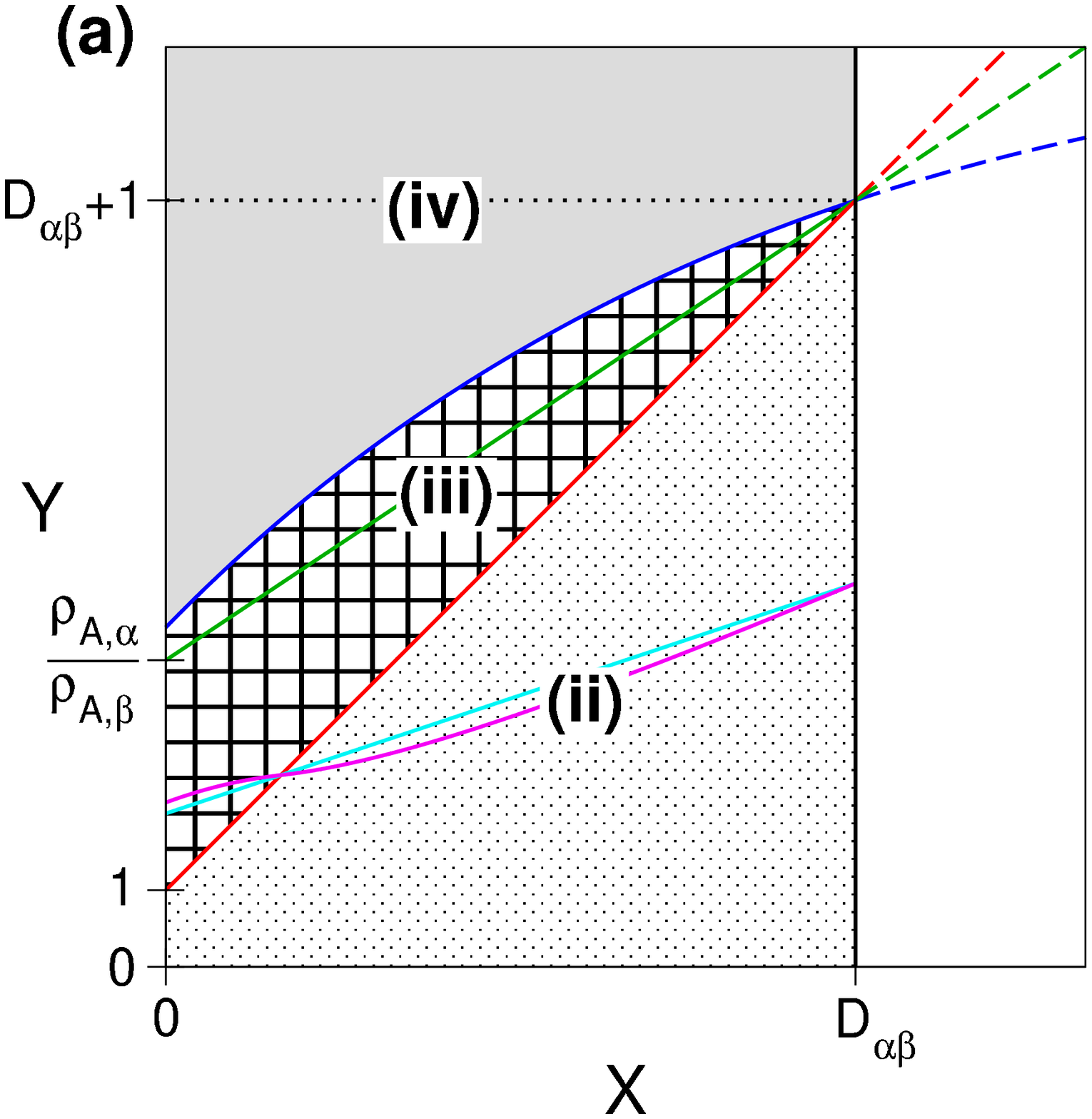}
   \includegraphics[trim={-1cm 0.5cm 0 0.5cm},width=0.4\textwidth]{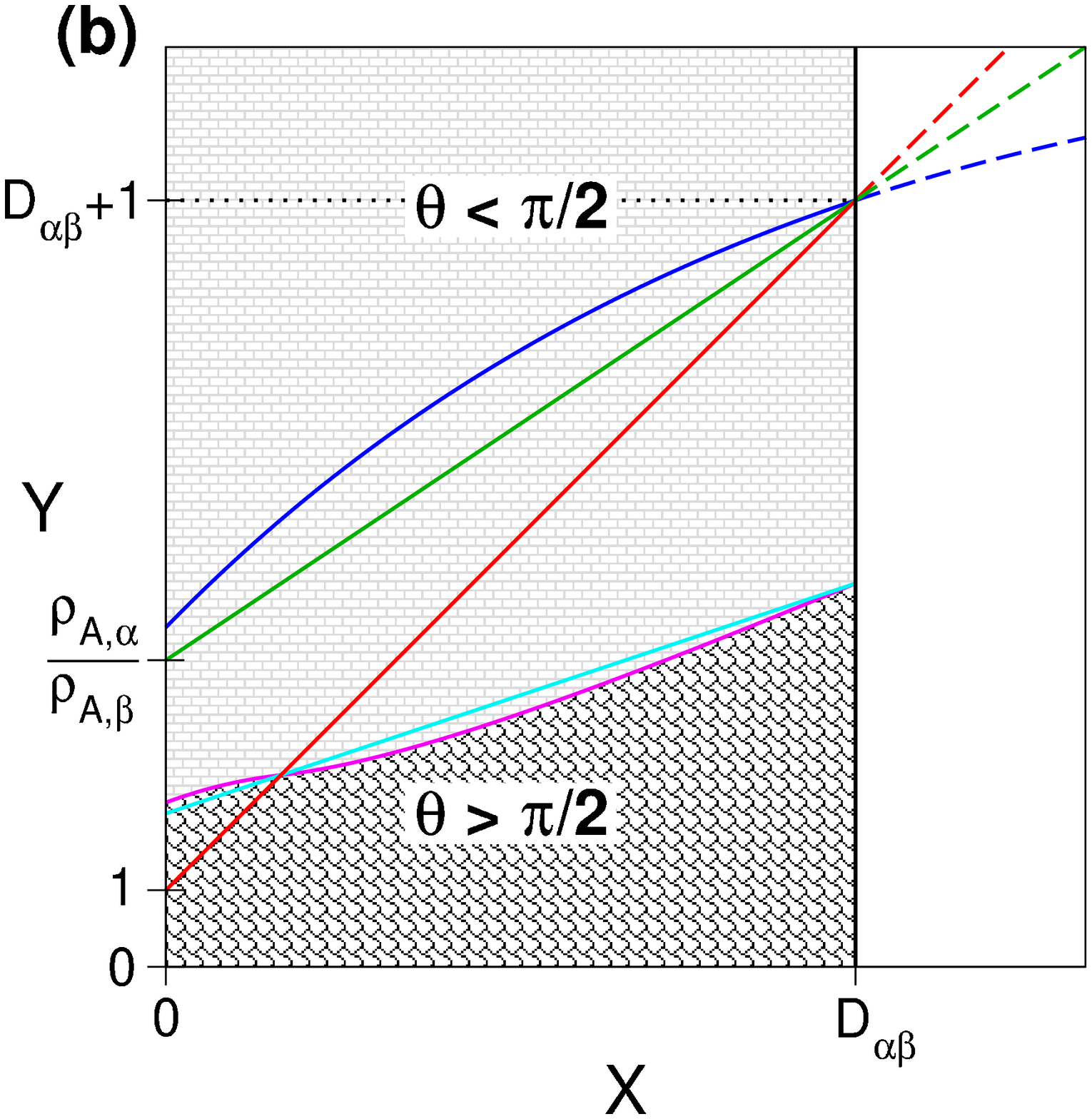}
   \includegraphics[trim={0 2cm -1cm 0},width=0.4\textwidth]{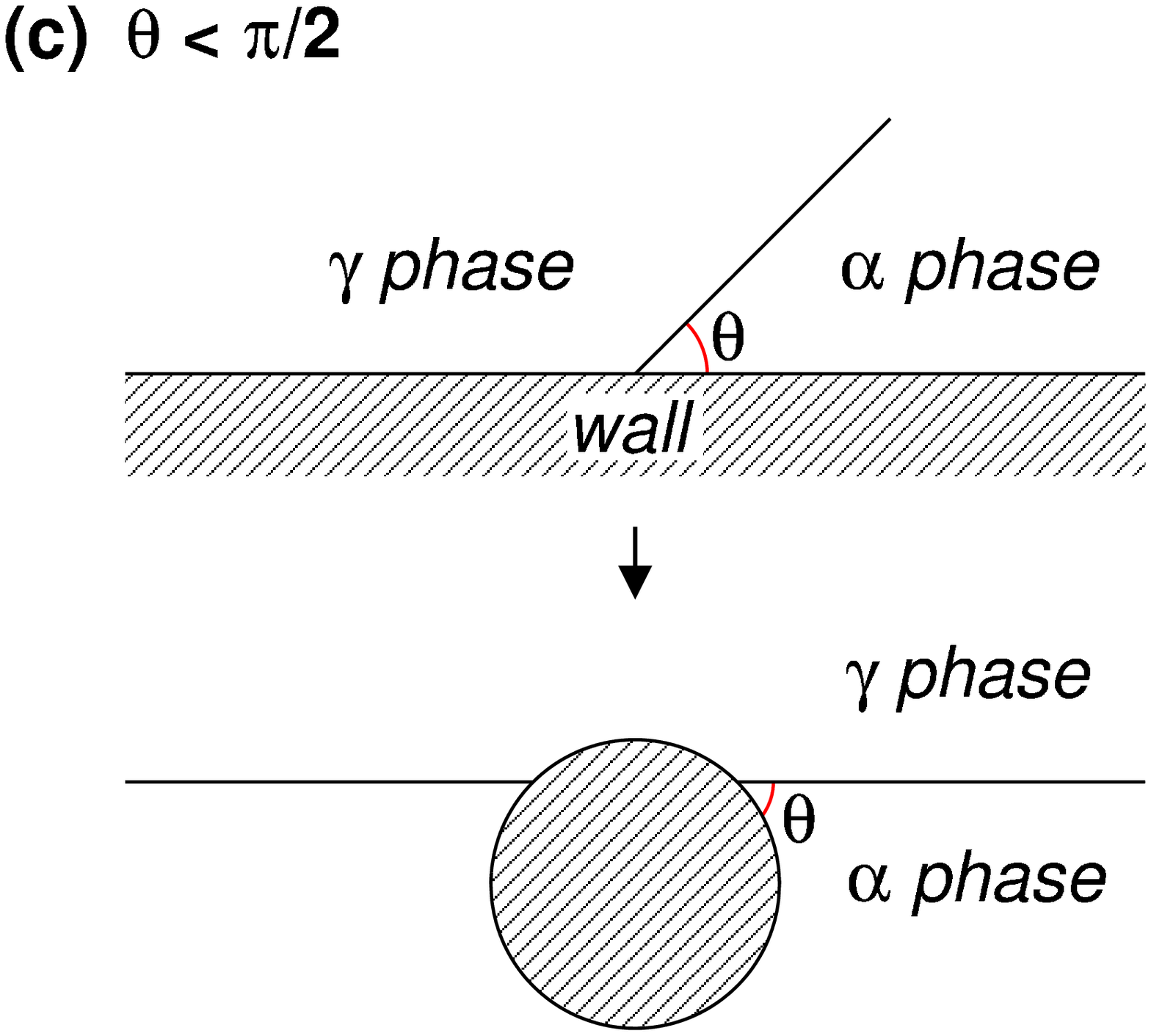}
   \includegraphics[trim={-1cm 2cm 0 0},width=0.4\textwidth]{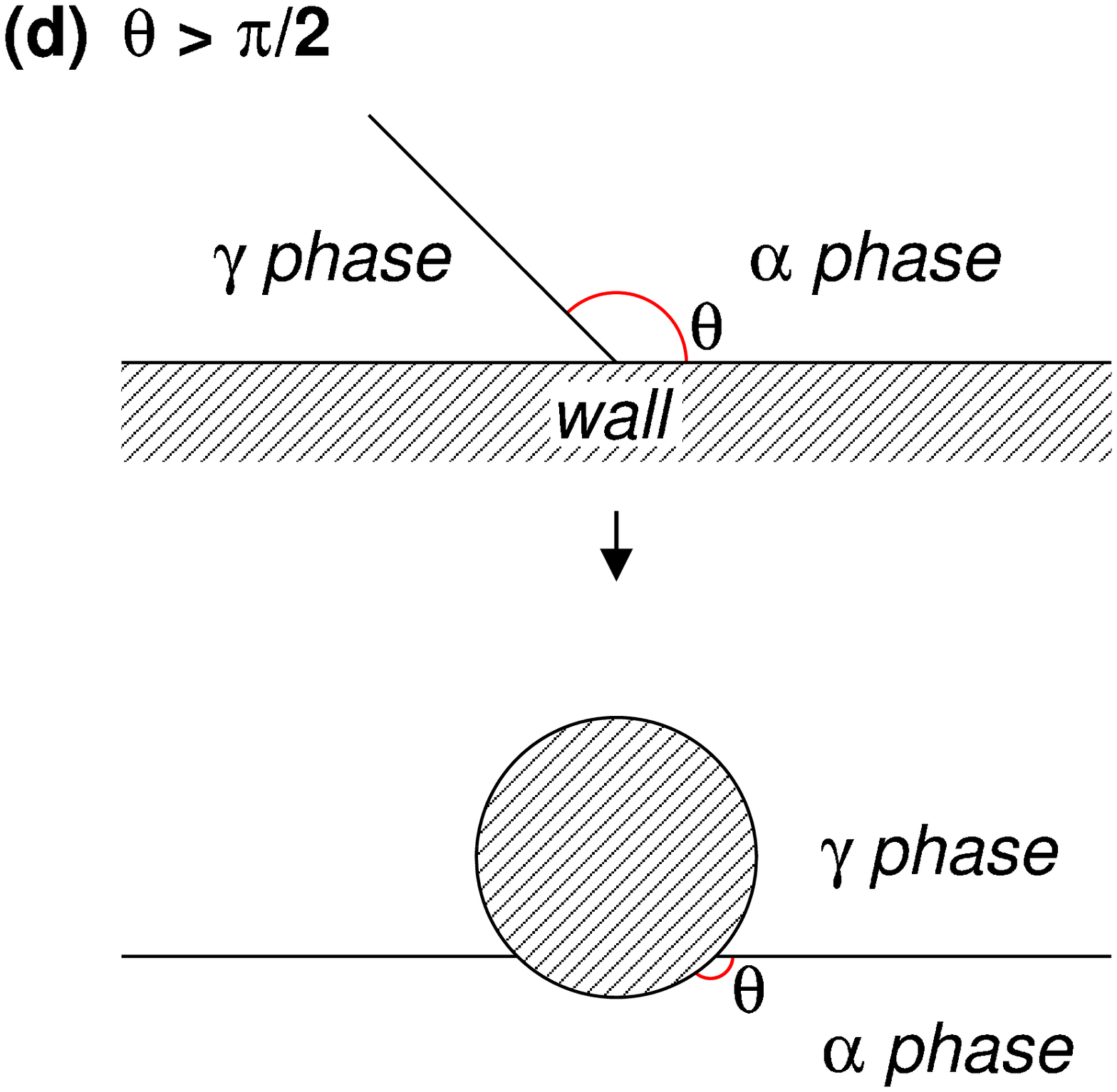}
   \caption{
      In (a) the three wetting domains depicted in Fig.~\ref{fig:strict} are shown again 
      together with $Y_{\text{ref}}$ (cyan straight line, Eq.~(\ref{eq:y_ref})) 
      and $Y_{G_{\kappa}(X,Y)}$ (magenta curve, Eqs.~(\ref{eq:y_g2}) and~(\ref{eq:y_g3}))
      for $\kappa=\text{(ii) and (iii)}$. 
      The latter is the separatrix between lyophobic ($\theta>\dfrac{\pi}{2}$)
      and lyophilic ($\theta<\dfrac{\pi}{2}$) behavior; the first is an 
      approximation to this separatrix. The model used here 
      (such as the constraints to interaction parameters, etc.) 
      is the same as the one explained in Fig.~\ref{fig:strict}.
      In (b) the domains of lyophobic and lyophilic behavior are shown.
      In (c) and (d) sketches of the two contact angle scenarios are given, for
      a planar wall and for a colloid floating at a liquid--vapor interface,
      neglecting gravity effects.  
      In (a) and (b) the curves correspond to the choices 
      $\rho_{A,\alpha}/\rho_{A,\beta}=4$, $\rho_{B,\alpha}/\rho_{B,\beta}=2/3$, 
      which implies $D_{\alpha\beta}=9$,
      $F_{AA}(l_{\text{w}\beta\gamma})=13/132$, and
      $\hat{F}_{AA}(l_{\text{w}\alpha\gamma})=13/66$.
   }
   \label{fig:strict_ca}
\end{figure*}

In order to locate the boundary $\theta_{\kappa} = \dfrac{\pi}{2}$
precisely, we study the full expressions on the right hand sides of 
Eqs.~(\ref{eq:psi_w2})-(\ref{eq:psi_w4}) and determine the separatrix 
$Y_{G_{\kappa}}$ between lyophilic behavior
($\theta_{\kappa}<\dfrac{\pi}{2}$, $\psi_{\kappa}>0$) and lyophobic behavior 
($\theta_{\kappa}>\dfrac{\pi}{2}$, $\psi_{\kappa}<0$) in each case.   

For the domain (ii) we find
\begin{widetext}
\begin{align}
\notag
   Y_{G_{\text{(ii)}}(X,Y)}
   =X+1
    +&\,\dfrac{13}{132}
        \left[\left(\dfrac{\rho_{B,\alpha}}
                          {\rho_{B,\beta}}
                   -\dfrac{\rho_{B,\gamma}}
                          {\rho_{B,\beta}}\right)X
             +\dfrac{\rho_{A,\alpha}}
                    {\rho_{A,\beta}}
             -\dfrac{\rho_{A,\gamma}}
                    {\rho_{A,\beta}}\right]\\
             &\,\times
             \bigg[\left(\dfrac{\rho_{B,\alpha}}
                          {\rho_{B,\beta}}
                   +\dfrac{\rho_{B,\gamma}}
                          {\rho_{B,\beta}}-2\right)X
             +\dfrac{\rho_{A,\alpha}}
                    {\rho_{A,\beta}}
             +\dfrac{\rho_{A,\gamma}}
                    {\rho_{A,\beta}}-2\bigg]
        \dfrac{1}{P_{G_{\text{(ii)}}(X,Y)}}
   \label{eq:y_g2}
\end{align}
with
\begin{align*} 
   P_{G_{\text{(ii)}}(X,Y)}
      =\dfrac{13}{66}
        \left[\left(1-\dfrac{\rho_{B,\gamma}}
                            {\rho_{B,\beta}}\right)X
             +1-\dfrac{\rho_{A,\gamma}}
                      {\rho_{A,\beta}}\right]
       +\left[\dfrac{13}{66}
             -F_{AA}(l_{\text{w}\beta\alpha})\right]
        \left[\left(\dfrac{\rho_{B,\alpha}}
                          {\rho_{B,\beta}}-1\right)X
             +\dfrac{\rho_{A,\alpha}}
                    {\rho_{A,\beta}}-1\right] \, .
\end{align*}
\end{widetext}
$P_{G_{\text{(ii)}}(X,Y)}$ is positive in the relevant interval $0<X<D_{\alpha\beta}$.
In these terms we can state that 
$\theta_{\text{(ii)}}<\dfrac{\pi}{2}$ for $Y>Y_{G_{\text{(ii)}}(X,Y)}$
and $\theta_{\text{(ii)}}>\dfrac{\pi}{2}$ for $0<Y<Y_{G_{\text{(ii)}}(X,Y)}$.

For the domain (iii) we find 
\begin{widetext}
\begin{align}
\notag
   Y_{G_{\text{(iii)}}(X,Y)}
      =X+1
      +&\,\dfrac{13}{132}
        \left[\left(\dfrac{\rho_{B,\alpha}}
                             {\rho_{B,\beta}}
                      -\dfrac{\rho_{B,\gamma}}
                             {\rho_{B,\beta}}\right)X
                +\dfrac{\rho_{A,\alpha}}
                       {\rho_{A,\beta}}
                -\dfrac{\rho_{A,\gamma}}
                       {\rho_{A,\beta}}\right]\\
          &\,\times
          \bigg[\left(\dfrac{\rho_{B,\alpha}}
                             {\rho_{B,\beta}}
                      +\dfrac{\rho_{B,\gamma}}
                             {\rho_{B,\beta}}-2\right)X
                +\dfrac{\rho_{A,\alpha}}
                       {\rho_{A,\beta}}
                +\dfrac{\rho_{A,\gamma}}
                       {\rho_{A,\beta}}-2\bigg]
           \dfrac{1}{P_{G_{\text{(iii)}}(X,Y)}}
   \label{eq:y_g3}
\end{align}
with
\begin{align*}
   P_{G_{\text{(iii)}}(X,Y)}
      =\dfrac{13}{66}
        \left[\left(\dfrac{\rho_{B,\alpha}}
                          {\rho_{B,\beta}}-1\right)X
             +\dfrac{\rho_{A,\alpha}}
                    {\rho_{A,\beta}}-1\right]
      +\left[\dfrac{13}{66}
             -F_{AA}(l_{\text{w}\beta\gamma})\right]
       \left[\left(1-\dfrac{\rho_{B,\gamma}}
                            {\rho_{B,\beta}}\right)X
             +1-\dfrac{\rho_{A,\gamma}}
                      {\rho_{A,\beta}}\right] \, .
\end{align*}
\end{widetext}
$P_{G_{\text{(iii)}}(X,Y)}$ is positive in the relevant interval $0<X<D_{\alpha\beta}$.
Accordingly, we obtain
$\theta_{\text{(iii)}}<\dfrac{\pi}{2}$ for $Y>Y_{G_{\text{(iii)}}(X,Y)}$
and $\theta_{\text{(iii)}}>\dfrac{\pi}{2}$ for $0<Y<Y_{G_{\text{(iii)}}(X,Y)}$.

Finally, for the domain (iv) we find 
\begin{widetext}
\begin{align}
   Y_{G_{\text{(iv)}}(X,Y)}
   =\dfrac{1}{2}\bigg\{
   \bigg[\left(\dfrac{\rho_{B,\alpha}}
                      {\rho_{B,\beta}}
               +\dfrac{\rho_{B,\gamma}}
                      {\rho_{B,\beta}}\right)X
         +\dfrac{\rho_{A,\alpha}}
                {\rho_{A,\beta}}
         +\dfrac{\rho_{A,\gamma}}
                {\rho_{A,\beta}}\bigg]
         +\bigg[\left(\dfrac{\rho_{B,\alpha}}
                      {\rho_{B,\beta}}
               -\dfrac{\rho_{B,\gamma}}
                      {\rho_{B,\beta}}\right)X
         +\dfrac{\rho_{A,\alpha}}
                {\rho_{A,\beta}}
         -\dfrac{\rho_{A,\gamma}}
                {\rho_{A,\beta}}\bigg]
    \hat{F}_{AA}(l_{\text{w}\alpha\gamma})\bigg\} \, .
   \label{eq:y_g4}
\end{align}
\end{widetext}
Because $\hat{F}_{AA}(l_{\text{w}\alpha\gamma}) \leq \dfrac{13}{66}$
and due to the inequalities in Eq.~(\ref{eq:rel_densities}) we find that
$Y_{G_{\text{(iv)}}(X,Y)}$ is always located below the green line
in Fig.~\ref{fig:strict} and thus below the domain (iv).
This confirms that $\theta_{\text{(iv)}}<\dfrac{\pi}{2}$ in the entire domain.

It is also interesting to note that 
$Y_{G_{\text{(ii)}}(X,Y)}$, $Y_{G_{\text{(iii)}}(X,Y)}$, and $Y=X+1$ meet
at the same point ($X_{\text{int}}$,$\,Y_{\text{int}}$) 
at which $Y_{\text{ref}}(X)$ and $Y=X+1$ intersect.
This intersection point 
($X_{\text{int}}$,$\,Y_{\text{int}}$) is given by
\begin{align}
\notag
   X_{\text{int}}=&\,\dfrac{\left(\dfrac{\rho_{A,\alpha}}
                                     {\rho_{A,\beta}}
                              +\dfrac{\rho_{A,\gamma}}
                                     {\rho_{A,\beta}}\right)
                         -2}
                        {2-\left(\dfrac{\rho_{B,\alpha}}
                                       {\rho_{B,\beta}}
                                +\dfrac{\rho_{B,\gamma}}
                                       {\rho_{B,\beta}}\right)}
   \quad\text{ and }\quad\\
   Y_{\text{int}}=&\,\dfrac{\left(\dfrac{\rho_{A,\alpha}}
                                      {\rho_{A,\beta}}
                               +\dfrac{\rho_{A,\gamma}}
                                      {\rho_{A,\beta}}\right)
                         -\left(\dfrac{\rho_{B,\alpha}}
                                      {\rho_{B,\beta}}
                               +\dfrac{\rho_{B,\gamma}}
                                      {\rho_{B,\beta}}\right)}
                         {2-\left(\dfrac{\rho_{B,\alpha}}
                                        {\rho_{B,\beta}}
                                 +\dfrac{\rho_{B,\gamma}}
                                        {\rho_{B,\beta}}\right)} \, .
   \label{eq:x_y_int}
\end{align}

Our findings are summarized in Fig.~\ref{fig:strict_ca} where
we show the three wetting domains in the parameter space ($X$,$Y$) together 
with the two domains corresponding to either lyophilic ($\theta<\dfrac{\pi}{2}$)
or lyophobic ($\theta>\dfrac{\pi}{2}$) behavior. 

\section{\label{section:conclusion_and_summary}Conclusions and Summary}

A region of the bulk phase diagram of a binary liquid mixture, composed of $A$ and $B$ particles,
has been considered in which the vapor phase ($\gamma$ phase) coexists with a stable $A$-rich liquid phase
($\alpha$ phase), whereas the $B$-rich phase ($\beta$ phase) is metastable. For such fluids
possible scenarios have been discussed, which may occur, if a composite liquid--vapor ($\alpha$--$\gamma$)
interface with an intruding film of the $\beta$ phase meets a solid wall. The study is based
on classical density functional theory, using the so-called sharp-kink approximation for the fluid density profiles.
Furthermore, certain inequalities~(Eq.~(\ref{eq:rel_densities})) have been assumed among the number densities of
the $A$ and $B$ particles in the three phases $\alpha$, $\beta$, and $\gamma$. These are valid for typical mixtures of two partially miscible liquids. 
Within the theoretical framework presented here, 
we have considered also cases in which one or two of these inequalities are reversed. 
We refrain from including these into this presentation in order to avoid an unnecessary complexity of the discussion 
and because these cases correspond to very special situations, 
which should be discussed separately by focusing on a particular system.
In order to simplify the analytical expressions, we have also assumed that
the various length parameters, which characterize the range of the repulsive core of the
various fluid--fluid and fluid--wall interactions, are all equal. It turns out that small deviations from this simplified
case do not change the general picture. 

In a first step we have assumed that the so-called mixing rule applies to both the fluid--fluid
and the fluid--wall interactions. In this case the strength of the $A$--$B$ interaction is the
geometric mean of the strengths of the $A$--$A$ and $B$--$B$ interactions, and the ratio of the wall--$A$
and the wall--$B$ interaction strengths is related to the ratio of the $A$--$A$ and $B$--$B$ interactions
by a corresponding relation. Given these relations, three different wetting scenarios are possible, which correspond
to three domains in a two-dimensional space of system parameters.
First, the $\beta$ film at the liquid--vapor interface extends 
into the wall--$\alpha$ interface,
but not into the wall--vapor (wall--$\gamma$) interface.
Second, the $\beta$ film extends into the wall--vapor interface, but not into the wall--$\alpha$-liquid interface.
Third, the $\beta$ film ends at the wall, but an $\alpha$ film forms at the wall--vapor interface.
A scenario, in which the $\beta$ film extends into both the wall--vapor and the wall--$\alpha$-liquid interface,
is not possible, given the relations between interaction strengths imposed by the mixing rules.
In case the wall is provided by a colloidal particle, 
floating at the considered composite liquid--vapor interface, 
the latter scenario corresponds to a colloid which is completely covered by 
a film of the $\beta$ phase. This scenario does not occur if the mixing rules apply. 

As further steps we have relaxed the mixing rules for the fluid--wall and for the fluid--fluid
interactions. If the ratio of the strengths of the wall--$A$ and of the wall--$B$ interactions
is reduced as compared to the mixing rule prescription, the wetting scenario corresponding to a colloid
completely covered by a film of the $\beta$ liquid can occur within a certain domain in
the parameter space. The same is true if the strength of the $A$--$B$ interaction is increased
beyond the mixing rule prescription. Up to six different wetting scenarios can occur,
if the mixing rules for the fluid--wall and for the fluid--fluid interactions are relaxed.
It depends on the sign of the deviations from the mixing rule prescriptions 
how the space of system parameters is divided into the corresponding domains and which
domains actually appear. 

For the special case that the mixing rules apply to both the fluid--fluid and the fluid--wall 
interactions, we searched for relations between these wetting domains and the contact angle
$\theta$. For the scenario in which the $\beta$ film ends at the wall, 
but an $\alpha$ film forms at the wall--vapor interface, one always finds $\theta<90^\mathrm{o}$.
Concerning the other two scenarios, the respective domains are subdivided into subdomains within which
$\theta>90^\mathrm{o}$ or $\theta<90^\mathrm{o}$, respectively. 

We note that the system parameters may be varied in two ways. Either via varying
the various interaction strengths (i.e., by using different liquids or a wall with modified
properties) or via changing the thermodynamic state and thus the bulk number densities, 
which enter into the definition of the dimensionless system parameters 
$X$ and $Y$ (Eqs.~(\ref{eq:X}) and~(\ref{eq:Y}), respectively).
Both routes facilitate to switch between the wetting domains. 
The insight we have gained concerning the wetting scenarios around a colloidal particle 
floating at a composite liquid--vapor interface, is potentially useful for tuning
the capillarity induced interactions between such colloidal particles. 
The particles could be fabricated from the same or from different materials. 
Such knowledge is essential for designing the self-assembly of colloidal particles
at liquid--vapor interfaces.

\begin{widetext}
\appendix

\section{\label{sec:solve_s_ff}
         Condition for $S_{\alpha\beta\gamma}<0$}
In this appendix, 
we determine the range of $X$ values within which $S_{\alpha\beta\gamma}<0$,
which is the condition for having an intruding wetting film of the $\beta$ phase
at the $\alpha$--$\gamma$ interface. 
We start from Eq.~(\ref{eq:s_abg2}), i.e.,
\begin{align*}
      S_{\alpha\beta\gamma}
      =\epsilon_{AA}\rho_{A,\beta}^{2}
        \bigg[&
        \left(1-\dfrac{\rho_{B,\alpha}}{\rho_{B,\beta}}\right)
        \left(1-\dfrac{\rho_{B,\gamma}}{\rho_{B,\beta}}\right)X^2
       +\left(1-\dfrac{\rho_{B,\alpha}}{\rho_{B,\beta}}\right)
        \left(1-\dfrac{\rho_{A,\gamma}}{\rho_{A,\beta}}\right)
        \xi_{f}X\\
      &\,+\left(1-\dfrac{\rho_{A,\alpha}}{\rho_{A,\beta}}\right)
        \left(1-\dfrac{\rho_{B,\gamma}}{\rho_{B,\beta}}\right)
        \xi_{f}X
       +\left(1-\dfrac{\rho_{A,\alpha}}{\rho_{A,\beta}}\right)
        \left(1-\dfrac{\rho_{A,\gamma}}{\rho_{A,\beta}}\right)
        \bigg] \, .
\end{align*}

The condition $S_{\alpha\beta\gamma}<0$ implies
\begin{align}
\notag
      &\left(1-\dfrac{\rho_{B,\alpha}}{\rho_{B,\beta}}\right)
       \left(1-\dfrac{\rho_{B,\gamma}}{\rho_{B,\beta}}\right)X^2
      +\left(1-\dfrac{\rho_{B,\alpha}}{\rho_{B,\beta}}\right)
       \left(1-\dfrac{\rho_{A,\gamma}}{\rho_{A,\beta}}\right)
       \xi_{f}X\\
     &\,+\left(1-\dfrac{\rho_{A,\alpha}}{\rho_{A,\beta}}\right)
       \left(1-\dfrac{\rho_{B,\gamma}}{\rho_{B,\beta}}\right)
       \xi_{f}X
      +\left(1-\dfrac{\rho_{A,\alpha}}{\rho_{A,\beta}}\right)
       \left(1-\dfrac{\rho_{A,\gamma}}{\rho_{A,\beta}}\right)<0 \, .
\label{eq:s_abg3_app}
\end{align}
In order to solve the inequality in Eq.~(\ref{eq:s_abg3_app}),
we consider $S_{\alpha\beta\gamma}=0$. 
There are two solutions
$X_{1}$ and $X_{2}>X_{1}$:
\begin{equation*}
   X_{1}=-\sqrt{\left[\dfrac{\xi_{f}}{2}
                   \left(D_{\alpha\beta}
                   +D_{\beta\gamma}\right)\right]^{2}
           -D_{\alpha\beta}D_{\beta\gamma}}
         +\dfrac{\xi_{f}}{2}
          \left(D_{\alpha\beta}+D_{\beta\gamma}\right)
\end{equation*}
and 
\begin{equation*}
   X_{2}=\sqrt{\left[\dfrac{\xi_{f}}{2}
                   \left(D_{\alpha\beta}
                   +D_{\beta\gamma}\right)\right]^{2}
           -D_{\alpha\beta}D_{\beta\gamma}}
         +\dfrac{\xi_{f}}{2}
          \left(D_{\alpha\beta}+D_{\beta\gamma}\right) \, ,
\end{equation*}
where 
$D_{\alpha\beta}={\left(\dfrac{\rho_{A,\alpha}}
                              {\rho_{A,\beta}}-1\right)}
                 \bigg/
                 {\left(1-\dfrac{\rho_{B,\alpha}}
                                {\rho_{B,\beta}}\right)}$ 
and 
$D_{\beta\gamma}={\left(1-\dfrac{\rho_{A,\gamma}}
                                {\rho_{A,\beta}}\right)}
                 \bigg/
                 {\left(\dfrac{\rho_{B,\gamma}}
                              {\rho_{B,\beta}}-1\right)}\, .$

Here we use the inequalities
$\rho_{A,\alpha}>\rho_{A,\beta}$ and 
$\rho_{B,\alpha}<\rho_{B,\beta}$,
from which we infer that the prefactor of $X^{2}$ 
in Eq.~(\ref{eq:s_abg3_app}) 
is positive. 
In order to have $S_{\alpha\beta\gamma}<0$,
$X$ must be in the range of $X_{1}<X<X_{2}$.
It is known that $X_{1}<0$ and $X_{2}>0$,
because $D_{\alpha\beta}>0$, $D_{\beta\gamma}<0$,
and $X>0$ by definition. 
As a result,
we have $S_{\alpha\beta\gamma}<0$ 
within the range $0<X<X_{2}$ (see Eq.~(\ref{eq:s_abg_rng})).

\section{\label{sec:wf_w_para}
         Wall--fluid wetting parameters}
In this appendix, 
the explicit expressions (see Eqs.~(\ref{eq:w_wba2})-(\ref{eq:w_wag2}))
for the wetting parameters
$W_{\text{w}\beta\alpha}$, 
$W_{\text{w}\beta\gamma}$, 
and $W_{\text{w}\alpha\gamma}$
are derived from the definitions
in Eqs.~(\ref{eq:w_wba})-(\ref{eq:w_wag}). 

By using relations similar 
to the ones in Eqs.~(\ref{eq:equil_cond1}) and~(\ref{eq:equil_cond2}),
we find the surface tensions
$\sigma_{\text{w}\beta\alpha}$,
$\sigma_{\text{w}\beta\gamma}$,
and 
$\sigma_{\text{w}\alpha\gamma}$
of the partially wet interfaces  
(see below). 
With the corresponding equilibrium wetting film thicknesses
$l=l_{\text{w}\beta\alpha}$, $l_{\text{w}\beta\gamma}$, 
and $l_{\text{w}\alpha\gamma}$
one has
\begin{align}
   \sigma_{\text{w}\beta\alpha}
   &=\Omega_{\text{s}}^{\text{w}\beta\alpha}
           (l_{\text{w}\beta\alpha},T,\{\mu_i\})
   \quad\text{  with  }\quad
   \dfrac{\partial{\Omega_{\text{s}}
                         ^{\text{w}\beta\alpha}}(l)}
        {\partial{l}}\bigg{|}_{l=l_{\text{w}\beta\alpha}}=0 \, ,
   \label{eq:equil_wba_app}\\
   \sigma_{\text{w}\beta\gamma}
   &=\Omega_{\text{s}}^{\text{w}\beta\gamma}
           (l_{\text{w}\beta\gamma},T,\{\mu_i\})
   \quad\text{  with  }\quad
   \dfrac{\partial{\Omega_{\text{s}}
                         ^{\text{w}\beta\gamma}}(l)}
        {\partial{l}}\bigg{|}_{l=l_{\text{w}\beta\gamma}}=0 \, , 
   \label{eq:equil_wbg_app}
\end{align}
and
\begin{equation}
   \sigma_{\text{w}\alpha\gamma}
   =\Omega_{\text{s}}^{\text{w}\alpha\gamma}
           (l_{\text{w}\alpha\gamma},T,\{\mu_i\})
   \quad\text{  with  }\quad
   \dfrac{\partial{\Omega_{\text{s}}
                         ^{\text{w}\alpha\gamma}}(l)}
        {\partial{l}}\bigg{|}_{l=l_{\text{w}\alpha\gamma}}=0 \, .
   \label{eq:equil_wag_app}
\end{equation}

Using the sharp-kink approximation, 
the surface contributions to the grand canonical potential 
for the various partially wet interfaces 
are given by
\begin{align}
   \Omega_{\text{s}}^{\text{w}\beta\alpha}(l)
                      =l(\Omega^{\beta}-\Omega^{\alpha})
                      +\omega_{\text{w}\beta\alpha}(l)
                      +\sigma_{\text{w}\beta}
                      +\sigma_{\beta\alpha},
   \label{eq:omega_surf_wba_app}\\
   \Omega_{\text{s}}^{\text{w}\beta\gamma}(l)
                      =l(\Omega^{\beta}-\Omega^{\gamma})
                      +\omega_{\text{w}\beta\gamma}(l)
                      +\sigma_{\text{w}\beta}
                      +\sigma_{\beta\gamma} \, ,
   \label{eq:omega_surf_wbg_app}
\end{align}
and
\begin{equation}
   \Omega_{\text{s}}^{\text{w}\alpha\gamma}(l)
                      =l(\Omega^{\alpha}-\Omega^{\gamma})
                      +\omega_{\text{w}\alpha\gamma}(l)
                      +\sigma_{\text{w}\alpha}
                      +\sigma_{\alpha\gamma} \, ,
   \label{eq:omega_surf_wag_app}
\end{equation} 
where $\sigma_{\text{w}\beta}$
is the surface tension of the plain wall--$\beta$ interface, without
any intruding wetting film.

Therefore
Eqs.~(\ref{eq:w_wba})-(\ref{eq:w_wag}),
$W_{\text{w}\beta\alpha}$,
$W_{\text{w}\beta\gamma}$,
and 
$W_{\text{w}\alpha\gamma}$,
can be expressed as 
\begin{align}
   W_{\text{w}\beta\alpha}&=l_{\text{w}\beta\alpha}
                           (\Omega^{\beta}-\Omega^{\alpha})
                          +\omega_{\text{w}\beta\alpha}
                           (l_{\text{w}\beta\alpha})
                          +\sigma_{\text{w}\beta}
                          +\sigma_{\beta\alpha}
                          -\sigma_{\text{w}\alpha} \, ,
   \label{eq:w_wba_app}\\
   W_{\text{w}\beta\gamma}&=l_{\text{w}\beta\gamma}
                           (\Omega^{\beta}-\Omega^{\gamma})
                          +\omega_{\text{w}\beta\gamma}
                           (l_{\text{w}\beta\gamma})
                          +\sigma_{\text{w}\beta}
                          +\sigma_{\beta\gamma}
                          -\sigma_{\text{w}\gamma} \, ,
   \label{eq:w_wbg_app}
\end{align}
and
\begin{equation}
   W_{\text{w}\alpha\gamma}=l_{\text{w}\alpha\gamma}
                            (\Omega^{\alpha}-\Omega^{\gamma})
                           +\omega_{\text{w}\alpha\gamma}
                            (l_{\text{w}\alpha\gamma})
                           +\sigma_{\text{w}\alpha}
                           +\sigma_{\alpha\gamma}
                           -\sigma_{\text{w}\gamma} \, .
   \label{eq:w_wag_app}
\end{equation} 
We also make use of the relations
\begin{align}
   \Omega^{\beta}-\Omega^{\alpha}
   =-\dfrac{\partial{\omega_{\text{w}\beta\alpha}(l)}}
          {\partial{l}}\bigg{|}_{l=l_{\text{w}\beta\alpha}} \, ,
   \label{eq:delta_omega_ba_app}\\
   \Omega^{\beta}-\Omega^{\gamma}
   =-\dfrac{\partial{\omega_{\text{w}\beta\gamma}(l)}}
           {\partial{l}}\bigg{|}_{l=l_{\text{w}\beta\gamma}} \, ,
   \label{eq:delta_omega_bg_app}
\end{align}
and
\begin{equation}
   \Omega^{\alpha}-\Omega^{\gamma}
   =-\dfrac{\partial{\omega_{\text{w}\alpha\gamma}(l)}}
           {\partial{l}}\bigg{|}_{l=l_{\text{w}\alpha\gamma}} \, .
   \label{eq:delta_omega_ag_app}
\end{equation}

The expressions for the surface tensions of
various fluid interfaces have already been derived
within the sharp-kink approximation (see Eqs.~(\ref{eq:sigma_ab})-(\ref{eq:sigma_ag})).
In addition, the surface tensions and the various interface potentials $\omega$ 
characterizing the wall--fluid interfaces are given by 
\begin{align}
   \sigma_{\text{w}\alpha}
     &=-\dfrac{1}{2}\sum_{i,j}\rho_{i,\alpha}\rho_{j,\alpha}
                             \int_{0}^{\infty}dy\,t_{ij}(y)
       +\sum_{i}\rho_{i,\alpha}\rho_{\text{w}}
                             \int_{0}^{\infty}dy\,V_{i}(y) \, ,
   \label{eq:sigma_wa_app}\\
   \sigma_{\text{w}\beta}
     &=-\dfrac{1}{2}\sum_{i,j}\rho_{i,\beta}\rho_{j,\beta}
                             \int_{0}^{\infty}dy\,t_{ij}(y)
       +\sum_{i}\rho_{i,\beta}\rho_{\text{w}}
                             \int_{0}^{\infty}dy\,V_{i}(y) \, ,
   \label{eq:sigma_wb_app}\\
   \sigma_{\text{w}\gamma}
     &=-\dfrac{1}{2}\sum_{i,j}\rho_{i,\gamma}\rho_{j,\gamma}
                             \int_{0}^{\infty}dy\,t_{ij}(y)
       +\sum_{i}\rho_{i,\gamma}\rho_{\text{w}}
                             \int_{0}^{\infty}dy\,V_{i}(y) \, ,
   \label{eq:sigma_wg_app}\\
   \omega_{\text{w}\beta\alpha}(l)
     &=\sum_{i,j}(\rho_{i,\beta}-\rho_{i,\alpha})
                  \rho_{j,\beta}\int_{l}^{\infty}dy\,t_{ij}(y)
      -\sum_{i}(\rho_{i,\beta}-\rho_{i,\alpha})
                \rho_{\text{w}}\int_{l}^{\infty}dy\,V_{i}(y) \, ,
   \label{eq:omega_wba_app}\\
   \omega_{\text{w}\beta\gamma}(l)
     &=\sum_{i,j}(\rho_{i,\beta}-\rho_{i,\gamma})
                  \rho_{j,\beta}\int_{l}^{\infty}dy\,t_{ij}(y)
      -\sum_{i}(\rho_{i,\beta}-\rho_{i,\gamma})
                \rho_{\text{w}}\int_{l}^{\infty}dy\,V_{i}(y) \, ,
   \label{eq:omega_wbg_app}
\end{align}
and
\begin{equation}
   \omega_{\text{w}\alpha\gamma}(l)
     =\sum_{i,j}(\rho_{i,\alpha}-\rho_{i,\gamma})
                  \rho_{j,\alpha}\int_{l}^{\infty}dy\,t_{ij}(y)
      -\sum_{i}(\rho_{i,\alpha}-\rho_{i,\gamma})
                \rho_{\text{w}}\int_{l}^{\infty}dy\,V_{i}(y) \, ,
   \label{eq:omega_wag_app}
\end{equation}
where 
\begin{equation*}
   t_{ij}(y)=\int_{y}^{\infty} dx\int d^2\mathbf{r}_{||}\,
             \tilde{w}_{ij}
            \left[(\mathbf{r}_{||}^2+x^2)^\frac{1}{2}\right]
\end{equation*}
and
\begin{equation*}
   V_{i}(y)=\int_{y}^{\infty} dx\int d^2\mathbf{r}_{||}\,
            \tilde{v}_{i}
            \left[(\mathbf{r}_{||}^2+x^2)^{\frac{1}{2}}
                                          \right] \, .
\end{equation*}

By taking the explicit expressions for 
$\tilde{w}_{ij}$ and $\tilde{v}_{i}$
(see Eqs.~(\ref{eq:modi1_lj}) and~(\ref{eq:modi2_lj})),
one obtains
for Eqs.~(\ref{eq:sigma_wa_app})-(\ref{eq:omega_wag_app})
\begin{align}
   \sigma_{\text{w}\alpha}
     &=\dfrac{13}{132}\pi\sum_{i,j}\epsilon_{ij}a_{ij}^{4}
                                   \rho_{i,\alpha}\rho_{j,\alpha}
      -\dfrac{13}{66}\pi\sum_{i}\epsilon_{\text{w}i}
                                a_{\text{w}i}^{4}
                                \rho_{i,\alpha}\rho_{\text{w}} \, ,
   \label{eq:sigma_wa2_app}\\
   \sigma_{\text{w}\beta}
     &=\dfrac{13}{132}\pi\sum_{i,j}\epsilon_{ij}a_{ij}^{4}
                                   \rho_{i,\beta}\rho_{j,\beta}
      -\dfrac{13}{66}\pi\sum_{i}\epsilon_{\text{w}i}
                                a_{\text{w}i}^{4}
                                \rho_{i,\beta}\rho_{\text{w}} \, ,
   \label{eq:sigma_wb2_app}\\
   \sigma_{\text{w}\gamma}
     &=\dfrac{13}{132}\pi\sum_{i,j}\epsilon_{ij}a_{ij}^{4}
                                   \rho_{i,\gamma}\rho_{j,\gamma}
      -\dfrac{13}{66}\pi\sum_{i}\epsilon_{\text{w}i}
                                a_{\text{w}i}^{4}
                                \rho_{i,\gamma}\rho_{\text{w}} \, ,
   \label{eq:sigma_wg2_app}\\
\notag
   \omega_{\text{w}\beta\alpha}(l)
     &=-\pi\sum_{i,j}\epsilon_{ij}a_{ij}^{4}
          (\rho_{i,\beta}-\rho_{i,\alpha})\rho_{j,\beta}
          \left[\dfrac{1}{3}\left(\dfrac{a_{ij}}{l}\right)^{2}
               -\dfrac{4}{5}\left(\dfrac{a_{ij}}{l}\right)^{3}
                                                      \right]\\
     &\quad+\pi\sum_{i}\epsilon_{\text{w}i} a_{\text{w}i}^{4}
                  (\rho_{i,\beta}-\rho_{i,\alpha})\rho_{\text{w}}
        \left[\dfrac{1}{3}\left(\dfrac{a_{\text{w}i}}{l}\right)^{2}
             -\dfrac{4}{5}\left(\dfrac{a_{\text{w}i}}{l}\right)^{3}
                                                      \right] \, ,
   \label{eq:omega_wba2_app}\\
\notag
   \omega_{\text{w}\beta\gamma}(l)
     &=-\pi\sum_{i,j}\epsilon_{ij}a_{ij}^{4}
          (\rho_{i,\beta}-\rho_{i,\gamma})\rho_{j,\beta}
          \left[\dfrac{1}{3}\left(\dfrac{a_{ij}}{l}\right)^{2}
               -\dfrac{4}{5}\left(\dfrac{a_{ij}}{l}\right)^{3}
                                                      \right]\\
     &\quad+\pi\sum_{i}\epsilon_{\text{w}i}a_{\text{w}i}^{4}
          (\rho_{i,\beta}-\rho_{i,\gamma})\rho_{\text{w}}
        \left[\dfrac{1}{3}\left(\dfrac{a_{\text{w}i}}{l}\right)^{2}
             -\dfrac{4}{5}\left(\dfrac{a_{\text{w}i}}{l}\right)^{3}
                                                      \right] \, ,
   \label{eq:omega_wbg2_app}
\end{align}
and
\begin{align}
\notag
   \omega_{\text{w}\alpha\gamma}(l)
     &=-\pi\sum_{i,j}\epsilon_{ij}a_{ij}^{4}
          (\rho_{i,\alpha}-\rho_{i,\gamma})\rho_{j,\alpha}
          \left[\dfrac{1}{3}\left(\dfrac{a_{ij}}{l}\right)^{2}
               -\dfrac{4}{5}\left(\dfrac{a_{ij}}{l}\right)^{3}
                                                      \right]\\
     &\quad+\pi\sum_{i}\epsilon_{\text{w}i}a_{\text{w}i}^{4}
          (\rho_{i,\alpha}-\rho_{i,\gamma})\rho_{\text{w}}
        \left[\dfrac{1}{3}\left(\dfrac{a_{\text{w}i}}{l}\right)^{2}
             -\dfrac{4}{5}\left(\dfrac{a_{\text{w}i}}{l}\right)^{3}
                                                      \right] \, .
   \label{eq:omega_wag2_app}
\end{align}

Using the derivatives of 
Eqs.~(\ref{eq:omega_wba2_app})-(\ref{eq:omega_wag2_app})
with respect to the wetting film thickness, 
Eqs.~(\ref{eq:delta_omega_ba_app})-(\ref{eq:delta_omega_ag_app})
turn into
\begin{align}
\notag
   \Omega^{\beta}-\Omega^{\alpha}
     &=-\pi\sum_{i,j}\epsilon_{ij}a_{ij}^{3}
          (\rho_{i,\beta}-\rho_{i,\alpha})\rho_{j,\beta}
         \left[\dfrac{2}{3}\left(\dfrac{a_{ij}}
                            {l_{\text{w}\beta\alpha}}\right)^{3}
               -\dfrac{12}{5}\left(\dfrac{a_{ij}}
                            {l_{\text{w}\beta\alpha}}\right)^{4}
                                                      \right]\\
     &\quad+\pi\sum_{i}\epsilon_{\text{w}i}a_{\text{w}i}^{3}
          (\rho_{i,\beta}-\rho_{i,\alpha})\rho_{\text{w}}
          \left[\dfrac{2}{3}\left(\dfrac{a_{\text{w}i}}
                            {l_{\text{w}\beta\alpha}}\right)^{3}
               -\dfrac{12}{5}\left(\dfrac{a_{\text{w}i}}
                            {l_{\text{w}\beta\alpha}}\right)^{4}
                                                      \right] \, ,
   \label{eq:delta_omega_ba2_app}\\
\notag
   \Omega^{\beta}-\Omega^{\gamma}
     &=-\pi\sum_{i,j}\epsilon_{ij}a_{ij}^{3}
          (\rho_{i,\beta}-\rho_{i,\gamma})\rho_{j,\beta}
         \left[\dfrac{2}{3}\left(\dfrac{a_{ij}}
                            {l_{\text{w}\beta\gamma}}\right)^{3}
               -\dfrac{12}{5}\left(\dfrac{a_{ij}}
                            {l_{\text{w}\beta\gamma}}\right)^{4}
                                                      \right]\\
     &\quad+\pi\sum_{i}\epsilon_{\text{w}i}a_{\text{w}i}^{3}
          (\rho_{i,\beta}-\rho_{i,\gamma})\rho_{\text{w}}
          \left[\dfrac{2}{3}\left(\dfrac{a_{\text{w}i}}
                            {l_{\text{w}\beta\gamma}}\right)^{3}
               -\dfrac{12}{5}\left(\dfrac{a_{\text{w}i}}
                            {l_{\text{w}\beta\gamma}}\right)^{4}
                                                      \right] \, ,
   \label{eq:delta_omega_bg2_app}
\end{align}
and
\begin{align}
\notag
   \Omega^{\alpha}-\Omega^{\gamma}
     &=-\pi\sum_{i,j}\epsilon_{ij}a_{ij}^{3}
          (\rho_{i,\alpha}-\rho_{i,\gamma})\rho_{j,\alpha}
         \left[\dfrac{2}{3}\left(\dfrac{a_{ij}}
                            {l_{\text{w}\alpha\gamma}}\right)^{3}
               -\dfrac{12}{5}\left(\dfrac{a_{ij}}
                            {l_{\text{w}\alpha\gamma}}\right)^{4}
                                                      \right]\\
     &\quad+\pi\sum_{i}\epsilon_{\text{w}i}a_{\text{w}i}^{3}
          (\rho_{i,\alpha}-\rho_{i,\gamma})\rho_{\text{w}}
          \left[\dfrac{2}{3}\left(\dfrac{a_{\text{w}i}}
                            {l_{\text{w}\alpha\gamma}}\right)^{3}
               -\dfrac{12}{5}\left(\dfrac{a_{\text{w}i}}
                            {l_{\text{w}\alpha\gamma}}\right)^{4}
                                                      \right] \, .
   \label{eq:delta_omega_ag2_app}
\end{align}
Here 
Eq.~(\ref{eq:delta_omega_ag2_app}) is zero 
because we consider the bulk phases $\alpha$ and $\gamma$ to be in thermal equilibrium.
 
Finally, based on 
Eqs.~(\ref{eq:sigma_wa2_app})-(\ref{eq:delta_omega_ag2_app}), 
Eqs.~(\ref{eq:w_wba_app})-(\ref{eq:w_wag_app}) 
can be expressed as
\begin{align*}
   W_{\text{w}\beta\alpha}
     =&\,\pi\sum_{i,j}\epsilon_{ij}a_{ij}^{4}
          (\rho_{i,\beta}-\rho_{i,\alpha})\rho_{j,\beta}
         \left[\dfrac{13}{66}-\left(\dfrac{a_{ij}}
                            {l_{\text{w}\beta\alpha}}\right)^{2}
              +\dfrac{16}{5}\left(\dfrac{a_{ij}}
                            {l_{\text{w}\beta\alpha}}\right)^{3}
                                                      \right]\\
     &-\pi\sum_{i}\epsilon_{\text{w}i}a_{\text{w}i}^{4}
          (\rho_{i,\beta}-\rho_{i,\alpha})\rho_{\text{w}}
          \left[\dfrac{13}{66}-\left(\dfrac{a_{\text{w}i}}
                            {l_{\text{w}\beta\alpha}}\right)^{2}
               +\dfrac{16}{5}\left(\dfrac{a_{\text{w}i}}
                            {l_{\text{w}\beta\alpha}}\right)^{3}
                                                      \right]\, ,\\
   W_{\text{w}\beta\gamma}
     =&\,\pi\sum_{i,j}\epsilon_{ij}a_{ij}^{4}
          (\rho_{i,\beta}-\rho_{i,\gamma})\rho_{j,\beta}
         \left[\dfrac{13}{66}-\left(\dfrac{a_{ij}}
                            {l_{\text{w}\beta\gamma}}\right)^{2}
              +\dfrac{16}{5}\left(\dfrac{a_{ij}}
                            {l_{\text{w}\beta\gamma}}\right)^{3}
                                                      \right]\\
     &-\pi\sum_{i}\epsilon_{\text{w}i}a_{\text{w}i}^{4}
          (\rho_{i,\beta}-\rho_{i,\gamma})\rho_{\text{w}}
          \left[\dfrac{13}{66}-\left(\dfrac{a_{\text{w}i}}
                            {l_{\text{w}\beta\gamma}}\right)^{2}
               +\dfrac{16}{5}\left(\dfrac{a_{\text{w}i}}
                            {l_{\text{w}\beta\gamma}}\right)^{3}
                                                      \right] \, ,
\end{align*}
and
\begin{align*}
   W_{\text{w}\alpha\gamma}
     =&\,\pi\sum_{i,j}\epsilon_{ij}a_{ij}^{4}
          (\rho_{i,\alpha}-\rho_{i,\gamma})\rho_{j,\alpha}
         \left[\dfrac{13}{66}
              -\dfrac{1}{3}\left(\dfrac{a_{ij}}
                            {l_{\text{w}\alpha\gamma}}\right)^{2}
              +\dfrac{4}{5}\left(\dfrac{a_{ij}}
                            {l_{\text{w}\alpha\gamma}}\right)^{3}
                                                      \right]\\
     &-\pi\sum_{i}\epsilon_{\text{w}i}a_{\text{w}i}^{4}
          (\rho_{i,\alpha}-\rho_{i,\gamma})\rho_{\text{w}}
          \left[\dfrac{13}{66}
               -\dfrac{1}{3}\left(\dfrac{a_{\text{w}i}}
                            {l_{\text{w}\alpha\gamma}}\right)^{2}
               +\dfrac{4}{5}\left(\dfrac{a_{\text{w}i}}
                            {l_{\text{w}\alpha\gamma}}\right)^{3}
                                                      \right] \, .
\end{align*}
Upon changing notation and by introducing
the Kronecker symbol $\delta$,
we arrive at the expressions 
in Eqs.~(\ref{eq:w_wba2})-(\ref{eq:w_wag2}). 

\section{\label{sec:solve_s_wf}
         Domains characterized by $S_{\text{w}\beta\alpha}<0$,
         $S_{\text{w}\beta\gamma}<0$, or $S_{\text{w}\alpha\gamma}<0$ }
In this appendix, we search for domains in the ($X$,$Y$) parameter space
within which the conditions $S_{\text{w}\beta\alpha}<0$,
$S_{\text{w}\beta\gamma}<0$, or $S_{\text{w}\alpha\gamma}<0$ hold.

We start this discussion with the condition $S_{\text{w}\beta\alpha}<0$.
To this end we introduce the notations
$\epsilon_{AB}=\xi_{f}\sqrt{\epsilon_{AA}\epsilon_{BB}}$,
$\epsilon_{\text{w}A}
=\xi_{\text{w}A}\sqrt{\epsilon_{\text{ww}}\epsilon_{AA}}$,
and $\epsilon_{\text{w}B}
=\xi_{\text{w}B}\sqrt{\epsilon_{\text{ww}}\epsilon_{BB}}$ so that $\epsilon_{\text{w}A}/\epsilon_{\text{w}B} = 
  (\xi_{\text{w}A}/\xi_{\text{w}B}) \sqrt{\epsilon_{AA}/\epsilon_{BB}}
  =: \xi_{\text{w}} \sqrt{\epsilon_{AA}/\epsilon_{BB}}$.
Accordingly we express $S_{\text{w}\beta\alpha}$ (Eq.~(\ref{eq:s_wba})) as  
\begin{align*}
      S_{\text{w}\beta\alpha}
            =&\,(\rho_{A,\beta}-\rho_{A,\alpha})
              (\epsilon_{AA}\rho_{A,\beta}
              -\xi_{\text{w}A}\sqrt{\epsilon_{\text{ww}}
               \epsilon_{AA}}\rho_{\text{w}}
              +\xi_{f}\sqrt{\epsilon_{AA}\epsilon_{BB}}
               \rho_{B,\beta})\\
            &+(\rho_{B,\beta}-\rho_{B,\alpha})
              (\xi_{f}\sqrt{\epsilon_{AA}\epsilon_{BB}}
               \rho_{A,\beta}
              +\epsilon_{BB}\rho_{B,\beta}
              -\xi_{\text{w}B}\sqrt{\epsilon_{\text{ww}}
               \epsilon_{BB}}\rho_{\text{w}}) \, .
\end{align*}
In terms of the dimensionless parameters
$X=\dfrac{\rho_{B,\beta}}{\rho_{A,\beta}}
   \sqrt{\dfrac{\epsilon_{BB}}{\epsilon_{AA}}}$ 
and $Y=\dfrac{\rho_{\text{w}}}{\rho_{A,\beta}}
       \dfrac{\epsilon_{\text{w}A}}{\epsilon_{AA}}$,
$S_{\text{w}\beta\alpha}$ can be written as
\begin{align}
\notag
      S_{\text{w}\beta\alpha}
      =\epsilon_{AA}\rho_{A,\beta}^{2}
       \bigg\{&
         \left[
          \left(1-\dfrac{\rho_{B,\alpha}}{\rho_{B,\beta}}\right)X
         +\left(1-\dfrac{\rho_{A,\alpha}}{\rho_{A,\beta}}\right)
         \right]\left(X+1\right)
      +\left[
         \left(1-\dfrac{\rho_{B,\alpha}}{\rho_{B,\beta}}\right)
        +\left(1-\dfrac{\rho_{A,\alpha}}{\rho_{A,\beta}}\right)
        \right]\left(\xi_{f}-1\right)X\\
      &\,-\left[
         \left(1-\dfrac{\rho_{B,\alpha}}{\rho_{B,\beta}}\right)
         \dfrac{X}{\xi_{\text{w}}}
        +\left(1-\dfrac{\rho_{A,\alpha}}{\rho_{A,\beta}}\right)
        \right]Y
        \bigg\} \, .
\label{eq:s_wba2_app}
\end{align}

The condition $S_{\text{w}\beta\alpha}<0$, 
together with Eq.~(\ref{eq:s_wba2_app}), 
leads to the inequality
\begin{align}
\notag
      \left[
         \left(1-\dfrac{\rho_{B,\alpha}}{\rho_{B,\beta}}\right)
         \dfrac{X}{\xi_{\text{w}}}
        +\left(1-\dfrac{\rho_{A,\alpha}}{\rho_{A,\beta}}\right)
        \right]Y>
        &\left[
          \left(1-\dfrac{\rho_{B,\alpha}}{\rho_{B,\beta}}\right)X
         +\left(1-\dfrac{\rho_{A,\alpha}}{\rho_{A,\beta}}\right)
         \right]\left(X+1\right)\\
      &\,+\left[
         \left(1-\dfrac{\rho_{B,\alpha}}{\rho_{B,\beta}}\right)
        +\left(1-\dfrac{\rho_{A,\alpha}}{\rho_{A,\beta}}\right)
        \right]\left(\xi_{f}-1\right)X \, .
\label{eq:s_wba3_app}
\end{align}
In order to proceed we analyze the sign of the prefactor of $Y$. 
It is positive, if
\begin{equation}
         \left(1-\dfrac{\rho_{B,\alpha}}{\rho_{B,\beta}}\right)
         \dfrac{X}{\xi_{\text{w}}}>
        -\left(1-\dfrac{\rho_{A,\alpha}}{\rho_{A,\beta}}\right) \, .
\label{eq:s_wba3_prefac_y_app}
\end{equation}
Taking into account the assumed inequalities 
$\rho_{A,\alpha}>\rho_{A,\beta}$ and 
$\rho_{B,\alpha}<\rho_{B,\beta}$,  
Eq.~(\ref{eq:s_wba3_prefac_y_app}) leads to
$X>\xi_{\text{w}}D_{\alpha\beta}$ with $D_{\alpha\beta}>0$.
Under this condition of a positive prefactor of $Y$, 
we find
\begin{align}
   Y>X+1
     +\dfrac{\left(1-\dfrac{\rho_{B,\alpha}}{\rho_{B,\beta}}\right)
      \left(1-\dfrac{1}{\xi_{\text{w}}}\right)X}
     {\left[\left(1-\dfrac{\rho_{B,\alpha}}{\rho_{B,\beta}}\right)
     \dfrac{X}{\xi_{\text{w}}}
     +\left(1-\dfrac{\rho_{A,\alpha}}{\rho_{A,\beta}}\right)\right]}
     \left(X+1\right)
     +\dfrac{\left[\left(1-\dfrac{\rho_{B,\alpha}}{\rho_{B,\beta}}\right)
     +\left(1-\dfrac{\rho_{A,\alpha}}{\rho_{A,\beta}}\right)\right]
      \left(\xi_{f}-1\right)}
     {\left[\left(1-\dfrac{\rho_{B,\alpha}}{\rho_{B,\beta}}\right)
     \dfrac{X}{\xi_{\text{w}}}+
     \left(1-\dfrac{\rho_{A,\alpha}}{\rho_{A,\beta}}\right)
     \right]}X \, ,
\end{align}
or, expressed in terms of $D_{\alpha\beta}$, 
\begin{align}
   Y>X+1
    +\left(\xi_{\text{w}}-1\right)
     \dfrac{X}{\left(X-\xi_{\text{w}}D_{\alpha\beta}\right)}
     \left(X+1\right)
     +\xi_{\text{w}}
     \left(\xi_{f}-1\right)
     \left(1-D_{\alpha\beta}\right)
     \dfrac{X}{\left(X-\xi_{\text{w}}D_{\alpha\beta}\right)} \, .
\end{align}

In the case of a negative prefactor of $Y$, i.e., within the $X$ interval
$0<X<\xi_{\text{w}}D_{\alpha\beta}$, 
we obtain the inequality 
\begin{align}
   0<Y<X+1
    +\left(\xi_{\text{w}}-1\right)
     \dfrac{X}{\left(X-\xi_{\text{w}}D_{\alpha\beta}\right)}
     \left(X+1\right)
    +\xi_{\text{w}}
     \left(\xi_{f}-1\right)
     \left(1-D_{\alpha\beta}\right)
     \dfrac{X}{\left(X-\xi_{\text{w}}D_{\alpha\beta}\right)} \, .
\end{align}

In summary, the conditions for wetting of the wall--$\alpha$ interface by 
a film of the $\beta$ phase are given by
\begin{align*}
   0<X<\xi_{\text{w}}D_{\alpha\beta}\,\,\,\mathrm{and}\,\,\, 
   0<Y<X+1
   +\left(\xi_{\text{w}}-1\right)
           \dfrac{X}{\left(X-\xi_{\text{w}}
           D_{\alpha\beta}\right)}\left(X+1\right)
   +\xi_{\text{w}}\left(\xi_{f}-1\right)
     \left(1-D_{\alpha\beta}\right)
           \dfrac{X}{\left(X-\xi_{\text{w}}
           D_{\alpha\beta}\right)}
\end{align*}
or if
\begin{align*}
   X>\xi_{\text{w}}D_{\alpha\beta}\,\,\,\mathrm{and}\,\,\, 
   Y>X+1
   +\left(\xi_{\text{w}}-1\right)
           \dfrac{X}{\left(X-\xi_{\text{w}}
           D_{\alpha\beta}\right)}\left(X+1\right)
   +\xi_{\text{w}}\left(\xi_{f}-1\right)
          \left(1-D_{\alpha\beta}\right)
          \dfrac{X}{\left(X-\xi_{\text{w}}
          D_{\alpha\beta}\right)} \, ,
\end{align*}
which coincide with
Eqs.~(\ref{eq:s_wba_rng1}) and~(\ref{eq:s_wba_rng2}).

Now we determine the domains in the ($X$,$Y$) parameter space 
within which the conditions $S_{\text{w}\beta\gamma}<0$ and 
$S_{\text{w}\alpha\gamma}<0$ are valid.
By using the same notation as introduced above we can rewrite
$S_{\text{w}\beta\gamma}$ and $S_{\text{w}\alpha\gamma}$ 
as
\begin{align}
\notag
      S_{\text{w}\beta\gamma}
      =\epsilon_{AA}\rho_{A,\beta}^{2}
       \bigg\{&
         \left[
          \left(1-\dfrac{\rho_{B,\gamma}}{\rho_{B,\beta}}\right)X
         +\left(1-\dfrac{\rho_{A,\gamma}}{\rho_{A,\beta}}\right)
         \right]\left(X+1\right)
      +\left[
         \left(1-\dfrac{\rho_{B,\gamma}}{\rho_{B,\beta}}\right)
        +\left(1-\dfrac{\rho_{A,\gamma}}{\rho_{A,\beta}}\right)
        \right]\left(\xi_{f}-1\right)X\\
      &\,-\left[
         \left(1-\dfrac{\rho_{B,\gamma}}{\rho_{B,\beta}}\right)
         \dfrac{X}{\xi_{\text{w}}}
        +\left(1-\dfrac{\rho_{A,\gamma}}{\rho_{A,\beta}}\right)
        \right]Y
        \bigg\} 
\label{eq:s_wbg2_app}
\end{align}
and
\begin{align}
\notag
      S_{\text{w}\alpha\gamma}
      =\epsilon_{AA}\rho_{A,\beta}^{2}
        \bigg\{&
         \left[
          \left(\dfrac{\rho_{B,\alpha}}{\rho_{B,\beta}}
               -\dfrac{\rho_{B,\gamma}}{\rho_{B,\beta}}\right)X
         +\left(\dfrac{\rho_{A,\alpha}}{\rho_{A,\beta}}
               -\dfrac{\rho_{A,\gamma}}{\rho_{A,\beta}}\right)
         \right]
         \left(\dfrac{\rho_{B,\alpha}}{\rho_{B,\beta}}X
              +\dfrac{\rho_{A,\alpha}}{\rho_{A,\beta}}\right)\\
\notag
      &\,+\left[
         \left(\dfrac{\rho_{B,\alpha}}{\rho_{B,\beta}}
              -\dfrac{\rho_{B,\gamma}}{\rho_{B,\beta}}\right)
         \dfrac{\rho_{A,\alpha}}{\rho_{A,\beta}}
        +\left(\dfrac{\rho_{A,\alpha}}{\rho_{A,\beta}}
              -\dfrac{\rho_{A,\gamma}}{\rho_{A,\beta}}\right)
         \dfrac{\rho_{B,\alpha}}{\rho_{B,\beta}}
        \right]\left(\xi_{f}-1\right)X\\
      &\,-\left[
         \left(\dfrac{\rho_{B,\alpha}}{\rho_{B,\beta}}
              -\dfrac{\rho_{B,\gamma}}{\rho_{B,\beta}}\right)
         \dfrac{X}{\xi_{\text{w}}}
        +\left(\dfrac{\rho_{A,\alpha}}{\rho_{A,\beta}}
              -\dfrac{\rho_{A,\gamma}}{\rho_{A,\beta}}\right)
        \right]Y
        \bigg\} \, .
\label{eq:s_wag2_app}
\end{align}
 
The condition $S_{\text{w}\beta\gamma}<0$ leads to the inequality
\begin{align}
\notag
      \left[
         \left(1-\dfrac{\rho_{B,\gamma}}{\rho_{B,\beta}}\right)
         \dfrac{X}{\xi_{\text{w}}}
        +\left(1-\dfrac{\rho_{A,\gamma}}{\rho_{A,\beta}}\right)
        \right]Y>
        &\left[
          \left(1-\dfrac{\rho_{B,\gamma}}{\rho_{B,\beta}}\right)X
         +\left(1-\dfrac{\rho_{A,\gamma}}{\rho_{A,\beta}}\right)
         \right]\left(X+1\right)\\
      &\,+\left[
         \left(1-\dfrac{\rho_{B,\gamma}}{\rho_{B,\beta}}\right)
        +\left(1-\dfrac{\rho_{A,\gamma}}{\rho_{A,\beta}}\right)
        \right]\left(\xi_{f}-1\right)X \, .
\label{eq:s_wbg3_app}
\end{align}
Analogously, the condition $S_{\text{w}\alpha\gamma}<0$ 
leads to the inequality
\begin{align}
\notag
      \bigg[
         \left(\dfrac{\rho_{B,\alpha}}{\rho_{B,\beta}}
              -\dfrac{\rho_{B,\gamma}}{\rho_{B,\beta}}\right)
         \dfrac{X}{\xi_{\text{w}}}
        +\left(\dfrac{\rho_{A,\alpha}}{\rho_{A,\beta}}
              -\dfrac{\rho_{A,\gamma}}{\rho_{A,\beta}}\right)
        \bigg]Y>
         &\left[
          \left(\dfrac{\rho_{B,\alpha}}{\rho_{B,\beta}}
               -\dfrac{\rho_{B,\gamma}}{\rho_{B,\beta}}\right)X
         +\left(\dfrac{\rho_{A,\alpha}}{\rho_{A,\beta}}
               -\dfrac{\rho_{A,\gamma}}{\rho_{A,\beta}}\right)
         \right]
         \left(\dfrac{\rho_{B,\alpha}}{\rho_{B,\beta}}X
              +\dfrac{\rho_{A,\alpha}}{\rho_{A,\beta}}\right)\\
      &\,+\left[
         \left(\dfrac{\rho_{B,\alpha}}{\rho_{B,\beta}}
              -\dfrac{\rho_{B,\gamma}}{\rho_{B,\beta}}\right)
         \dfrac{\rho_{A,\alpha}}{\rho_{A,\beta}}
        +\left(\dfrac{\rho_{A,\alpha}}{\rho_{A,\beta}}
              -\dfrac{\rho_{A,\gamma}}{\rho_{A,\beta}}\right)
         \dfrac{\rho_{B,\alpha}}{\rho_{B,\beta}}
        \right]\left(\xi_{f}-1\right)X \, .
\label{eq:s_wag3_app}
\end{align}
In both inequalities 
(Eqs.~(\ref{eq:s_wbg3_app}) and~(\ref{eq:s_wag3_app})), 
the prefactors of $Y$ are positive for all $X>0$, due to the
assumed inequalities between the number densities (Eq.~(\ref{eq:rel_densities})).
Therefore one can reformulate
Eqs.~(\ref{eq:s_wbg3_app}) and~(\ref{eq:s_wag3_app}) 
as
\begin{align}
   Y>X+1
    +\dfrac{\left(1-\dfrac{\rho_{B,\gamma}}{\rho_{B,\beta}}\right)
      \left(1-\dfrac{1}{\xi_{\text{w}}}\right)X}
     {\left[\left(1-\dfrac{\rho_{B,\gamma}}{\rho_{B,\beta}}\right)
     \dfrac{X}{\xi_{\text{w}}}
     +\left(1-\dfrac{\rho_{A,\gamma}}{\rho_{A,\beta}}\right)\right]}
     \left(X+1\right)
     +\dfrac{\left[\left(1-\dfrac{\rho_{B,\gamma}}{\rho_{B,\beta}}\right)
     +\left(1-\dfrac{\rho_{A,\gamma}}{\rho_{A,\beta}}\right)\right]
      \left(\xi_{f}-1\right)}
     {\left[\left(1-\dfrac{\rho_{B,\gamma}}{\rho_{B,\beta}}\right)
     \dfrac{X}{\xi_{\text{w}}}+
     \left(1-\dfrac{\rho_{A,\gamma}}{\rho_{A,\beta}}\right)
     \right]}X
\label{eq:s_wbg4_app}
\end{align}
and 
\begin{align}
\notag
   Y>&\,\dfrac{\rho_{B,\alpha}}{\rho_{B,\beta}}X+
     \dfrac{\rho_{A,\alpha}}{\rho_{A,\beta}}
    +\dfrac{\left(\dfrac{\rho_{B,\alpha}}{\rho_{B,\beta}}
                  -\dfrac{\rho_{B,\gamma}}{\rho_{B,\beta}}\right)
      \left(1-\dfrac{1}{\xi_{\text{w}}}\right)X}
     {\left[\left(\dfrac{\rho_{B,\alpha}}{\rho_{B,\beta}}
                 -\dfrac{\rho_{B,\gamma}}{\rho_{B,\beta}}\right)
     \dfrac{X}{\xi_{\text{w}}}
     +\left(\dfrac{\rho_{A,\alpha}}{\rho_{A,\beta}}
           -\dfrac{\rho_{A,\gamma}}{\rho_{A,\beta}}\right)\right]}
     \left(\dfrac{\rho_{B,\alpha}}{\rho_{B,\beta}}X+
           \dfrac{\rho_{A,\alpha}}{\rho_{A,\beta}}\right)\\
    &\,+\,\dfrac{\left[\left(\dfrac{\rho_{B,\alpha}}{\rho_{B,\beta}}
                        -\dfrac{\rho_{B,\gamma}}{\rho_{B,\beta}}
                   \right)
                   \dfrac{\rho_{A,\alpha}}{\rho_{A,\beta}}
                  +\left(\dfrac{\rho_{A,\alpha}}{\rho_{A,\beta}}
                        -\dfrac{\rho_{A,\gamma}}{\rho_{A,\beta}}
                   \right)
                   \dfrac{\rho_{B,\alpha}}{\rho_{B,\beta}}\right]
      \left(\xi_{f}-1\right)}
     {\left[\left(\dfrac{\rho_{B,\alpha}}{\rho_{B,\beta}}
                 -\dfrac{\rho_{B,\gamma}}{\rho_{B,\beta}}\right)
     \dfrac{X}{\xi_{\text{w}}}+
     \left(\dfrac{\rho_{A,\alpha}}{\rho_{A,\beta}}
          -\dfrac{\rho_{A,\gamma}}{\rho_{A,\beta}}\right)
     \right]}X \, .
\label{eq:s_wag4_app}
\end{align}
By introducing the expressions for 
$D_{\beta\gamma}$ and $D_{\alpha\gamma}$,
Eqs.~(\ref{eq:s_wbg4_app}) and~(\ref{eq:s_wag4_app}) 
render the conditions expressed via
Eqs.~(\ref{eq:s_wbg_rng}) and~(\ref{eq:s_wag_rng}), 
respectively.   

\section{\label{sec:equil_wet_film_thick}
         Equilibrium wetting film thicknesses
}
In this appendix, we investigate the equilibrium wetting film thicknesses 
at fluid--fluid or wall--fluid interfaces. 
$\Omega_{\text{s}}(l)$ attains its minimum at the equilibrium wetting film thickness.

First, we consider the case of a planar $\alpha$--$\gamma$ interface 
with an intruding $\beta$ wetting film.
From Eqs.~(\ref{eq:omega_surf}) and~(\ref{eq:equil_cond2}), 
we find
\begin{equation}
   \dfrac{\partial{\Omega_{\text{s}}
                         ^{\alpha\beta\gamma}(l)}}
         {\partial{l}}
   \bigg{|}_{l=l_{\alpha\beta\gamma}}
   =\Omega^{\beta}-\Omega^{\gamma}
   +\dfrac{\partial{\omega_{\alpha\beta\gamma}(l)}}
          {\partial{l}}
    \bigg{|}_{l=l_{\alpha\beta\gamma}}=0 \, .
   \label{eq:thick_abg1_app}
\end{equation}
Here, $\Omega^{\beta}-\Omega^{\gamma}$ is positive,
given that the $\alpha$ phase is the stable phase and 
the $\beta$ phase is slightly off coexistence.
Therefore $\dfrac{\partial{\omega_{\alpha\beta\gamma}(l)}}
{\partial{l}}$ must be negative at $l=l_{\alpha\beta\gamma}$.
By using Eq.~(\ref{eq:omega_abg2}) and 
taking all length parameters to be equal, 
one obtains
\begin{equation}
   \dfrac{\partial{\omega_{\alpha\beta\gamma}(l)}}
         {\partial{l}}
   =\pi{a}_{AA}^{3}S_{\alpha\beta\gamma}
    \left[\dfrac{2}{3}
          \left(\dfrac{a_{AA}}{l}\right)^{3}
         -\dfrac{12}{5}
          \left(\dfrac{a_{AA}}{l}\right)^{4}
    \right] \, .
   \label{eq:thick_abg2_app}
\end{equation}
We have to consider only the case that $S_{\alpha\beta\gamma}$ is
negative, because only then a sufficiently thick $\beta$ wetting film
can occur.
This implies that the expression in square brackets in Eq.~(\ref{eq:thick_abg2_app}) must be positive. 
As a result, $l_{\alpha\beta\gamma}$ is definitely larger than $\dfrac{18}{5}a_{AA}$.

By inserting Eq.~(\ref{eq:thick_abg2_app})
into Eq.~(\ref{eq:thick_abg1_app}),
one finds that it is possible that 
Eq.~(\ref{eq:thick_abg1_app}) has 
no solution, only one solution, or two solutions, 
depending on the magnitude of $\Omega^{\beta}-\Omega^{\gamma}$.
If Eq.~(\ref{eq:thick_abg1_app}) has no solution or one solution,
$\Omega_{\text{s}}^{\alpha\beta\gamma}(l)$
has a minimum at $l\rightarrow0$, i.e., there is no $\beta$ wetting film.

If Eq.~(\ref{eq:thick_abg1_app}) has two solutions
$l_{1}$ and $l_{2}\,>l_{1}$, 
one knows that $\dfrac{18}{5}a_{AA}<l_{1}<l_{2}$ (see above).
In order to find out which of the two solutions corresponds to a minimum or rather to a 
maximum, we explore the sign of  
$\dfrac{\partial{\Omega_{\text{s}}
                       ^{\alpha\beta\gamma}(l)}}
{\partial{l}}$ near $l_{1}$ and $l_{2}$.
\begin{itemize}
\item If $l<l_{1}$, one has
      $\dfrac{\partial{\Omega_{\text{s}}
                                     ^{\alpha\beta\gamma}(l)}}
                     {\partial{l}}>0$,
      i.e., the slope of  
      $\Omega_{\text{s}}^{\alpha\beta\gamma}(l)$
      is positive for $l<l_{1}$.  
\item If $l_{1}<l<l_{2}$,  
      $\dfrac{\partial{\Omega_{\text{s}}
                                     ^{\alpha\beta\gamma}(l)}}
                     {\partial{l}}<0$,
      i.e., the slope of 
      $\Omega_{\text{s}}^{\alpha\beta\gamma}(l)$
      is negative for $l_{1}<l<l_{2}$.  
\item If $l>l_{2}$,  
      $\dfrac{\partial{\Omega_{\text{s}}
                                     ^{\alpha\beta\gamma}(l)}}
                     {\partial{l}}>0$,
      i.e., the slope of 
      $\Omega_{\text{s}}^{\alpha\beta\gamma}(l)$
      is again positive for $l>l_{2}$. 
\end{itemize}
Thus, $\Omega_{\text{s}}^{\alpha\beta\gamma}(l)$ has 
its maximum at $l=l_{1}$ and 
its minimum at $l=l_{2}$.
Therefore the equilibrium
film thickness is given by $l_{2} =: l_{\alpha\beta\gamma}$.
At three-phase coexistence, we have $\Omega^{\beta}-\Omega^{\gamma}\rightarrow0$
so that $l_{1}\rightarrow\dfrac{18}{5}a_{AA}$ and $l_{2}\rightarrow\infty$,
i.e., $l_{\alpha\beta\gamma}\rightarrow\infty$.

Analogously the equilibrium thickness of the $\beta$ wetting film
at a planar wall--$\alpha$ interface is determined. 
Based on Eqs.~(\ref{eq:equil_wba_app}) and~(\ref{eq:omega_surf_wba_app}),
one finds 
\begin{equation*}
   \dfrac{\partial{\Omega_{\text{s}}
                         ^{\text{w}\beta\alpha}(l)}}
         {\partial{l}}
   \bigg{|}_{l=l_{\text{w}\beta\alpha}}
   =\Omega^{\beta}-\Omega^{\alpha}
   +\dfrac{\partial{\omega_{\text{w}\beta\alpha}(l)}}
          {\partial{l}}
    \bigg{|}_{l=l_{\text{w}\beta\alpha}}=0 \, .
\end{equation*}
Here, $\Omega^{\beta}-\Omega^{\alpha}$ is positive
and therefore
$\dfrac{\partial{\omega_{\text{w}\beta\alpha}(l)}}
       {\partial{l}}$ must be negative at 
$l=l_{\text{w}\beta\alpha}$.
Using Eq.~(\ref{eq:omega_wba2_app}) with equal length parameters
for all interactions, 
we find
\begin{equation*}
   \dfrac{\partial{\omega_{\text{w}\beta\alpha}(l)}}
         {\partial{l}}
   =\pi{a}_{AA}^{3}S_{\text{w}\beta\alpha}
    \left[\dfrac{2}{3}
          \left(\dfrac{a_{AA}}{l}\right)^{3}
         -\dfrac{12}{5}
          \left(\dfrac{a_{AA}}{l}\right)^{4}
    \right] \, .
\end{equation*}
Again, we have to consider only $S_{\text{w}\beta\alpha}<0$
and as a result we have 
$l_{\text{w}\beta\alpha}>\dfrac{18}{5}a_{AA}$.
At three-phase coexistence, 
i.e., $\Omega^{\beta}-\Omega^{\alpha}\rightarrow0$,
again we have $l_{\text{w}\beta\alpha}\rightarrow\infty$.

Based on Eqs.~(\ref{eq:equil_wbg_app}) and~(\ref{eq:omega_surf_wbg_app}),
in the case of a planar wall--$\gamma$ interface wetted by a
film of the $\beta$ phase one has 
\begin{equation*}
   \dfrac{\partial{\Omega_{\text{s}}
                         ^{\text{w}\beta\gamma}(l)}}
         {\partial{l}}
   \bigg{|}_{l=l_{\text{w}\beta\gamma}}
   =\Omega^{\beta}-\Omega^{\gamma}
   +\dfrac{\partial{\omega_{\text{w}\beta\gamma}(l)}}
          {\partial{l}}
    \bigg{|}_{l=l_{\text{w}\beta\gamma}}=0 \, .
\end{equation*}
Using Eq.~(\ref{eq:omega_wbg2_app}) and choosing
all length parameters to be equal,
we have
\begin{equation*}
   \dfrac{\partial{\omega_{\text{w}\beta\gamma}(l)}}
         {\partial{l}}
   =\pi{a}_{AA}^{3}S_{\text{w}\beta\gamma}
    \left[\dfrac{2}{3}
          \left(\dfrac{a_{AA}}{l}\right)^{3}
         -\dfrac{12}{5}
          \left(\dfrac{a_{AA}}{l}\right)^{4}
    \right] \, .
\end{equation*}
For $\Omega^{\beta}-\Omega^{\gamma}>0$ and 
$S_{\text{w}\beta\gamma}<0$ we find again
$l_{\text{w}\beta\gamma}>\dfrac{18}{5}a_{AA}$
and $l_{\text{w}\beta\gamma}\rightarrow\infty$ at
three-phase coexistence.

For the planar wall--$\gamma$ interface 
with an intruding $\alpha$ wetting film, it follows
from Eqs.~(\ref{eq:equil_wag_app}),
~(\ref{eq:omega_surf_wag_app}), 
and~(\ref{eq:omega_wag2_app}) 
that
\begin{equation}
   \dfrac{\partial{\Omega_{\text{s}}
                         ^{\text{w}\alpha\gamma}(l)}}
         {\partial{l}}
   \bigg{|}_{l=l_{\text{w}\alpha\gamma}}
   =\Omega^{\alpha}-\Omega^{\gamma}
   +\dfrac{\partial{\omega_{\text{w}\alpha\gamma}(l)}}
          {\partial{l}}
    \bigg{|}_{l=l_{\text{w}\alpha\gamma}}=0
   \label{eq:thick_wag1_app}
\end{equation}
where
\begin{equation}
   \dfrac{\partial{\omega_{\text{w}\alpha\gamma}(l)}}
         {\partial{l}}
   =\pi{a}_{AA}^{3}S_{\text{w}\alpha\gamma}
    \left[\dfrac{2}{3}
          \left(\dfrac{a_{AA}}{l}\right)^{3}
         -\dfrac{12}{5}
          \left(\dfrac{a_{AA}}{l}\right)^{4}
    \right] \, .
   \label{eq:thick_wag2_app}
\end{equation}
Again, all length parameters are taken to be equal. 
Here, $\Omega^{\alpha}$ and $\Omega^{\gamma}$ are equal
because the system is at $\alpha$--$\gamma$ coexistence. 
With $S_{\text{w}\alpha\gamma}<0$, the equilibrium wetting film thickness
$l_{\text{w}\alpha\gamma}\rightarrow\infty$ is found.
 
\section{\label{sec:cond_for_sigma_wa_sigma_wg}
                             Condition for 
                             $\sigma_{\text{w}\alpha}
                             <\sigma_{\text{w}\gamma}$}
In this appendix, we determine the domain in the 
($X$,$Y$) parameter space for which 
$\sigma_{\text{w}\alpha}<\sigma_{\text{w}\gamma}$ is satisfied. 
By using Eqs.~(\ref{eq:sigma_wa2_app}) and~(\ref{eq:sigma_wg2_app}),
we find
\begin{equation}
   \sigma_{\text{w}\alpha}-\sigma_{\text{w}\gamma}
     =\dfrac{13}{132}\pi
      \left[\sum_{i,j}\epsilon_{ij}a_{ij}^{4}
                      \left(\rho_{i,\alpha}\rho_{j,\alpha}
                           -\rho_{i,\gamma}\rho_{j,\gamma}\right)
           -2\sum_{i}\epsilon_{\text{w}i}
                     a_{\text{w}i}^{4}
                     \left(\rho_{i,\alpha}-\rho_{i,\gamma}\right)
                     \rho_{\text{w}}\right]<0 \, .
   \label{eq:sigma_wa_wg_app}
\end{equation}

Assuming that all length parameters in Eq.~(\ref{eq:sigma_wa_wg_app})
are equal, i.e., $a_{AA}=a_{AB}=a_{BB}=a_{\text{w}A}=a_{\text{w}B}$,
we find
\begin{align}
\notag
       &\left(\rho_{A,\alpha}-\rho_{A,\gamma}\right)
        \left[\left(\rho_{A,\alpha}+\rho_{A,\gamma}\right)
              \epsilon_{AA}
             -2\epsilon_{\text{w}A}\rho_{\text{w}}\right]
       +\left(\rho_{A,\alpha}\rho_{B,\alpha}
             -\rho_{A,\gamma}\rho_{B,\gamma}\right)
        \left(2\epsilon_{AB}\right)\\
       &+\left(\rho_{B,\alpha}-\rho_{B,\gamma}\right)
         \left[\left(\rho_{B,\alpha}+\rho_{B,\gamma}\right)
               \epsilon_{BB}
              -2\epsilon_{\text{w}B}\rho_{\text{w}}\right]<0 \, .
   \label{eq:sigma_wa_wg2_app}
\end{align}

By introducing the dimensionless parameters $X$ and $Y$, 
Eq.~(\ref{eq:sigma_wa_wg2_app}) can be written as
\begin{align}
\notag
       \left[\left(\dfrac{\rho_{B,\alpha}}{\rho_{B,\beta}}
             -\dfrac{\rho_{B,\gamma}}{\rho_{B,\beta}}\right)X
       +\dfrac{\rho_{A,\alpha}}{\rho_{A,\beta}}
             -\dfrac{\rho_{A,\gamma}}{\rho_{A,\beta}}
       \right]
       \left[\left(\dfrac{\rho_{B,\alpha}}{\rho_{B,\beta}}
             +\dfrac{\rho_{B,\gamma}}{\rho_{B,\beta}}\right)X
       +\dfrac{\rho_{A,\alpha}}{\rho_{A,\beta}}
       +\dfrac{\rho_{A,\gamma}}{\rho_{A,\beta}}
       -2Y\right]<0.
   \label{eq:sigma_wa_wg3_app}
\end{align}
Here one has
$\left(\dfrac{\rho_{B,\alpha}}{\rho_{B,\beta}}
      -\dfrac{\rho_{B,\gamma}}{\rho_{B,\beta}}\right)X
+\dfrac{\rho_{A,\alpha}}{\rho_{A,\beta}}
      -\dfrac{\rho_{A,\gamma}}{\rho_{A,\beta}}>0$ 
for all $X>0$.

In conclusion,
$\sigma_{\text{w}\alpha}<\sigma_{\text{w}\gamma}$ 
is satisfied if 
\begin{equation*}
   Y>\dfrac{1}{2}
      \left[\left(\dfrac{\rho_{B,\alpha}}
                        {\rho_{B,\beta}}
                 +\dfrac{\rho_{B,\gamma}}
                        {\rho_{B,\beta}}\right)X
           +\dfrac{\rho_{A,\alpha}}
                        {\rho_{A,\beta}}
                 +\dfrac{\rho_{A,\gamma}}
                        {\rho_{A,\beta}}
      \right] \, .
\end{equation*}
Otherwise, we have 
$\sigma_{\text{w}\alpha}>\sigma_{\text{w}\gamma}$.
\end{widetext}

\end{document}